\documentclass[12pt, twoside]{article}
\usepackage[utf8]{inputenc} 
\usepackage[ text={160mm,225mm},centering]{geometry}
\usepackage{amstext,amsmath,latexsym,amsbsy,amssymb}
\usepackage{url}
\usepackage[colorlinks=true,linkcolor=blue,citecolor=blue,urlcolor=blue, backref=page]{hyperref}
\usepackage{amsfonts}
\usepackage[final]{graphicx}
\usepackage{amsthm}
\usepackage{mathtools}
\usepackage{graphics}
\usepackage{psfrag}
\usepackage{times}

\usepackage[titletoc]{appendix}
\usepackage{multirow}
\usepackage{hhline}
\usepackage{bbm}
\usepackage{color}
\usepackage{bigints}
\usepackage{caption}
\usepackage{subcaption}
\usepackage{booktabs}
\usepackage{float}

\usepackage{tikz}
\usepackage{enumitem}
\usepackage{bbm}
\usepackage{dsfont}
\usetikzlibrary{arrows.meta,positioning}
\usepackage[ruled,vlined]{algorithm2e}
\usepackage{setspace}

\SetKwInOut{Input}{Input}
\SetKwInOut{Output}{Output}
\SetKwFunction{FMap}{Map}
\SetKwFunction{FReduce}{Reduce}
\SetKwProg{Fn}{Function}{:}{}
\SetKw{KwGlobal}{Global} 

\RequirePackage{natbib}
\long\def\comment#1{}
\definecolor{battleshipgrey}{rgb}{0.52, 0.52, 0.51}
\definecolor{darkgray}{rgb}{0.66, 0.66, 0.66}
\definecolor{darkgreen}{rgb}{0.0, 0.2, 0.13}
\definecolor{darkspringgreen}{rgb}{0.09, 0.45, 0.27}
\definecolor{dukeblue}{rgb}{0.0, 0.0, 0.61}
\definecolor{olivedrab7}{rgb}{0.24, 0.2, 0.12}
\definecolor{darkblue}{rgb}{0.0, 0.0, 0.55}
\definecolor{darkscarlet}{rgb}{0.34, 0.01, 0.1}
\definecolor{candyapplered}{rgb}{1.0, 0.03, 0.0}
\definecolor{ao(english)}{rgb}{0.0, 0.5, 0.0}
\definecolor{applegreen}{rgb}{0.55, 0.71, 0.0}

\comment{
\setlength{\topmargin}{0 in}
\setlength{\textwidth}{5.5 in}
\setlength{\textheight}{8 in}
\setlength{\oddsidemargin}{0.5 in}
}

\newif\ifhidetext
\hidetextfalse 

\newcommand{\hidetext}[1]{%
  \ifhidetext
  \else
    #1
  \fi
}

\theoremstyle{plain}

\newtheorem*{lemma*}{Lemma}

\newtheorem{theorem}{Theorem}
\newtheorem*{theorem*}{Theorem}
\newtheorem*{remark*}{Remark}
\numberwithin{theorem}{section}

\numberwithin{proposition}{section}

\newtheorem{lemma}{Lemma}

\numberwithin{lemma}{section}

\newtheorem{definition}{Definition}

\numberwithin{definition}{section}

\numberwithin{condition}{section}

\numberwithin{problem}{section}

\numberwithin{corollary}{section}

\numberwithin{assumption}{section}

\numberwithin{example}{section}

\numberwithin{conjecture}{section}

\theoremstyle{definition}

\numberwithin{remark}{section}

\allowdisplaybreaks

\renewenvironment{abstract}
 {\small
  \begin{center}
  \bfseries \abstractname\vspace{-.5em}\vspace{0pt}
  \end{center}
  \list{}{%
    \setlength{\leftmargin}{15mm}
    \setlength{\rightmargin}{\leftmargin}%
  }%
  \item\relax}
 {\endlist}
 
\begin{document}

\title{\bf A Bayesian approach to learning mixtures of nonparametric components }
  \author{
    \large
    Yilei Zhang$^{1}$,\quad
    Yun Wei$^{2,}$\thanks{corresponding authors},\quad
    Aritra Guha$^{3}$,\quad
    XuanLong Nguyen$^{1, *}$ \\[0.6cm]
    \large
    $^{1}$ Department of Statistics, University of Michigan--Ann Arbor\\
    $^{2}$ Department of Mathematical Science, University of Texas at Dallas\\
    $^{3}$ AT\&T Data Science and AI Research\\[0.6cm]
    }
  \date{}
  \maketitle
\smallskip
\begin{abstract}
\begin{spacing}{1.2}
Mixture models are widely used in modeling heterogeneous data populations. A standard approach of mixture modeling assumes that the mixture component takes a parametric kernel form. In many applications, making parametric assumptions on the latent subpopulation distributions may be unrealistic, which motivates the need for nonparametric modeling of the mixture components themselves. In this paper, we study finite mixtures with nonparametric mixture components, using a Bayesian nonparametric modeling approach. In particular, it is assumed that the data population is generated according to a finite mixture of latent component distributions, where each component is endowed with a Bayesian nonparametric prior such as the Dirichlet process mixture. We present conditions under which the individual mixture component's distribution can be identified, and establish posterior contraction behavior for the data population's density, as well as densities of the latent mixture components. We develop an efficient MCMC algorithm for posterior inference and demonstrate via simulation studies and real-world data illustrations that it is possible to efficiently learn complex distributions for the latent subpopulations. In theory, the posterior contraction rate of the component densities is nearly polynomial, which is a significant improvement over the logarithmic convergence rates of estimating mixing measures via deconvolution. 
\end{spacing}

\end{abstract}

\noindent%
\bigskip

\newpage
\begin{spacing}{0.9}
\tableofcontents
\end{spacing}

\begin{spacing}{1.2}
\section{Introduction}
\label{Section:introduction}

Large heterogeneous datasets often contain multiple underlying subpopulations of interest. Accurately identifying and characterizing them is a fundamental yet difficult task. In practice, these subpopulation labels are typically unavailable---either because they are costly to obtain or because they are inherently unobservable---so the task falls within the domain of unsupervised learning. Clustering methods that focus on identifying subpopulations typically learn a partition of the data domain, aiming to group samples from the same subpopulation into the same cluster. However, they usually do not provide a distributional characterization for each subpopulation. Examples include K-means clustering \citep{mcqueen1967some, lloyd1982least}, hierarchical clustering \citep{hartigan1981consistency, thomann2015towards}, mode clustering \citep{chacon2013data, chen2016comprehensive}, spectral clustering \citep{schiebinger2015geometry}, and density-based clustering \citep{rinaldo2010generalized}. In comparison, model-based clustering explicitly characterizes the underlying distribution of each subpopulation and represents the overall population as a mixture—a weighted sum—of these subpopulation distributions (components), thereby yielding a generative probabilistic model for each subpopulation. Typically, each component of a mixture model is assumed to be from a parametric family, and a popular example is the Gaussian mixture model (GMM). When the parametric family satisfies identifiability conditions and the model is well-specified, meaning that the true subpopulation distributions indeed belong to the specified family, the latent structure can be consistently recovered and the convergence rates of the latent mixing measure have been established in a series of work \citep{chen1995optimal, long2013, ho2016convergence,heinrich2018strong, wei2022convergence,wei2023minimum}. However, if the model is misspecified, the relationship between the learned and true mixing measures becomes elusive \citep{guha2021posterior, miller2019robust}, and the learned mixing measure usually fails to convey meaningful information about the underlying latent structure. In practice, the parametric families are almost always misspecified, which therefore leaves a considerable gap in the practical applications of parametric mixture models.  

A variety of mixture models have been proposed to model each subpopulation with more flexible distribution families. Examples include the t-distribution \citep{peel2000robust, stephens2000bayesian,andrews2012model}, skew-t distribution \citep{lin2007robust}, skewed power-exponential distribution \citep{dang2023model}, shifted asymmetric Laplace distribution \citep{franczak2013mixtures}, and normal inverse Gaussian distribution \citep{karlis2009model}. \cite{levine2011maximum} assumes each component density factorizes as the product of its marginal densities, and \cite{rodriguez2014univariate} devises a family of nonparametric unimodal distributions. See also \cite{xu2025tree} and the references therein for nonparametric clustering. These models enrich the component family, enabling it to accommodate heavy-tailed behavior, asymmetry, and non-elliptical structures that Gaussian mixture models cannot capture. However, each of these distribution families is tailored to capture only certain types of patterns, such as heavy tails or skewness; no single parametric family is sufficiently flexible to accommodate all of the above patterns. Moreover, these approaches require all components to come from the same specified family, an assumption that may not hold in heterogeneous real-world data. Finally, switching between different component families generally demands completely different EM or MCMC algorithms, making model selection and component-family adjustment computationally costly and practically inconvenient. 

Instead of specifying a more flexible parametric family for each component, \cite{aragam2020identifiability} proposes fitting nonparametric components by merging components from an overfitted Gaussian mixture model, and establishes key identifiability conditions for nonparametric mixture models under this framework. However, this approach relies on a strong component-separability condition and doesn't perform well when components have overlapping tails. In other contexts, the general idea of merging components from an overfitted model has been utilized in estimating the number of components in finite mixture models \citep{manole2021estimating, guha2021posterior}, to learn hierarchical clustering structures \citep{do2024dendrogram}, and to mitigate model misspecification \citep{dombowsky2024bayesian}. \cite{bryon_2023} obtain theoretical convergence rates for nonparametric components in a mixture model by assuming that the support of the underlying mixing measure is fixed and known. However, their estimation method is not directly applicable in practice when the support is typically unknown. \cite{aragam2024model} employs Gaussian convolutions with varying scale parameters to uncover latent data structure without assuming a mixture-model form for the density. More recently, \cite{chakraborty_thesis} investigated the identifiability of a broad class of mixture models with nonparametric components, each defined as a convolution with a probability measure supported on a low-dimensional affine manifold, and developed methods for modeling a parametric subclass of these components. These approaches represent important steps toward uncovering clustering structures in data with complex subpopulations. However, they do not provide a systematic and practical framework for estimating the underlying component distributions in a fully nonparametric setting.

In this paper we develop a first and practical Bayesian method with theoretical guarantees to estimate the mixture models of nonparametric components. We focus on two classes of nonparametric mixture models: the first class is characterized by the assumption that each component has its probability mass spatially concentrated around a connected region, with distinct components concentrated in different regions. We allow overlap between component tails. Apart from this mild spatial separation condition, we do not impose any additional assumptions on the component distributions. The second class of models are spike-and-slab–type mixtures, in which one component exhibits one or more high-density “spikes”, while the other component has a flat, relatively low-density distribution over the entire support. The two components may share fully overlapping support. Moreover, we develop a unified framework that uses mixtures of Dirichlet process mixtures with Gaussian kernels to model both types of mixture structures. The hierarchical design leverages conjugacy to enable an efficient MCMC algorithm for fitting the posterior distributions. We apply our method to both univariate and multivariate settings through several simulation studies, and further demonstrate its scalability and practical performance in two real-data applications: disentangling astronomical sources in two-dimensional space (from approximately 0.8 million observed events) and analyzing the dynamic behavioral patterns of an oceanic whitetip shark using movement-metric data.

To enable identifiability of nonparametric mixture models while allowing for the overlap of the mixture components' support, we develop a new separation condition defined in terms of the distances between connected regions within the support of the latent mixing measure. 
We note that these settings are not covered by the geometric identifiability conditions in the modeling approach of \cite{chakraborty_thesis}.
Furthermore, we establish theoretical guarantees showing the posterior contraction rate of the estimated component densities under appropriate conditions. To our knowledge, this is perhaps the first theoretical guarantee for a practical Bayesian method that learns nonparametric component densities within a finite mixture modeling framework. 
In our work, the posterior contraction rate for the component densities is shown to be of the same minimax order of convergence as that of a point estimate given by \cite{bryon_2023}. Since the minimax order does not have an analytical form, we derive a closed-form upper bound for this rate, which shows that it is nearly polynomial in nature. 

The paper is organized as follows. Preliminary background 
is introduced in Section~\ref{section:preliminary}. Section~\ref{section:Problem_Setting} introduces the problem setting for the two classes of mixture models described above. In Section~\ref{section:Mixture_of_DP}, we present the framework of mixtures of Dirichlet process mixtures (MDPM) and illustrate our method through several simulation examples. Section~\ref{section:Multivariate} extends the method to multivariate distributions and provides an additional simulation study. In Section~\ref{section:Identifiability}, we introduce the separation conditions based on which the identifiability of nonparametric mixture components can be established. Section~\ref{section:posterior_contraction} presents the posterior contraction results for both the overall mixture density and the individual component densities under certain conditions. Finally, Section~\ref{section:applications} applies our method to the oceanic whitetip shark acceleration data and the XMM-Newton astronomical dataset.

\section{Preliminaries} 
\label{section:preliminary}
In this section, we recall basic notions of nonparametric mixture models, and introduce the definition of topological connectedness that will be useful in determining the identifiability of mixture models.

\paragraph*{Mixture models} 
Let $\mathcal{P}(\mathbb{R}^m)$ denote the set of Borel probability measures on $\mathbb{R}^m$. Suppose $F\in\mathcal{P}(\mathbb{R}^m)$. We say that $F$ admits a finite mixture representation if, for any Borel set $A\subseteq\mathbb{R}^m$, $F(A)$ takes the form:
\begin{equation}\label{eq:mixture_model}
F( A) = \sum _{i=1}^{K} w_{i} G_{i}( A) = \int G( A) d\Lambda( G),\ K\coloneq |\operatorname{supp}( \Lambda) |,\ ( w_{1} ,\cdots ,w_{K}) \in \Delta ^{K-1},
\end{equation}
where $G_i$'s are distinct probability measures from a subset $\widetilde{\mathcal{P}}\subseteq \mathcal{P}(\mathbb{R}^{m})$, and $\Lambda$ is a probability measure over $\widetilde{\mathcal{P}}$, supported on a finite set of $K$ component distributions $G_1,\cdots, G_K$. The choice of $G_1,\cdots, G_K$ and $w_1,\ldots,w_K$ do not depend on $A$. In this case, we refer to $F$ as a finite mixture model with $K$ components, where each $G_i$ represents a distinct subpopulation of $F$. If the component class $\widetilde{\mathcal{P}}$ is parametric, e.g. $m$-dimensional Gaussian distributions, then $F$ is a mixture model of parametric components. In this work, however, we focus on a more general setting where $K$ is finite and known, but $\widetilde{\mathcal{P}}$ is a nonparametric family. The first question that arises is the identifiability of the representation \eqref{eq:mixture_model}. Denote by $\mathcal{P}(\widetilde{\mathcal{P}})$ a set of probability measures on $\widetilde{\mathcal{P}}$. The classic definition of \cite{teicher1961identifiability} specifies the probability measure $F$ to be identifiable in $\mathcal{P}(\widetilde{\mathcal{P}})$ if for any $\Lambda, \Lambda'\in \mathcal{P}(\widetilde{\mathcal{P}})$ the condition  $F=\int Gd\Lambda ( G) =\int Gd\Lambda '( G)$  entails that $\Lambda=\Lambda'$. 

In this work, we study the conditions under which $F$ is identifiable, and develop a Bayesian method to estimate each component probability measure $G_i$, as well as its corresponding weight $w_i$ for $i=1,\cdots, K$, while we only have observations from $F$. A fundamental concept underlying our identifiability conditions is connectedness, therefore we first introduce the notion of \emph{connectedness} for a topological space $\mathbb{X}$ (cf. \cite{munkres2020topology}). 

\begin{definition}[\textbf{Connectedness}]\label{def:connectedness}
Let $\mathbb{X}$ be a topological space. A separation of $\mathbb{X}$ is a pair of disjoint nonempty open subsets $A$ and $B$ of $\mathbb{X}$ s.t. $\mathbb{X}=A\cup B$. $\mathbb{X}$ is connected if there does not exist a separation of $\mathbb{X}$, otherwise it is disconnected.
\end{definition}

The above definition makes precise the natural intuition that a topological space is connected if it is “all in one piece”, meaning it cannot be split into distinct parts that do not touch each other. We now recall two classical Gaussian mixture families on $\mathbb{R}$. These serve as building blocks for the mixtures with nonparametric components studied in this paper. Later we extend the definitions to the multivariate setting $\mathbb{R}^m$.

\paragraph*{Location mixture of normals} Let $g_{u,\sigma}$ stand for the normal density on $\mathbb{R}$ with mean $u$ and standard deviation $\sigma$:
$g_{u,\sigma }( x) =\frac{1}{\sqrt{2\pi } \sigma }\exp\left( -\frac{( x-u)^{2}}{2\sigma ^{2}}\right)$.
The density $p(x)$ is a location mixture of normals if
\begin{equation}\label{location-mixture}
p( x) =p_{\sigma,V}( x) \coloneq \int g_{u,\sigma }( x) dV( u)
\end{equation}
for some $\sigma>0$, where $V(\cdot)$ is a mixing probability distribution on $\mathbb{R}$. 

\paragraph*{Location-scale mixture of normals} $p(x)$ is a location-scale mixture of normals if
\begin{equation}\label{location-scale-mixture}
p( x) =p_{H}( x) \coloneq \int g_{u,\sigma }( x) dH( u,\sigma),
\end{equation}
where $H(\cdot,\cdot)$ is a mixing probability distribution on $\mathbb{R}\times (0,\infty)$.


\section{Problem and modeling assumptions}
\label{section:Problem_Setting}

In this work, we study finite mixture models with nonparametric component densities. For clarity, we begin with mixture models on $\mathbb{R}$. The multivariate case follows in subsequent sections. Our goal is to learn both the overall population distribution and the latent subpopulation distributions within the mixture model framework \eqref{eq:mixture_model}, assuming that the number of components $K$ is finite, fixed and known. To this end, we develop our methodology under two different settings:

\begin{enumerate}[label=(S\arabic*), ref=S\arabic*, series=setting]
    \item \label{setting1} The component distribution class $\widetilde{\mathcal{P}}$ consists of location mixtures of normals, as defined in \eqref{location-mixture}. The variance parameter $\sigma>0$ is fixed but possibly unknown, and the mixing distribution $V(u)$ is assumed to have bounded support. We further restrict $\mathcal{P}(\widetilde{\mathcal{P}})$ to probability measures supported on $K$ components satisfying the following \emph{separation} condition: for any two components $G_i=p_{\sigma, V_i}$ and $G_j=p_{\sigma,V_j}$, $i\neq j$, there exist disjoint bounded connected sets $I_i, I_j\subset \mathbb{R}$ (i.e., intervals on $\mathbb{R}$) such that $\operatorname{supp}(V_i)\subseteq I_i$, and $\operatorname{supp}(V_j)\subseteq I_j$.
    \item \label{setting2} The component distribution class $\widetilde{\mathcal{P}}$ consists of location-scale mixtures of normals, as defined in \eqref{location-scale-mixture}. The mixing distribution $H$ has bounded support on $(u,\sigma)$. Accordingly, $\mathcal{P}(\widetilde{\mathcal{P}})$ represents a set of probability measures supported on $K$ components satisfying the separation condition: there exists $x\in\{u,\sigma\}$, such that for any two components $G_i=p_{H_i}$, and $G_j=p_{H_j}$, $i\neq j$, we can find disjoint bounded connected sets $I_i, I_j\subset S_x$ with $\operatorname{supp}(\pi_x\#H_i )\subseteq I_i$, and $\operatorname{supp}(\pi_x\#H_j)\subseteq I_j$, where $S_u=\mathbb{R}$, $S_\sigma=(0,\infty)$. Here, $\pi_x\#H$ stands the $x$-marginal distribution of $H$, where $x\in\{u,\sigma\}$.
\end{enumerate}

 It is important to note that the assumptions in the two settings apply only to our models and are not imposed on the actual data-generating distribution. In particular, the true component densities do not have to be a location mixture or a location–scale mixture of Gaussians. In \eqref{setting1}, we assume that each component in the finite mixture model is a location mixture of normals. This family is already quite rich, since Gaussian kernel density estimators are themselves location mixture of normals, and it is known that Gaussian kernel density estimators can approximate any continuous density arbitrarily well as 
 $\sigma\to 0$ (\cite{parzen1962estimation}). By restricting the support of mixing distribution $V$ to be bounded and imposing a separation condition on components, \eqref{setting1} characterizes mixture models in which each component has a dominant support region where most of its mass is concentrated, while still allowing distinct components to overlap in their tails. \eqref{setting2}, on the other hand, generalizes the component distribution family $\widetilde{\mathcal{P}}$ to location-scale mixture of normals. In \eqref{setting2} the separation condition is imposed on either the $u$-marginal or the $\sigma$-marginal; thus, in addition to the spatially differentiated components, this class also accommodates spike-and-slab–type components that overlap in location but differ in scale: one component exhibits one or more high-density “spikes”, while the other has a relatively flat, low-density distribution over the entire support. The two components may have fully overlapping support. In this work, we develop a Bayesian framework for learning mixture models applicable to the general setting  \eqref{setting2} in Section \ref{section:Mixture_of_DP}. We establish theoretical results under setting \eqref{setting1}, which consists of \textup{(i)} an identifiability theory in Section \ref{section:Identifiability} and \textup{(ii)} posterior contraction rates of both the mixture density and the component densities in Section \ref{section:posterior_contraction}.


\section{Mixture of Dirichlet Process mixtures}
\label{section:Mixture_of_DP}

Dirichlet process (DP), introduced by \citet{Ferguson_1973,Ferguson_1974}, is a foundational prior for Bayesian nonparametric modeling and has become widely used due to its analytical tractability. To model data arising from multiple distinct yet related subpopulations, a rich class of structured extensions of the DP has been developed to accommodate different data-generating mechanisms. Notable examples include the hierarchical DP \citep{teh2006hierarchical}, nested DP \citep{rodriguez2008nested}, additive DP \citep{muller2004method}, and dependent DP \citep{maceachern2000dependent}, among many others. A unifying perspective on these constructions is provided by \citet{franzolini2025multivariate}. 

As we shall see, the hierarchical Bayesian modeling framework may be extended and adapted to incorporate the model setting \eqref{setting2}. In particular, we shall describe a formulation of a mixture of Dirichlet process mixtures (MDPM), an early version of which goes back to \cite{mix_DP_Antoniak_1974}. We begin by recalling the Dirichlet process mixture (DPM); see, e.g., \cite{ghosal2017fundamentals} and \cite{muller2015bayesian}. Consider the location-scale mixture of normals in \eqref{location-scale-mixture}: $p_{H}( x) \coloneq \int g_{u,\sigma }( x) dH( u,\sigma )$. If we place a Dirichlet process prior on \(H\), i.e., \(H \sim \operatorname{DP}(\alpha H_0)\) with base measure \(H_0\) on $\mathbb{R}\times(0,\infty)$ and concentration parameter \(\alpha>0\),
then the induced random density \(p_H\) is called a Dirichlet process mixture (DPM). 

\begin{figure}[htbp]
    \centering
    \captionsetup{width=1.0\linewidth}
    \includegraphics[width=0.7\linewidth]{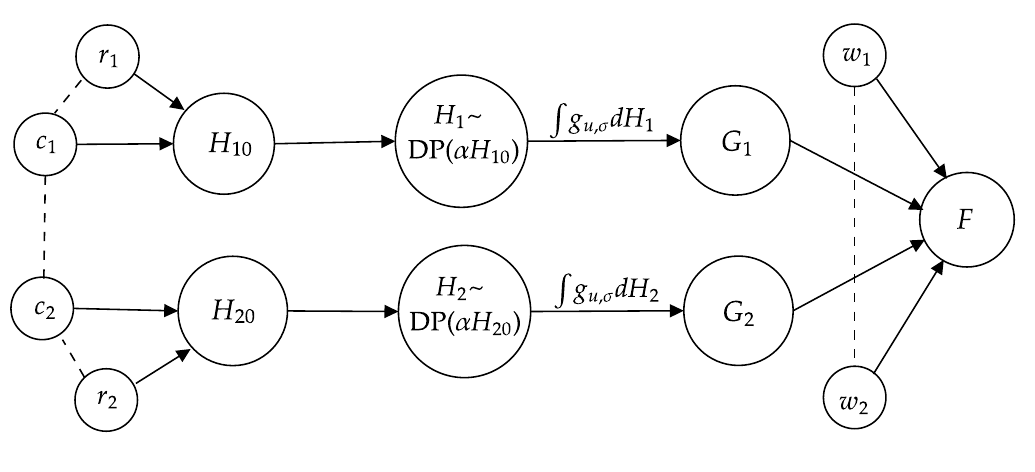}
    \caption{Graphical model representation of the MDPM \eqref{hierarchical_model} for $K=2$. The dashed edges among $c_1,r_1,c_2,r_2$ represent the repulsive prior on $(\boldsymbol{c},\boldsymbol{r})$, which enforces disjoint intervals $I_1$ and $I_2$. For $i=1,2$, the base measure $H_{i0}(u,\sigma)$ is supported on $I_i\times(0,\infty)$. The dashed edge between $w_1$ and $w_2$ indicates the truncated Beta prior (equivalently, the two-parameter truncated Dirichlet) on the mixture weights $\boldsymbol{w}=(w_1,w_2)$. Finally, the mixture weights $\boldsymbol{w}=(w_1,w_2)$ combine the component distributions into the overall mixture $F=w_1G_1+w_2G_2$.}
    \label{fig:diagram_two_comp}
\end{figure}

In setting \eqref{setting2}, we impose a separation condition either on $u$ or $\sigma$. For clarity, we begin with the condition on $u$. Since any nonempty bounded connected subset $I$ of $\mathbb{R}$ is an interval, we parametrize it as $[ c-r,c+r]$, where $c\in\mathbb{R}$ is the midpoint and $r>0$ is the half-length. We then place priors on $(c,r)$ to model the unknown intervals. By the separation condition in \eqref{setting2}, there exist $K$ pairwise disjoint intervals $I_1,\cdots, I_K$ such that, for each $i=1,\cdots, K$, the $i$-th component $G_i=p_{H_i}$ satisfies $\operatorname{supp}(\pi_u\# H_i)\subseteq I_i$. To ensure the $K$ intervals $I_1,\cdots, I_K$ are pairwise disjoint, we adapt the repulsive priors of \cite{repulsive_2012, repulsive_2017} to our setting and place the following prior on $(\boldsymbol{c} ,\boldsymbol{r})$ where $\boldsymbol{c} = ( c_{1} ,\cdots ,c_{K})$ and $\boldsymbol{r} =(r_{1} ,\cdots ,r_{K})$:
\begin{equation}\label{repulsive_prior}
    p_{\text{rp}}(\boldsymbol{c} ,\boldsymbol{r}) \propto \left(\prod _{i=1}^{K} g_{\mu_0,\eta}( c_{i})\cdot\mathrm{Gamma}( r_{i})\right)  \cdot \underset{1\leq i< j\leq K}{\min}\exp\left( -\frac{\tau }{\max( |c_{i} -c_{j} |-r_{i} -r_{j} ,0)^{\nu }}\right),
\end{equation}
where $\tau>0$ is a scale parameter and $\nu$ is a positive integer controlling the rate at which the exponential term approaches zero as $\max( |c_{i} -c_{j} |-r_{i} -r_{j} ,0)$ decreases, and $\eta>0$ controls the prior spread of the centers $\boldsymbol{c}$. We omit the parameters of the Gamma distribution for notational convenience. The repulsion term is zero whenever intervals overlap (i.e.,
\(|c_i-c_j|\le r_i+r_j\)), thereby assigning zero prior mass to
non-disjoint configurations. Then for the $i$-th component, the base measure $H_{i0}$ on $(u,\sigma)$ assigns to $u$ a normal distribution truncated to $[c_i-r_i,c_i+r_i]$, and to $\sigma^2$ an inverse-gamma distribution on $(0,\infty)$, with $u$ and $\sigma$ independent. Equivalently, $H_{i0}$ has density
\begin{equation}\label{base_measure}
h_{i0}(u,\sigma)\ \propto\ 
g_{c_i,\sigma_0}(u)\,\mathbf{1}_{[c_i-r_i,\,c_i+r_i]}(u)\cdot\mathrm{InvGamma}(\sigma^2)\cdot2\sigma,
\end{equation}
where $\sigma_0>0$ is a hyperparameter. The inverse-gamma parameters are omitted for notational convenience.  We place a Dirichlet process prior $\text{DP}(\alpha H_{i0})$ on $H_i$, for some $\alpha>0$. With the choice of $H_{i0}$ above, the DPM is conditionally conjugate at the component level, enabling closed-form updates and substantially improving computational efficiency. Finally, we place a truncated Dirichlet prior \citep{fang2000statistical} on the mixture weights $\boldsymbol{w}=(w_1,\cdots, w_K)\in\Delta^{K-1}$. A random vector $\boldsymbol{x}\in \Delta^{K-1}$ is said to follow a truncated Dirichlet distribution $\mathrm{TDir}(\gamma_1,\cdots, \gamma_K;\boldsymbol{a}, \boldsymbol{b})$, if its density is given by
\begin{equation*}
p( \boldsymbol{x}) =
c\prod _{i=1}^{K} x_{i}^{\gamma _{i} -1} , \ \ \ \ \boldsymbol{x} \in S_{K-1}(\boldsymbol{a} ,\boldsymbol{b})
\end{equation*}
where $c$ is a normalizing constant, and $S_{K-1}(\boldsymbol{a} ,\boldsymbol{b}) =\left\{\boldsymbol{x} \in \Delta ^{K-1} :a_{i} \leq x_{i} \leq b_{i} ,\ i=1,\cdots ,K\right\}$.
The truncation set $S_{K-1}(\boldsymbol{a} ,\boldsymbol{b})$ is nonempty if and only if $\sum _{i=1}^{K} a_{i} \leq 1\leq \sum _{i=1}^{K} b_{i}$. In the special case $K=2$, this distribution reduces to the truncated beta distribution. We specify the prior on the mixture weights $\boldsymbol{w}$ as $\boldsymbol{w}\sim \mathrm{TDir}(1/K,\cdots, 1/K; \underline{w}\mathbf{1}_K, \overline{w}\mathbf{1}_K)$, where $\underline{w}$ and $\overline{w}$ are lower and upper bounds on each component weight, satisfying $0< K\underline{w} < 1$ and $K\overline{w}>1$. 
For simplicity we assume that all component weights share the same lower and upper bounds. In practice 
one may assign component-specific bounds $\underline{w_i}>0$ and $\overline{w_i}>0$, as long as $0< \sum _{i=1}^{K}\underline{w_{i}} < 1$ and $\sum _{i=1}^{K}\overline{w_{i}}  >1$. Based on the construction above, we obtain the hierarchical model specified as follows:
\begin{equation}\label{hierarchical_model}
\begin{gathered}
\boldsymbol{w} =( w_{1} ,\cdots ,w_{K}) \sim \mathrm{TDir}(1/K,\cdots, 1/K; \underline{w}\mathbf{1}_K, \overline{w}\mathbf{1}_K),\ \text{given} \ K \\
z_{1} ,\cdots ,z_{n} \ \overset{\mathrm{i.i.d.}}{\sim } \ \mathrm{Categorical}(w_1,\cdots, w_K) ,\\
(\boldsymbol{c} ,\boldsymbol{r}) \ \sim \ p_{\text{rp}}(\boldsymbol{c} ,\boldsymbol{r}) \\
H_{i0}( u,\sigma ) \varpropto g_{c_{i} ,\sigma _{0}^{2}}( u) \cdot \mathbf{1}_{[ c_{i} -r_{i} ,c_{i} +r_{i}]}( u) \cdot \text{InvGamma}\left( \sigma ^{2}\right) \cdot 2\sigma \\
H_{i} \ \sim \ \mathrm{DP}( \alpha H_{i0}) \\
G_{i} =p_{H_{i}} =\int g_{u,\sigma } dH_{i}( u,\sigma ) \ \text{for} \ i=1,\cdots ,K \\
x_{j} \ \sim \ G_{z_{j}} \ \text{for} \ j=1,\cdots ,n,\ \text{given} \ G_{1:K} ,\ z_{1:n}. 
\end{gathered}
\end{equation}
Here, $x_1,\cdots, x_n$ are the $n$ samples observed, and $z_i,i=1,\cdots,n$ are the latent variables indicating the component generating $x_i$. Figure \ref{fig:diagram_two_comp} illustrates the hierarchical model for the two-component case ($K=2$).

In some application, we may have prior knowledge about the connected regions in which most of the component mass is concentrated. In such cases, the repulsive prior on $(\boldsymbol{c} ,\boldsymbol{r})$ can be replaced by fixed parameters $(\boldsymbol{c} ,\boldsymbol{r})$. When the separation condition is imposed on $\sigma$, we treat $(\boldsymbol{c} ,\boldsymbol{r})$ as fixed and known, and modify the base measure $H_{i0}(u,\sigma)$, for $i=1,\ldots, K$ to
\begin{equation*}
H_{i0}( u,\sigma ) \varpropto g_{c_{0} ,\sigma _{0}^{2}}( u) \cdot \mathrm{InvGamma}\left( \sigma ^{2}\right) \cdot 2\sigma \cdot \mathbf{1}_{[ c_{i} -r_{i} ,\ c_{i} +r_{i}]}( \sigma ).
\end{equation*}

Note that the structure of \eqref{hierarchical_model} resembles the classical Bayesian hierarchical model for finite mixture of parametric components (\cite{diebolt1994estimation, richardson1997bayesian, miller2018mixture}), apart from replacing the prior of component parameters with a DPM prior. Given that a single DPM is already highly flexible--- in particular, it achieves consistent estimation for a broad class of smooth densities at a nearly optimal rate (\cite{ghosal2007posterior})---one may wonder whether employing a mixture of DPMs (MDPM) is necessary. Accordingly, we explain why an MDPM is warranted and highlight its benefits over a single DPM. The first advantage lies in increased modeling effectiveness and computational efficiency. While one might expect a single DPM prior to be simpler, in practice this requires the base measure itself to be a mixture supported on several disjoint connected sets. Modeling and posterior inference with respect to complex forms of the latent base measure may be highly inefficient. In particular, constructing a base measure supported on disjoint subsets breaks conjugacy and greatly reduces the efficiency of the MCMC sampling. By contrast, employing a mixture of DPMs---each with a truncated normal-inverse Gamma base measure---preserves conjugacy at the component level and substantially improves sampling efficiency. Second, the structural similarity between MDPM and the classical Bayesian hierarchical model for finite parametric mixtures allows us to leverage the extensive toolbox of methods developed for finite mixture modeling over the past several decades. For example, repulsive priors were originally proposed to reduce redundancy among components in Bayesian mixture models (\cite{repulsive_2012}) and are here adapted to model disjoint intervals. Therefore, the MDPM framework not only enables efficient computation but also offers considerable potential for extension using the rich set of tools developed for parametric mixture modeling. In addition, in Section~\ref{section:posterior_contraction} it will be shown that, under appropriate conditions, an MDPM prior achieves the same posterior contraction rate to the true mixture density as a single DPM prior. This further justifies the MDPM framework---counterintuitively, it does not add complexity but rather provides a practical framework even if one is only interested in modeling the data population, not its latent subpopulations. This leads to the fourth, and perhaps the most interesting feature of the MDPM framework, which is that it enables the inference of individual components' distributions even when they have complex forms for which no parametric class of probability kernels can be adequate.

Inference algorithms for DPMs are typically of two types, corresponding to two representations of the DPM: the P\'{o}lya--urn representation (\cite{blackwell1973ferguson}) and the stick-breaking representation (\cite{sethuraman1994constructive}). The P\'{o}lya--urn representation integrates out the random measures from the Dirichlet process and characterizes the marginal distribution of the observed data. This perspective underlies a class of Gibbs samplers \citep{escobar1994estimating, maceachern1998estimating, neal2000markov, jain2004split}. In contrast, the stick-breaking representation provides an explicit constructive form of the random distributions. To deal with the potentially infinite number of components in the random distribution functions, \cite{walker2007sampling, kalli2011slice} developed the slice sampler. Because we aim to infer the mixture weights $\boldsymbol{w}$, the component distributions $\{G_i\}_{i=1}^K$, and the overall mixture $F$ jointly, an explicit representation of the random distribution functions is preferred. Accordingly, we adapt a slice sampler to the MDPM. In order to accommodate large-dataset applications, we adopt the slice-sampling approach proposed by \cite{ge2015distributed}, which leverages a MapReduce framework to parallelize the sampler’s computations. Our algorithm for MDPM is detailed in Appendix \ref{section:algorithm}. We implement the algorithm in Julia, leveraging its fast loops and built-in support for parallel programming. The source code is available at \href{https://github.com/yilei-zhang/NPMixture.git}{GitHub}. As noted earlier, conjugacy yields closed-form conditionals, which makes the sampler highly efficient. 
\begin{figure}[htbp]
  \centering
  \begin{subfigure}[htbp]{0.49\linewidth}
    \centering
    \includegraphics[width=\linewidth]{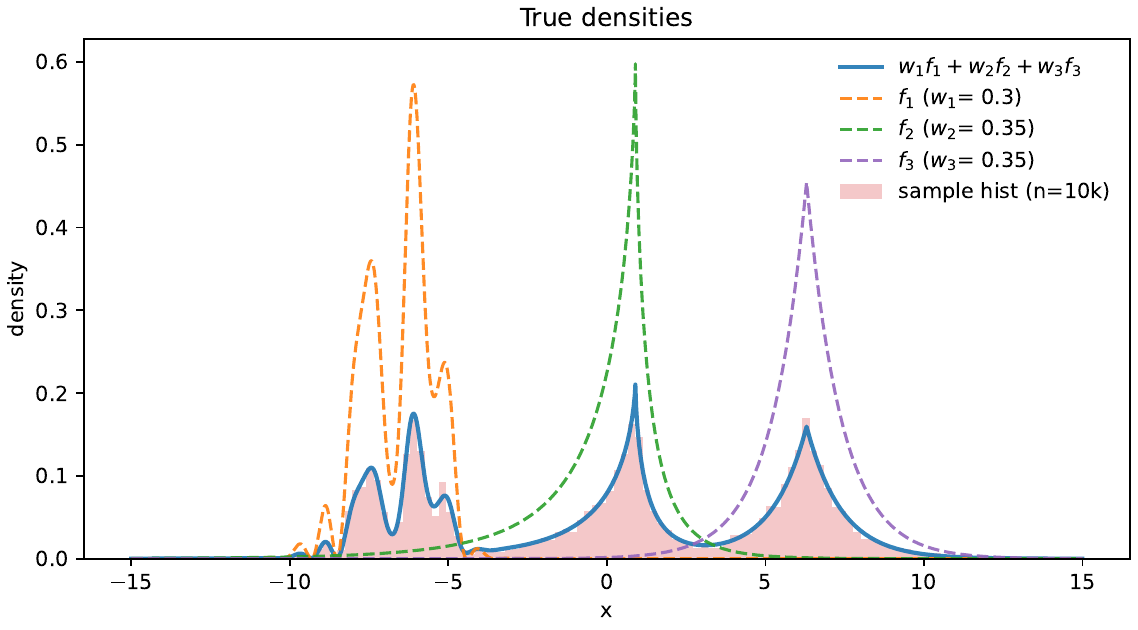}
    \caption{Example 1: True densities}
    \label{fig:a}
  \end{subfigure}\hfill
  \begin{subfigure}[htbp]{0.49\linewidth}
    \centering
    \includegraphics[width=\linewidth]{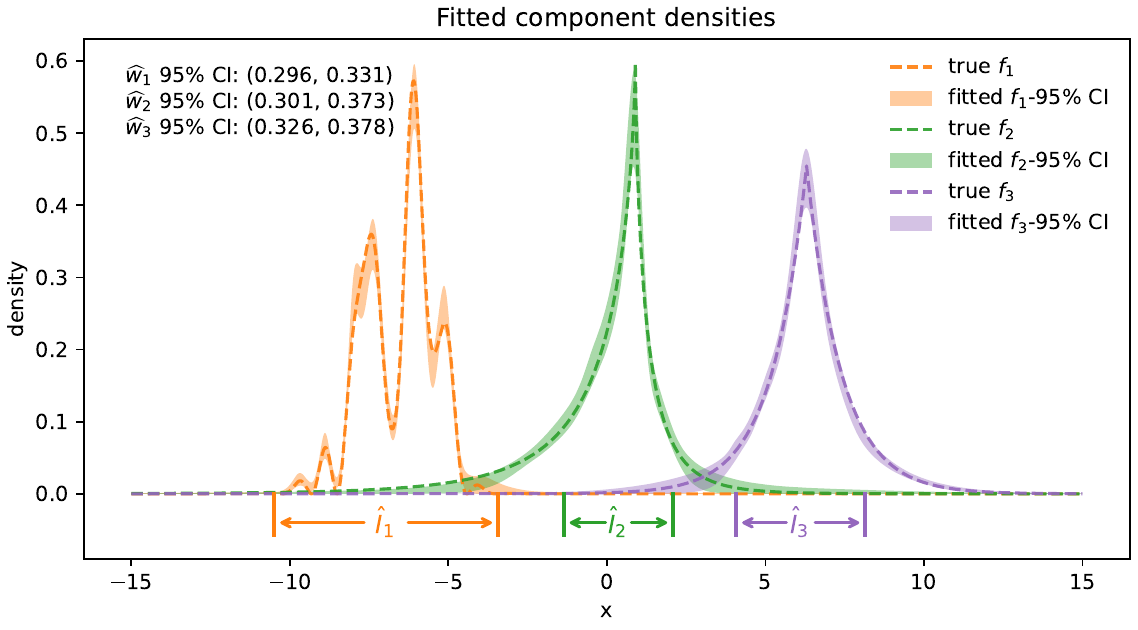}
    \caption{Example 1: Component density estimations}
    \label{fig:b}
  \end{subfigure}

  \vspace{0.8em}

  \begin{subfigure}[htbp]{0.49\linewidth}
    \centering
    \includegraphics[width=\linewidth]{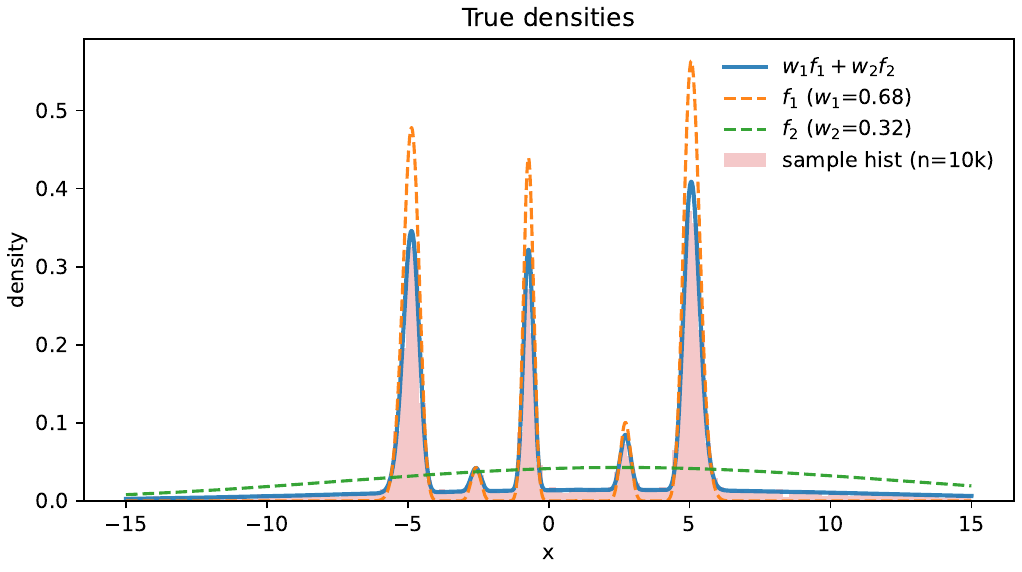}
    \caption{Example 2: True densities}
    \label{fig:c}
  \end{subfigure}\hfill
  \begin{subfigure}[htbp]{0.49\linewidth}
    \centering
    \includegraphics[width=\linewidth]{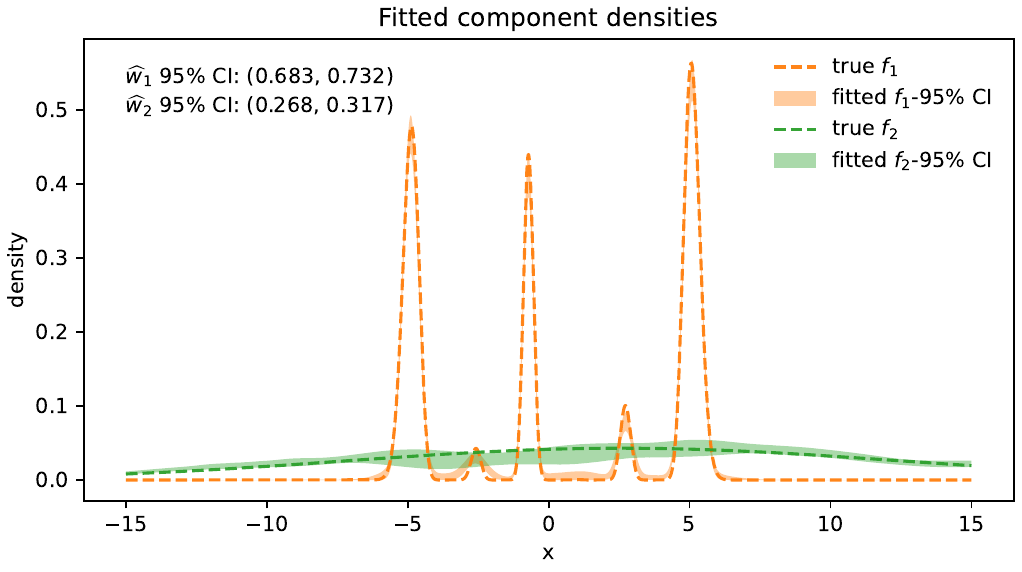}
    \caption{Example 2: Component density estimations}
    \label{fig:d}
  \end{subfigure}
  \caption{Example 1: a three-component mixture with the separation condition imposed on the location parameter $u$; $f_1$ is generated from a random combination of Hermite functions; $f_2$ is a skewed exponential-power distribution, and $f_3$ is a Laplace distribution. Example 2: a two component mixture with the separation condition imposed on the scale parameter $\sigma$; Both component densities are generated from random combinations of Hermite functions. }
  \label{fig:twobytwo}
\end{figure}

Two simulation examples are shown in Figure \ref{fig:twobytwo}. Example 1 in panels \eqref{fig:a} and \eqref{fig:b} illustrates a setting in which the separation condition is imposed on the location parameter $u$. In \eqref{fig:b}, the pointwise 95\% posterior credible intervals for the component densities closely follow the true component densities, including within the overlapping tail region between $f_2$ and $f_3$. The three estimated latent intervals are displayed beneath the density curves and approximately identify the regions where each component places most of its mass, and the estimated weights are shown in the upper-left corner in \eqref{fig:b}. Panels \eqref{fig:c} and \eqref{fig:d} show a second example in which the separation condition is imposed on the scale parameter $\sigma$. In both scenarios, the proposed method demonstrates strong performance. For the simulation in Example 1 with $n=10{,}000$ observations and $K=3$ components, $10{,}000$ MCMC iterations without multi-threading take about 1.5 minutes (wall-clock) on a MacBook Pro (Apple M4, 10-core; 24~GB unified memory). For large datasets, enabling multi-threading typically yields an additional 10–20\% speedup. We further demonstrate the scalability of our method by applying a multivariate version of our algorithm to an astronomical dataset containing approximately 0.8 million two-dimensional events in Section \ref{section:applications}.


\section{Multivariate distributions}
\label{section:Multivariate}
This section extends the problem setting in section~\ref{section:Problem_Setting} and the method in section~\ref{section:Mixture_of_DP} to the multivariate case. We first consider location-scale mixture of normals on $\mathbb{R}^m$: For any $\boldsymbol{\mu}\in \mathbb{R}^m$, $\Sigma\in \mathbb{R}^{m\times m}$, $\Sigma\succ 0$, let $g_{\boldsymbol{\mu},\Sigma}$ denote the $m$-dimensional normal density with mean $\boldsymbol{\mu}$ and covariance matrix $\Sigma$: $g_{\boldsymbol{\mu } ,\Sigma }(\boldsymbol{x}) ={\left(\sqrt{2\pi }\right)^{-m} |\Sigma |^{-1/2}}\exp\left( -(\boldsymbol{x} -\boldsymbol{\mu })^{T} \Sigma ^{-1}(\boldsymbol{x} -\boldsymbol{\mu })\right).$

We say $p(\boldsymbol{x})$ is a location-scale mixture of normals on $\mathbb{R}^m$ if there exists a mixing distribution \(H(\cdot,\cdot)\) on \(\mathbb{R}^m \times \mathbb{S}_{++}^m\) such that
\begin{equation}\label{mv-location-scale-mixture}
p(\boldsymbol{x}) =p_{H}(\boldsymbol{x}) \coloneq \int g_{\boldsymbol{\mu } ,\Sigma }(\boldsymbol{x}) dH(\boldsymbol{\mu } ,\Sigma ),
\end{equation}
where $\mathbb{S}_{++}^m$ denotes the set of all \(m\times m\) positive-definite matrices. In this section, we consider mixture models under the following setting:
\begin{enumerate}[resume*=setting]  
  \item \label{S3} The component distribution class $\widetilde{\mathcal{P}}$ consists of location-scale mixtures of normals on $\mathbb{R}^m$, as defined in \eqref{mv-location-scale-mixture}. The mixing measure $H$ satisfies: \textup{(i)} its $\boldsymbol{\mu}$-marginal $\pi_{\boldsymbol{\mu}}\#H$ has bounded support on $\mathbb{R}^m$; and \textup{(ii)} its $\Sigma$-marginal $\pi_\Sigma\#H$ is supported on covariance matrices with eigenvalues uniformly bounded above and away from $0$. $\mathcal{P}(\widetilde{\mathcal{P}})$ are probability measures supported on finitely many components satisfying the separation condition: for any two components $G_i=p_{H_i}$, and $G_j=p_{H_j}$, $i\neq j$, we can find two disjoint bounded connected sets $I_i, I_j\subset \mathbb{R}^m$ such that $\operatorname{supp}(\pi_{\boldsymbol{\mu}}\# H_i)\subseteq I_i$, and $\operatorname{supp}(\pi_{\boldsymbol{\mu}}\# H_j)\subseteq I_j$.
\end{enumerate}

To extend the hierarchical model \eqref{hierarchical_model} to the multivariate setting \eqref{S3}, the major modification is the modeling of pairwise disjoint connected sets $I_1,\cdots, I_K\subset \mathbb{R}^m$. In this work, we model $I_1,\cdots, I_K$ as axis-aligned hypercubes centered at $c_k$ with half-side length $r_k>0$, equivalently,
\[
I_k=\{x\in\mathbb{R}^m:\ \|x-c_k\|_\infty \le r_k\},\quad k=1,\ldots,K.
\]
Although modeling $I_k$'s with more flexible shapes are possible, the main computational bottleneck is efficient sampling from a
truncated multivariate normal. Most existing methods address linear constraints
(\cite{geweke1991efficient,kotecha1999gibbs,fernandez2007perfectly,chopin2011fast,botev2017normal}), and an exact Hamiltonian Monte Carlo sampler \cite{pakman2014exact} has been developed to handle both linear and quadratic inequality constraints. However, scalable samplers for more flexible constraints remain limited. Thus, extending from hypercubes to more general H-polytopes is straightforward (\cite{botev2017normal}), but efficient sampling under nonconvex constraints remains underexplored. Developing scalable algorithms for such cases is an interesting direction for future work. 

\begin{figure}[H]
\centering
\includegraphics[width=1.\linewidth]{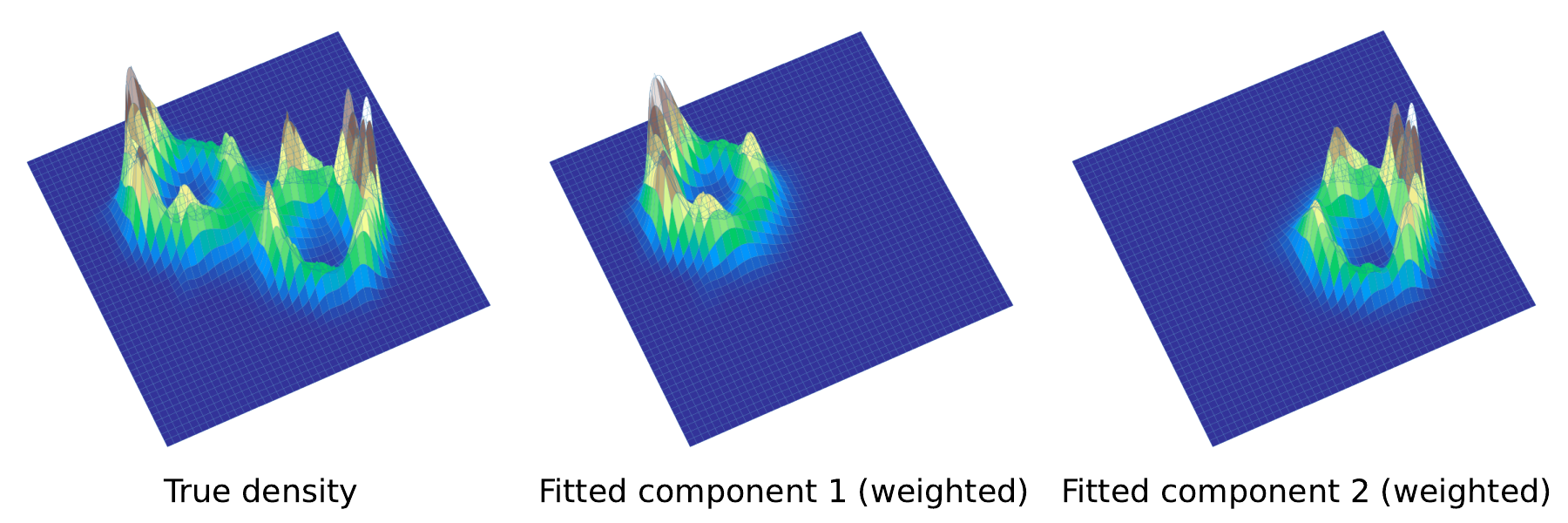}
\captionsetup{width=1.\linewidth}
\caption{A bivariate two-component mixture model. Each component is a Gaussian mixture located on a circle with random covariance matrices. The left panel displays the true density, while the middle and right panels show the pointwise posterior means of the fitted component densities, weighted by their respective posterior mixture weights.}
\label{fig:threecol}
\end{figure}

Accordingly, we extend the repulsive prior (\ref{repulsive_prior}) to $\mathbb{R}^m$,
\begin{equation}\label{mv_repulsive_prior}
p_{mrp}(\boldsymbol{c} ,\boldsymbol{r}) \propto \left(\prod _{i=1}^{K} g_{\boldsymbol{\mu }_{0} ,\Sigma _{0}}(\boldsymbol{c}_{i}) \cdot \mathrm{Gamma}( r_{i})\right) \cdot \underset{1\leq i< j\leq K}{\min}\exp\left( -\frac{\tau }{\max( \| \boldsymbol{c}_{i} -\boldsymbol{c}_{j} \| _{\infty } -r_{i} -r_{j} ,0)^{\nu }}\right),
\end{equation}
and change $H_{i0}$ to its multivariate counterpart:
\begin{equation}\label{mu_Sigma_prior}
H_{i0}(\boldsymbol{\mu } ,\Sigma ) \propto g_{\boldsymbol{c}_{i} ,\Sigma _{1}}(\boldsymbol{\mu }) \cdot \mathbf{1}_{[\boldsymbol{c}_{i} -r_{i} ,\boldsymbol{c}_{i} +r_{i}]}(\boldsymbol{\mu }) \cdot \mathrm{InvWishart}( \Sigma ),
\end{equation}
where $\mathrm{InvWishart}$ denotes the inverse-Wishart distribution, and hyperparameters are omitted for brevity. The rest of model (\ref{hierarchical_model}) remain unchanged. A simulation example is shown in Figure \ref{fig:threecol}.

In some applications, reasonable \emph{a priori} estimates of where each component 
lies are available; in such settings we can fix \(I_1,\ldots,I_K\) and drop the repulsive prior over regions, yielding a simpler 
model. Code for the fixed-region algorithm is also provided. An application example is in Section \ref{section:applications}.


\section{Identifiability}
\label{section:Identifiability}
In this section, we assume that the true mixture distribution satisfies \eqref{setting1} and establish conditions for the identifiability of the mixture model's nonparametric components. We begin by introducing notions of distance between connected sets in the topological space $\mathbb{X}$. 

Let $I_1$ and $I_2$ be two connected subsets of $\mathbb{X}$ such that $I_1\cup I_2$ is disconnected.  If the space $\mathbb{X}$ is equipped with some metric $d$, we can define the distance $d_c(I_1,I_2)$ between the two connected sets $I_1$ and $I_2$ as
\begin{equation}\label{dc}
d_{c}( I_{1} ,I_{2}) =\underset{x\in I_{1} ,y\in I_{2}}{\inf} d( x,y).
\end{equation}

Any non-empty $I\subseteq\mathbb{X}$ has a unique separation $S_I$ such that each element $I_i\in S_I$ is non-empty and connected, and if $S_I$ contains more than 2 elements, then for any subset $T\subseteq S_{I}$ with $|T|\geq 2$, the union $\underset{I_{i} \in T}{\cup } I_{i}$ is disconnected, and moreover $I=\cup_{I_i\in S_I} I_{i}$. This separation is unique because if there is a different separation $S_I'$ that satisfies the same conditions, there must be a connected set $I^*\in S_I$ such that $I^*\notin S_I'$, then $\underset{A\in S'_{I}}{\cup }\left( I^{*} \cap A\right) =I^{*}$ is connected, which contradicts the conditions of $S'_I$. Note that $S_I$ may be an infinite set. In particular, a singleton is itself a connected set. 

For distinct $I_i, I_j\in S_{I}$, we say $I_i$ and $I_j$ are \emph{neighbors} in $S_I$ if there is no $I_k\in S_I\backslash\{I_i,I_j\}$ such that $d_c(I_i,I_k)<d_c(I_i,I_j)$ and $d_c(I_j,I_k)<d_c(I_i, I_j)$. Denote the set of all such unordered neighbor pairs by $\mathcal{N}(S_I)$. Then we can define the within-space distance $d_w$ of a space $I$ as
\begin{equation}\label{dist: within-space}
d_{w}( I) =\begin{cases}
0, & |S_{I} |=1\\
\underset{( I_{i} ,I_{j}) \in \mathcal{N}( S_{I})}{\sup } d_{c}( I_{i} ,I_{j}) , & |S_{I} |\geq 2
\end{cases}
\end{equation}
Furthermore, the between-space distance $d_b$ of two spaces $I$ and $J$ can be defined as
\begin{equation}\label{dist: between-space}
d_{b}( I ,J) = \underset{I_{i} \in S_{I} ,J_{i} \in S_{J}}{\inf} d_{c}( I_{i} ,J_{i}).
\end{equation}

In this section, we provide theoretical results for the identifiability of mixture models under setting \eqref{setting1}. Let $I_{1} ,\cdots ,I_{K}$ be pairwise disjoint compact intervals on $\mathbb{R}$, where $I_{k}=[c_k-r_k,c_k+r_k]$ for some $c_k\in\mathbb{R}$, and $r_k>0$ ( $k=1,\cdots,K$). Without loss of generality, index these intervals so that $c_k+r_k<c_{k+1}-r_{k+1}$, $\forall 1\leq k\leq K-1$. Consider the mixture model
\begin{equation}\label{interval_location_mixture}
f=\sum _{k=1}^{K} w_{k} f_{k} ,\ \text{s.t.} \ f_{k} =\int g_{u,\sigma }( x) dV_{k}( u) ,\ \operatorname{supp}( V_{k}) \subseteq I_{k} ,\ \text{and} \ ( w_{1} ,\cdots ,w_{K}) \in \Delta ^{K-1},
\end{equation}
where $\sigma\in (0,\infty)$ is a fixed but probably unknown standard deviation. For $k=1,\cdots,K$, $V_k$ can be any probability measure supported within $I_k$. To ensure the identifiability of (\ref{interval_location_mixture}), we consider the two situations:
\begin{enumerate}[label=(C\arabic*), ref=C\arabic*, series=condition]
    \item \label{C1} The sets $I_1,\cdots, I_K$ are fixed and known.\label{condition_C1}
    \item \label{C2} $K$ is fixed and known, but the sets $I_1,\cdots, I_K$ are unknown, and 
\begin{equation*}
0\leq \underset{1\leq k\leq K}{\max} d_{w}( \operatorname{supp}(V_{k})) < \underset{1\leq i< j\leq K}{\min} d_{b}( \operatorname{supp}(V_{i}) ,\operatorname{supp}(V_{j})).
\end{equation*}\label{conditon_C2}
\end{enumerate}
The following theorem shows that if either condition \eqref{C1} or \eqref{C2} is satisfied, the function $f$ defined as in (\ref{interval_location_mixture}) has a unique representation as a mixture model.
\begin{theorem}\label{Theorem:identifiability}
    For any $f$ as defined in (\ref{interval_location_mixture}) such that \eqref{C1} or \eqref{C2} is satisfied, there exist a unique $\sigma$, a unique collection of densities $f_{1} ,\cdots ,f_{K}$, and a corresponding weight vector $(w_1,\cdots, w_K)\in \Delta^{K-1}$ that satisfy the conditions in (\ref{interval_location_mixture}) such that $f=\sum _{k=1}^{K} w_{k} f_{k}$.
\end{theorem}


\section{Posterior contraction rates}
\label{section:posterior_contraction}

In this section, we obtain the contraction rates of the Bayesian posteriors arising from the models described in the previous sections. Our theoretical analysis focuses on the following class of mixture densities:

Let $\mathcal{F}_{K,\sigma }(\boldsymbol{c}, \boldsymbol{r}, \underline{w})$ denote the class of densities $f$ of the form $f=\sum _{k=1}^{K} w_{k} f_{k}$, where $(w_1,\cdots, w_K)\in\Delta^{K-1}$, $K\geq2$ satisfies $w_k\in [\underline{w}, \frac{1-\underline{w}}{K-1}]$, $k=1,\cdots, K$ for some $0<\underline{w}<1/K^2$. For each $k=1,\cdots, K$, the component density is given by $f_{k}( x) =\int g_{u,\sigma }( x) dV_{k}( u)$, $\forall x\in \mathbb{R}$, where $V_i$ is a probability measure with $\operatorname{supp}( V_{k}) \subset [ c_{k} -r_{k} ,c_{k} +r_{k}]$, for some $c_k\in\mathbb{R}$, $r_k\in (0,+\infty)$. Here $\boldsymbol{c}=(c_1,\cdots, c_K)$ and $\boldsymbol{r}=(r_1,\cdots, r_K)$. Without loss of generality, we assume that $c_k+r_k<c_{k+1}-r_{k+1}$ for $1\leq k\leq K-1$. Throughout this section,  $K,\sigma, \boldsymbol{c}, \boldsymbol{r}, \underline{w}$ are assumed fixed and known, and the true mixture density $f_{0} =\sum _{k=1}^{K} w_{k0} f_{k0}$ is assumed to lie in $ \mathcal{F}_{K,\sigma }(\boldsymbol{c} ,\boldsymbol{r} ,\underline{w})$.

We study posterior contraction rates for Bayesian density estimation under a mixture of Dirichlet process mixture (MDPM) prior imposed on the density class $\mathcal{F}_{K,\sigma }(\boldsymbol{c}, \boldsymbol{r}, \underline{w})$. The prior specification is given by the following assumptions:
\begin{enumerate}[label=(A\arabic*), ref=A\arabic*, series=priorassump]
\item \label{A1} Let $\mathbf{W}\coloneqq(W_1,\cdots,W_K)$ follow the distribution $\mathrm{Tdir}(\alpha_1,\cdots, \alpha_K; \underline{w}\mathbf{1}_K, \frac{1-\underline{w}}{K-1}\mathbf{1}_K)$. We further assume that the concentration parameters satisfy $A \leq\alpha_k\leq 1$, $k=1,\cdots,K$, for some constant $0<A<1/K$.
\item \label{A2} For $1\leq k\leq K$, let $H_k$ be a finite, positive measure on $[c_k-r_k,c_k+r_k]$. $D_{H_k}$ is a Dirichlet process with base measure $H_k$. 
\item \label{A3} We assume $\mathbf{W}$ and $D_{H_k}$'s are mutually independent. Let $\Pi_k\coloneqq D_{H_k}$ and $\Pi\coloneqq \sum_{k=1}^K W_k D_{H_k}$.
\end{enumerate}

$\Pi$ defined above induces a prior on the mixing measure $\sum_{k=1}^{K}W_kV_k$. By the identifiability result established in Theorem \ref{Theorem:identifiability}, there is a one-on-one correspondence between the mixing measure and the associated mixture density $f\in \mathcal{F}_{K,\sigma }(\boldsymbol{c} ,\boldsymbol{r} ,\underline{w})$. Consequently, $\Pi$ also uniquely induces a prior on $\mathcal{F}_{K,\sigma }(\boldsymbol{c} ,\boldsymbol{r} ,\underline{w})$. With a slight abuse of notation, we henceforth regard $\Pi$ as a prior on $\mathcal{F}_{K,\sigma }(\boldsymbol{c} ,\boldsymbol{r} ,\underline{w})$.

Section~\ref{subsection: mixture_density} establishes the posterior contraction rate of the overall mixture density under Assumptions~\eqref{A1}–\eqref{A3}, while Section~\ref{subsection: component_density} derives contraction rates for the component densities when $K=2$ under additional separation conditions.

Throughout this section, we introduce several notions of distances and divergences between probability densities. The Hellinger distance $d$ between two densities $f$ and $g$ is defined as:
$d( f,g) =\left(\int \left( f^{1/2}( x) -g^{1/2}( x)\right)^{2} dx\right)^{1/2}$,
and the $L_1$-norm $\Vert f-g\Vert_1=\int|f(x)-g(x)|dx$. $L_1$ and Hellinger distances are related by
\begin{equation}\label{ineq: l1_hellinger}
d^{2}( f,g) \leq \| f-g\| _{1} \leq 2d( f,g).
\end{equation}
Let $p_0\in \mathcal{P}$, a class of densities and let $P_0$ be the probability measure with density $p_0$. The Kullback-Leibler (KL) divergence is $K( p_{0} ,p) =\int \log( p_{0} /p) dP_{0}$, and the second Kullback-Leibler variation is $K_2( p_{0} ,p) =\int (\log( p_{0} /p))^{2} dP_{0}$. Also relevant is the KL ball
$B_{2}( p_{0} ,\epsilon ) =\left\{p:K( p_{0} ,p) \leq \epsilon ^{2} ,\ K_2( p_{0} ,p) \leq \epsilon ^{2}\right\}$.

\subsection{Posterior contraction of the overall mixture density}\label{subsection: mixture_density}
In this section we establish the posterior contraction rates of the overall mixture density.
\begin{theorem}\label{thm: density_posterior}
Let $f_{0} \in \mathcal{F}_{K,\sigma }(\boldsymbol{c} ,\boldsymbol{r} ,\underline{w})$, where $K, \sigma, \boldsymbol{c}, \boldsymbol{r}, \underline{w}$ are fixed and known. Denote by $P_{f_0}$ the probability measure with density $f_0$. Let $X^{(n)}\coloneqq (X_1,\cdots, X_n)$ be $n$ i.i.d samples from $P_{f_0}$. Consider the prior $\Pi$ constructed under specifications \eqref{A1}, \eqref{A2}, and \eqref{A3}, and let $\Pi^{(n)}(\cdot| X^{(n)})$ be the posterior distribution induced by the prior $\Pi$. Then for a large enough constant $M$,
\begin{equation}\label{f0_posterior}
\Pi ^{( n)}\left( d( f,f_{0})  >M\frac{\log n}{\sqrt{n}} |X_{1} ,\cdots ,X_{n}\right)\rightarrow 0
\end{equation}
in $P_{f_0}^n$-probability. 
\end{theorem}

Theorem~\ref{thm: density_posterior} establishes that, under the assumptions \eqref{A1}-\eqref{A3} for a MDPM prior, the posterior contraction rate of our estimator for the overall mixture density matches that obtained under a single Dirichlet process mixture prior. 

The proof of the above theorem follows the general framework laid out by Theorem 2.1 in \cite{ghosal2001entropies}. To apply this framework, we will find a positive sequence $\tilde{\epsilon}_n\rightarrow 0$ such that $\Pi \left( B_{2}\left( f_{0} ,\tilde{\epsilon }_{n}\right)\right) \geq c_{3} e^{-c_{4} n\tilde{\epsilon }_{n}^{2}}$, for some constants $c_3,c_4>0$. Lemma \ref{lemma: mix_of_dp_prior_conctr} identifies such $\tilde{\epsilon}_n$ for the mixture of Dirichlet processes prior under the assumptions \eqref{A1}-\eqref{A3}, and the proof of Theorem \ref{thm: density_posterior} follows.

\subsection{Posterior contraction of component densities} \label{subsection: component_density}

The novel and considerably more demanding part of our theory lies in establishing posterior contraction rates for the individual component densities. For the ease of technical exposition we shall restrict our theorem's claim to the case $K=2$ with the following conditions:

\begin{enumerate}[resume*=condition] 
    \item \label{C3} Consider the density class $\mathcal{F}=\mathcal{F}_{2,\sigma}(\boldsymbol{c}, \boldsymbol{r}, \underline{w})$, for some $0<\underline{w}<\frac{1}{4}$. Assume that $\boldsymbol{c}$ and $\boldsymbol{r}$ satisfy $|c_1-c_2|>8\max(r_1,r_2)$. Denote $r=|c_1-c_2|$. $\sigma, \boldsymbol{c}, \boldsymbol{r}, \underline{w}$ are fixed and known.
    \item \label{C4} Let $f_{0} =w_{10} f_{10} +w_{20} f_{20}\in \mathcal{F}_{2,\sigma}(\boldsymbol{c}, \boldsymbol{r}, \underline{w})$.
    \item \label{C5} Consider an MDPM prior $\Pi$ on $\mathcal{F}$ satisfying \eqref{A1}, \eqref{A2} and \eqref{A3} with $K=2$. 
\end{enumerate}

\begin{theorem}\label{thm:fi_post_rate}
    Let $\mathcal{F}$, $f_0$, $\Pi$ be specified as in \eqref{C3}, \eqref{C4}, and \eqref{C5}. Denote by $P_{f_0}$ the probability measure with density $f_0$. Let $X^{(n)}\coloneqq (X_1,\cdots, X_n)$ be $n$ i.i.d. samples from $P_{f_0}$. Let $\Pi^{(n)}(\cdot| X^{(n)})$ denote the posterior distribution induced by the prior $\Pi$. For $i=1,2$, if $\epsilon _{n} \geq n^{-\frac{c_{i}}{\log\log n}}$
for some positive constant $c_{i}$, then we have 
\begin{equation*}
\Pi ^{( n)}(\Vert f_{i} -f_{i0}\Vert _{1}  >M_n\epsilon _{n} |X_{1} ,\cdots ,X_{n})\rightarrow 0
\end{equation*}
in $P_{f_0}^n$-probability, for every $M_n\rightarrow\infty$.
\end{theorem}

\begin{remark*}
    Because $\log\log n$ grows extremely slowly and may be treated as almost constant as $n\rightarrow \infty$, $\epsilon_n$ is almost a polynomial rate. This rate is of the same minimax order as that established for component density estimation in \cite{bryon_2023}.
\end{remark*}

A critical portion of this proof requires an existence of test argument. To establish the existence of such a test with desired controls on its type I and type II errors, we need a component density estimator $\widehat{f}_i$, and establish an upper bound on $\| \widehat{f_i} -f_i \| _{1}$ that vanishes as the sample size $n\rightarrow \infty$. The estimator is due to \cite{bryon_2023}, who also provided a convergence analysis under the $l_1$ norm $\| \widehat{f_i} -f_i \| _{1}$. For simplicity, they fixed $r_1=r_2=\frac{1}{2}$, and $\sigma=1$, and provided only the order of the convergence rate. For our purposes we extend their results to the setting where $r_i\in(0,+\infty)$, $i=1,2$, $\sigma \in ( 0,+\infty )$, and we derive the convergence rate with explicit constants. This refinement is useful, as it allows us to establish a concentration inequality for the estimate of the component densities, from which the desired existence of a test will follow.
\begin{lemma}\label{lemma: E_phi_maintext}
   Let $f\in\mathcal{F}$ and $f_{0}$ be as defined in \eqref{C3} and \eqref{C4}. Fix a constant $M>2$, and let $\{\epsilon_n\}$ be a positive sequence such that $\lim _{n\rightarrow \infty } \epsilon _{n} =0$. For any component $i=1$ (or $2$), there exists a test function $\phi _{n} :\mathbb{R}^{n}\rightarrow \{ 0,1\}$ such that for sufficiently large $n$ and small $\epsilon_n$,
\begin{equation*}
E_{f_{0}}[ \phi _{n}] \leq \exp\left( -\tilde{c} n^{4/5}\left(\frac{\epsilon _{n}}{\overline{c}_{1i}}\right)^{c_{2i}\log\log\left(\frac{\overline{c}_{1i}}{\epsilon _{n}}\right) +c_{3i}}\right),
\end{equation*}
and
\begin{equation*}
\underset{f\in \mathcal{F} :\ \| f_{i} -f_{i0} \| _{1}  >M\epsilon _{n}}{\sup } E_{f}[ 1-\phi _{n}] \leq \exp\left( -\tilde{c} n^{4/5}\left(\frac{( M-1) \epsilon _{n}}{\overline{c}_{1i}}\right)^{c_{2i}\log\log\left(\frac{\overline{c}_{1i}}{( M-1) \epsilon _{n}}\right) +c_{3i}}\right),
\end{equation*}
where $\tilde{c}$ is a positive absolute constant, $\overline{c}_{1i}$ is a positive constant depending on $i, r, \sigma, r_1,r_2,\underline{w}$, $c_{2i}$ is a positive constant depending on $i, r, \sigma, r_1,r_2$, and $c_{3i}$ is a constant depending on $i, r, \sigma, r_1,r_2$.
\end{lemma}

In Appendix \ref{appendix: construction of fi_hat}, we detail the construction of such a component density estimator $\widehat{f}_i$, and in Appendix \ref{section:appendix_A}, we give an upper bound of $\| \widehat{f_{i}} -f_{i} \| _{1}$, which is a refinement of a result of \cite{bryon_2023}. In Appendix \ref{section:appendix_B}, we further derive the tail bound of $\widehat{f}_i$. In Appendix \ref{section:appendix_C}, we establish the existence of a test function $\phi_n$ with desired controls on type I error $E_{f_{0}}[ \phi _{n}]$ and type II error $ E_{f}[ 1-\phi _{n}] $. Note that this test function $\phi_n$ cannot be used in practice, but the existence of such a test function is sufficient for the proof of the main theorem (Theorem \ref{thm:fi_post_rate}) on the posterior contraction of the component density $f_i$.


\section{Applications}
\label{section:applications}

\subsection{Disentangling astronomical sources}\label{section:disentagle_astro_sources}
\paragraph*{XMM\allowbreak-Newton dataset.}This dataset (\cite{jones2015disentangling} ) consists of X-ray events from the nearby star pair FK Aqr and FL Aqr (a visual binary) observed by ESA's XMM-Newton telesccope. The detectors recorded about 540,000 individual X-ray photons (a total of 758,869 events including noise) over a 47 ks ($\sim$13 hr) exposure. For each photon, we have its position $(X,Y)$ on the detector and a spectral energy value. The two sources partly overlap in the image and the background is weak, making this a clean test case for methods that aim to separate blended sources. 

\begin{figure}[htbp]
\centering
\includegraphics[width=1.\linewidth]{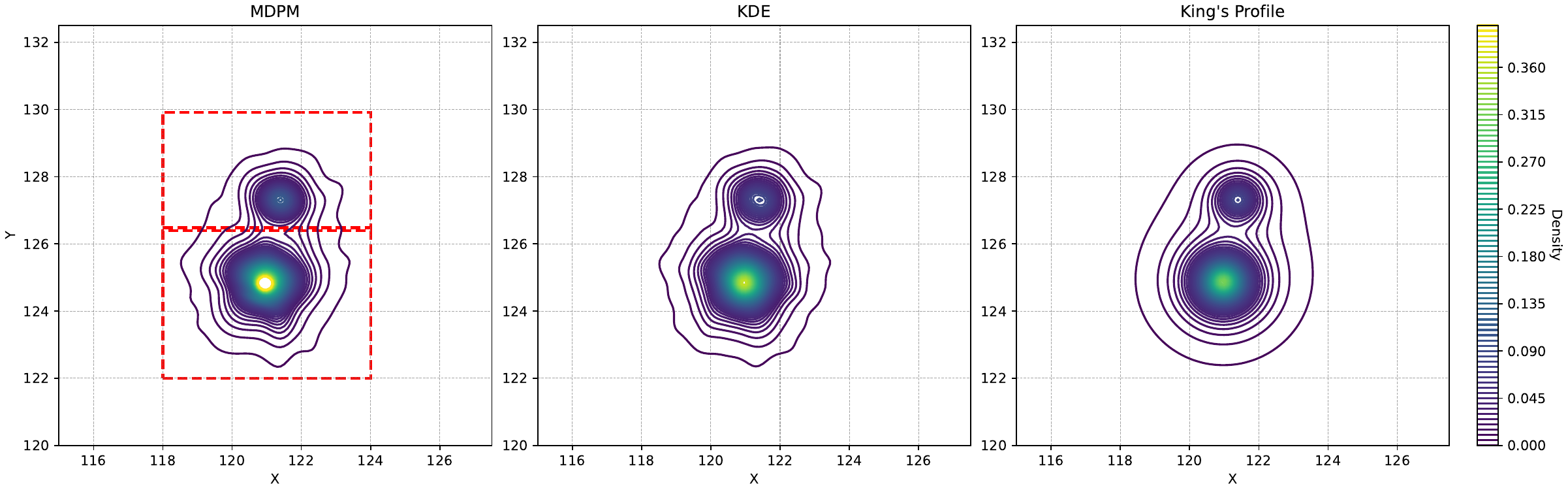}
\captionsetup{width=1.\linewidth}
\captionof{figure}{Density estimation contours from MDPM, KDE, and mixture of King's profiles, respectively.}
\label{figure:astro_3fig}
\end{figure}

Mixture models have been widely used to model X-ray events originating from different astronomical sources.
\cite{jones2015disentangling} modeled the spatial information ($X$ and $Y$) using a mixture of King profiles. This and subsequent work by \cite{meyer2021ebascs} extended this spatial model by incorporating additional spectral and temporal information. In this section, we employ a multivariate MDPM with fixed regions to model the spatial distribution of X-ray events. The regions are held fixed to avoid identifiability issues and are pre-defined. The two regions are determined based on the kernel density estimate (KDE) of the observed data and serve as approximate estimates of the locations of the two astronomical sources. The two rectangle regions are illustrated in the left panel of Figure \ref{figure:astro_3fig}. In addition to the two source components, we include a third component—a uniform distribution over the observation window—to capture the diffuse background noise.

We compare our density estimation result from MDPM (left panel), Kernel density estimate (middle panel) and Mixture of King's profiles from \cite{jones2015disentangling} (right panel) in Figure \ref{figure:astro_3fig}. King's profile density is a parametric model and assumes the regular elliptical contours of each astronomical source density, however, from its comparison with KDE we know that it failed to capture the tail structure of real data sharply. In contrast, MDPM can closely recover the real data density. 
\begin{figure}[H]
\centering
\includegraphics[width=1.\linewidth]{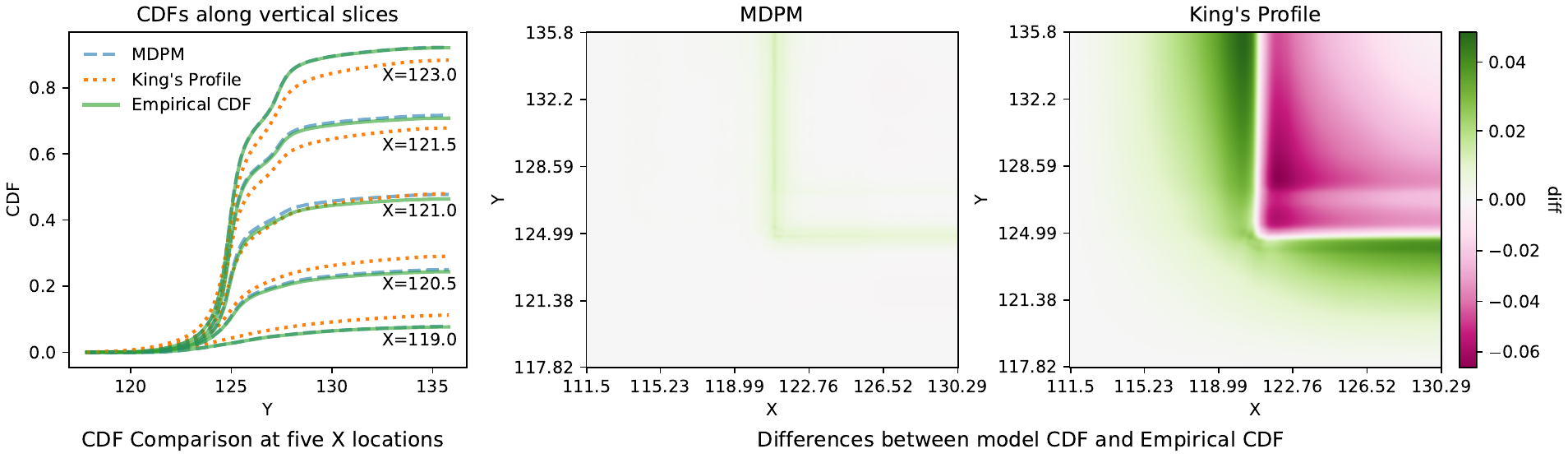}
\captionsetup{width=1.\linewidth}
\captionof{figure}{Comparison of CDFs.}
\label{figure:astro_cdfs}
\end{figure}

\begin{table}[H]
\centering
\setlength{\tabcolsep}{8pt} 
\small
\begin{tabular}{c|cc|cc}
\toprule
& \multicolumn{2}{c|}{\textbf{MDPM}} 
& \multicolumn{2}{c}{\textbf{King's Profile}} \\
\cmidrule(lr){2-3} \cmidrule(lr){4-5}
\textbf{Weights} 
& \textbf{posterior mean} & \textbf{68\% CI} 
& \textbf{posterior mean} & \textbf{68\% CI} \\
\midrule
$w_1$ (lower component) & 0.643 & (0.641, 0.646) & 0.717 & (0.716, 0.718) \\
$w_2$ (upper component)& 0.229 & (0.227, 0.231) & 0.182 & (0.181, 0.182) \\
$w_0$ (background noise)& 0.128 & (0.127, 0.128) & 0.102 & (0.101, 0.102) \\
\bottomrule
\end{tabular}
\caption{Comparison of the estimated component weights. The 68\% credible intervals are computed using 16\% and 84\% posterior quantiles.}
\label{table:astro_weights}
\end{table}

Figure \ref{figure:astro_cdfs} compares the cumulative distribution functions (CDFs) obtained from the MDPM estimator, the mixture of King’s profiles estimator, and the empirical CDF. For any location $(x,y)$, the estimated CDF is defined as $\hat{F}(x,y)=\textit{Pr}_{\hat{f}}(X\leq x, Y\leq y)$. In the left panel of Figure \ref{figure:astro_cdfs}, we examine five $X$ values—119.0, 120.5, 121.0, 121.5, and 123.0—and, for each value, plot the CDF as $Y$ varies from 117.5 to 135.0. The resulting curves are shown using distinct colors and line styles for clarity. Across all five $X$-values, the MDPM estimator closely tracks the empirical CDF, whereas the Mixture of King's profiles consistently overestimates the tail mass. The middle and right panels of Figure \ref{figure:astro_cdfs} display the pointwise differences between each model’s estimated CDF and the empirical CDF over the spatial domain, with $X$ ranging from 111.5 to 130.29 and $Y$ ranging from 117.8 to 135.8. The MDPM estimator matches the empirical CDF closely across most of the plane, with only minor deviations near the centers of the two astronomical sources. In contrast, the mixture of King’s profiles exhibits noticeably larger discrepancies throughout the region. These results indicate that the MDPM estimator captures the subtle tail behavior of both sources much more effectively. The estimated component weights are compared in Table \ref{table:astro_weights}.

\subsection{Analysis of oceanic whitetip shark acceleration data}
\label{section: oceanic_shark}
Studying the dynamic patterns of an animal's behavioral states from animal movement metrics is a topic of interest in ecology. The Overall Dynamic Body Acceleration (ODBA) data were collected for an oceanic whitetip shark at a rate of 16 Hz over 4 days (\cite{langrock2018spline}). The movements of the shark were recorded by accelerometers along three axes at very fine temporal scales (multiple times per second) over multiple days. The acceleration measurements from the three axes were then combined into a single representative metric. A line of work (\cite{langrock2015nonparametric, langrock2018spline, chen2024bayesian}) has modeled such data with spline-based hidden Markov models (HMM) to infer latent movement states (e.g., resting, foraging or migrating). \cite{chen2024bayesian} use a smaller dataset over a timespan of 24 hours, and the raw ODBA values are averaged over nonoveralpping windows of length 15 seconds and log transformed (lODBA), resulting in a total of 5{,}760 observations. Prior work typically assumes that latent movement states follow a
time-homogeneous Markov chain with a fixed transition matrix. This
implies that, as sample size increases, the marginal distribution of the
states approaches the chain’s stationary distribution (assuming
ergodicity). To assess our approach on real data, we apply the MDPM to the raw
observations, estimate a component (emission) density for each movement state,
and compare the results with those from existing HMM-based analyses.
\begin{figure}[H]
\centering
\includegraphics[width=1.\linewidth]{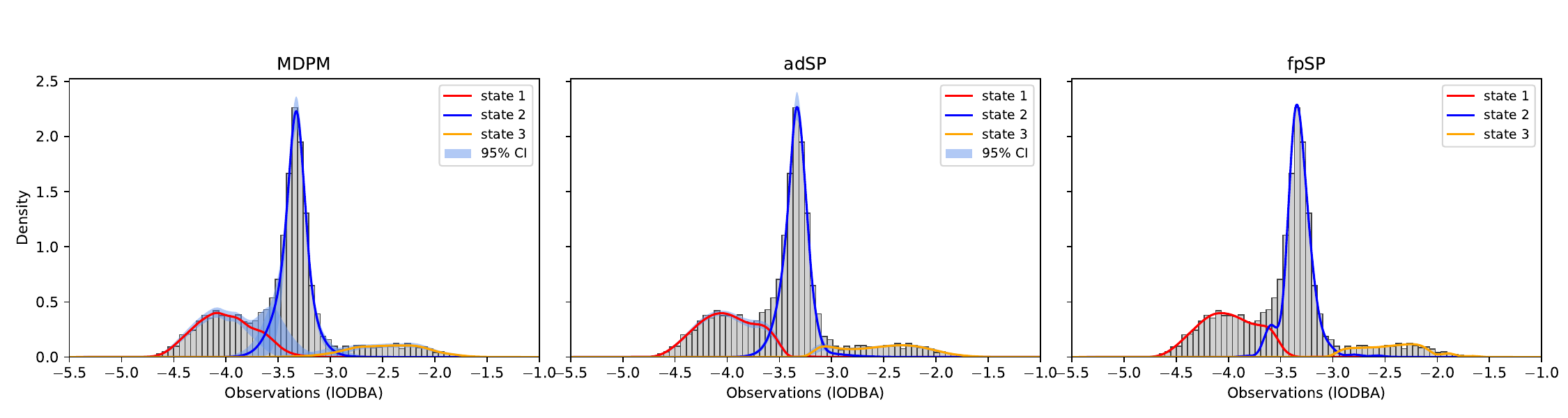}
\captionsetup{width=1.\linewidth}
\caption{Left, middle, and right panels show the lODBA histogram with the estimated
weighted emission densities from MDPM, adSP \citep{chen2024bayesian}, and
fpSP \citep{langrock2015nonparametric}, respectively.}
\label{fig:lDOBA}
\end{figure}
We analyze the lODBA data with MDPM and set \(K=3\), as justified on biological
grounds by \citet{langrock2018spline,chen2024bayesian}. In
Figure~\ref{fig:lDOBA}, the left panel shows the MDPM posterior mean emission
densities, weighted by estimated state occupancy, with pointwise 95\% credible intervals; the
middle panel reports the corresponding estimates from the Bayesian
spline-based HMM (adSP; \citet{chen2024bayesian}); and the right panel shows
the earlier P-spline HMM (fpSP; \citet{langrock2015nonparametric}). Although our approach does not exploit the temporal structure of the HMM and relies only on the marginal distribution of the observations, it produces state-dependent emission density estimates that are very similar to those obtained by methods that leverage the Markov structure. However, because only marginal information is used, the pointwise 95\% confidence intervals in regions where the state densities overlap tend to be wider than those produced by adSP. These results validate our approach’s ability to recover the underlying subpopulation distributions accurately using only population-level observations.


\section{Conclusions}
\label{section:conclusions}

In this paper, we develop a Bayesian framework with theoretical guarantees to study finite mixture models of nonparametric components. Our approach employs a mixture of Dirichlet process mixtures prior and leverages conjugacy to enable an efficient MCMC algorithm for posterior inference. Through several simulation studies and real-data applications in both univariate and multivariate settings, we demonstrate that the proposed method achieves strong empirical performance and scalability. To address the fundamental challenge of identifiability in nonparametric mixture models while allowing for overlapping component support, we introduce a novel separation condition defined in terms of distances between connected regions in the support of the latent mixing measure. Under this framework, we establish posterior contraction rates for both the overall mixture density and the individual component densities. In particular, we show that the contraction rate for component density estimation attains the minimax order of convergence. While this minimax rate does not admit a closed-form expression, we derive an explicit upper bound demonstrating that the rate is nearly polynomial, representing a substantial improvement over the logarithmic convergence rates typically associated with deconvolution-based estimation of mixing measures. To our best knowledge, this work provides the first theoretical guarantees for a practical Bayesian method that consistently estimates nonparametric component densities within a finite mixture modeling framework. 

\section*{Acknowledgments}
\label{section:acknowledgment}

 We are grateful to \hidetext{Yang Chen} for valuable discussions and helpful suggestions on the domain applications which helped to motivate this work.
We thank David E. Jones for providing the raw observations of FK Aqr and FL Aqr obtained with the XMM-Newton Observatory, an European Space Agency (ESA) science mission with instruments and contributions directly funded by ESA Member States and NASA; these data are analyzed in Section~\ref{section:disentagle_astro_sources}. We thank Sida Chen for sharing the lODBA dataset analyzed in Section~\ref{section: oceanic_shark}.
\end{spacing}

\begin{spacing}{1.1}
\bibliography{Yilei}
\end{spacing}

\newpage


\appendix 

\section{Algorithm}
\label{section:algorithm}
\begin{spacing}{1.2}
This section describes the slice sampler algorithm for the univariate MDPM hierarchical model in \eqref{hierarchical_model}. The multivariate version, presented in Section~\ref{section:Multivariate}, is a natural generalization of this algorithm. For each component $1\leq i\leq K$, the random measure $H_i\sim \mathrm{DP}(\alpha H_{i0})$ can be generated via the stick-breaking construction (\cite{sethuraman1994constructive}):
\begin{align}
H_i &= \sum_{j=1}^{\infty} \beta_{ij}\,\delta_{(u_{ij},\sigma_{ij})}, \notag\\
\beta_{ij} &= \nu_{ij}\!\!\prod_{k=1}^{j-1}(1-\nu_{ik}),\ \nu_{ij}\overset{\mathrm{i.i.d.}}{\sim}\mathrm{Beta}(1,\alpha),\ (u_{ij},\sigma_{ij})\sim H_{i0}, \label{draw_new_beta}
\end{align}
where \(\delta_{(u_{ij},\sigma_{ij})}\) denotes the Dirac measure at 
\((u_{ij},\sigma_{ij})\). For each sample $x_i$ ($1\leq i\leq n$), $z_i\in\{1,\cdots, K\}$ denotes its component label. Conditional on $z_i=k$, define $s_i\in\mathbb{N}$ to be the cluster (atom) label of $x_i$ within component $k$, i.e.,  $P(s_i=j|z_i=k)=\beta_{kj}$, and $x_i\mid z_i=k,s_i=j\sim g_{u_{kj},\sigma_{kj}}$. Introduce a slice variable for each observation $x_i$: given the atom index \(s_i\) in component \(z_i\), denote $\rho_i$ the slice variable $\rho _{i} \ \sim \ \mathrm{Uniform}( 0,\beta _{z_{i} s_{i}})$ and find the minimum $\rho ^{*} \coloneq \underset{i}{\min} \rho _{i}$. The slice variables restrict attention to atoms whose weights exceed the global threshold \(\rho^*\),
thereby yielding a finite, iteration–specific active set. Define the truncation index in component \(i\) as $m_i = \min \left\{\, j \geq 1 : 1 - \sum_{k=1}^j \beta_{ik} < \rho^* \,\right\}$, so that only atoms \(j=1,\dots,m_i\) need to be considered when updating allocations for data assigned to component \(i\).

To sample $(\boldsymbol{c},\boldsymbol{r})$ from the repulsive prior (\ref{repulsive_prior}), we introduce another auxiliary slice variable $\xi$ following \cite{repulsive_2012}:
\begin{equation}\label{draw_xi}
\begin{aligned}
\zeta  & =\begin{cases}
0, & \text{if} \ \underset{1\leq i< j\leq K}{\min} |c_{i} -c_{j} |-r_{i} -r_{j} \leq 0\\
\underset{1\leq i< j\leq K}{\min}\exp\left( -\frac{\tau }{( |c_{i} -c_{j} |-r_{i} -r_{j})^{\nu }}\right) , & \text{else}
\end{cases} \\
\xi  & \ \sim \ \mathrm{Uniform}( 0,\ \zeta )
\end{aligned}
\end{equation}
Let $\boldsymbol{z} = (z_i)_{i=1}^n$, $\boldsymbol{s} = (s_i)_{i=1}^n$, $\boldsymbol{u} = \bigl( (u_{ij})_{j=1}^{m_i} \bigr)_{i=1}^K$, $\boldsymbol{x} = (x_i)_{i=1}^n$, $\boldsymbol{c}_{-i} = (c_j)_{j \neq i}$, $\boldsymbol{r}_{-i} = (r_j)_{j \neq i}$, and define $\mathcal{J}_i = \{\, j : \exists\, 1\leq l \leq n, \; z_l=i, \; s_l=j \,\}$. Then the full conditional of $c_i$ is
\begin{align}
p( c_{i} |\boldsymbol{c}_{-i} ,\boldsymbol{r} ,\xi, \boldsymbol{u} ,\boldsymbol{z} ,\boldsymbol{s}) & \varpropto g_{\mu _{0} ,\eta }( c_{i}) \cdot \mathbf{1}\{\xi < \zeta \} \cdot \prod _{j\in \mathcal{J}_{i}} g_{c_{i} ,\sigma _{0}}( u_{ij})\mathbf{1}_{[ c_{i} -r_{i} ,\ c_{i} +r_{i}]}( u_{ij})\notag\\
 & \varpropto g_{\mu ',\sigma '}( c_{i}) \cdot \mathbf{1}\{\xi < \zeta \}\prod _{j\in \mathcal{J}_{i}}\mathbf{1}_{[ c_{i} -r_{i} ,\ c_{i} +r_{i}]}( u_{ij}) \label{draw_ci}
\end{align}
where $\mu '=\frac{\mu _{0} \sigma _{0}^{2} +\left(\sum _{j\in \mathcal{J}_{i}} u_{ij}\right) \cdot \eta ^{2}}{\sigma _{0}^{2} +|\mathcal{J}_{i} |\cdot \eta ^{2}} ,\ \sigma '=\frac{1}{\sqrt{1/\eta ^{2} +|\mathcal{J}_{i} |/\sigma _{0}^{2}}}$. Furthermore, the full conditional of $r_i$ is 
\begin{equation}\label{draw_ri}
p( r_{i} |\boldsymbol{c} ,\boldsymbol{r}_{-i} ,\xi, \boldsymbol{u} ,\boldsymbol{z} ,\boldsymbol{s}) \varpropto \mathrm{Gamma}( r_{i}) \cdot \frac{1}{\left( 1-2\Phi \left( -\frac{r_{i}}{\sigma _{0}}\right)\right){^{|\mathcal{J}_{i} |}}} \cdot \mathbf{1}\{\xi < \zeta \} \cdot \prod _{j\in \mathcal{J}_{i}}\mathbf{1}_{[ c_{i} -r_{i} ,\ c_{i} +r_{i}]}( u_{ij}),
\end{equation}
where $\Phi(\cdot)$ is the standard normal cdf. The posterior of $r_i$ is no longer conjugate, and this is the only step that requires a Metropolis-Hastings update.

Let $\boldsymbol{\beta} = \bigl( (\beta_{ij})_{j=1}^{m_i} \bigr)_{i=1}^K$, and $\boldsymbol{\sigma} = \bigl( (\sigma_{ij})_{j=1}^{m_i} \bigr)_{i=1}^K$. We update $(z_i,s_i)$ as a block. First, $z_i$ is sampled through the following rule:
\begin{equation}\label{draw_zi}
p( z_{i} =k|\boldsymbol{x} ,\boldsymbol{w} ,\rho _{i} ,\boldsymbol{\beta } ,\boldsymbol{u} ,\boldsymbol{\sigma }) \varpropto w_{k} \cdot \left(\sum _{j=1}^{m_{k}}\mathbf{1}\{\beta _{kj}  >\rho _{i}\} \cdot g_{u_{kj} ,\sigma _{kj}}( x_{i})\right) ,\ k=1,\cdots ,K.
\end{equation}
Given $z_i=k$, sample $s_i$ from
\begin{equation}\label{draw_si}
p( s_{i} =j|z_{i} =k,\boldsymbol{x} ,\rho _{i} ,\boldsymbol{\beta } ,\boldsymbol{u} ,\mathbf{\sigma }) \varpropto \mathbf{1}\{\beta _{kj}  >\rho _{i}\} \cdot g_{u_{kj} ,\sigma _{kj}}( x_{i}) ,\ j=1,\cdots ,m_{k}.
\end{equation}
Let $n_k=\#\{i:z_i=k\}$ for $k=1,\cdots,K$. The posterior of $\boldsymbol{w}$ given $\boldsymbol{z}$ is
\begin{equation}\label{draw_w}
\boldsymbol{w}\ |\ \boldsymbol{z} \ \sim \ \mathrm{Dir}( 1/K+n_{1} ,\cdots ,1/K+n_{K}).
\end{equation}
Now given $z_i=k,\ s_i=j$, let $\mathcal{I}_{kj} =\{i:z_{i} =k,s_{i} =j\}$, then the full conditional of $u_{kj}$ is
\begin{equation}\label{draw_ukj}
p( u_{kj} |\boldsymbol{x} ,\boldsymbol{z} ,\boldsymbol{s} ,c_{k} ,r_{k} ,\sigma _{kj}) \varpropto g_{\mu ',\sigma '}( u_{kj}) \cdot \mathbf{1}_{[ c_{k} -r_{k} ,\ c_{k} +r_{k}]}( u_{kj}),
\end{equation}
where $\mu '=\frac{c_{k} \sigma _{kj}^{2} +\left(\sum _{i\in \mathcal{I}_{kj}} x_{i}\right) \cdot \sigma _{0}^{2}}{\sigma _{kj}^{2} +|\mathcal{I}_{kj} |\cdot \sigma _{0}^{2}} ,\ \sigma '=\frac{1}{\sqrt{1/\sigma _{0}^{2} +|\mathcal{I}_{kj} |/\sigma _{kj}^{2}}}$, and given $u_{kj}$, the full conditional of $\sigma_{kj}^2$ is
\begin{equation}\label{draw_sigmakj]}
\sigma _{kj}^{2} |\boldsymbol{x} ,\boldsymbol{z} ,\boldsymbol{s} ,u_{kj} \ \sim \ \mathrm{InvGamma}\left( \theta _{1} +|\mathcal{I}_{kj} |/2,\ \theta _{2} +\sum _{i\in \mathcal{I}_{kj}}( x_{i} -u_{kj})^{2} /2\right),
\end{equation}
where $\theta_1$ and $\theta_2$ are the hyperparameters in the inverse-Gamma prior distribution in (\ref{base_measure}). Finally, define the vector of active-atom weights for component $k$ as $\boldsymbol{\beta }_{k} =( \beta _{k1} ,\cdots ,\beta _{kj_{k}} ,\beta _{k0}) ,\ \beta _{k0} =1-\sum _{j=1}^{j_{k}} \beta _{kj_{k}}$. Following \cite{ge2015distributed}, the posterior of $\boldsymbol{\beta}_k$ is
\begin{equation}\label{draw_betak}
\boldsymbol{\beta }_{k} |\boldsymbol{z} ,\boldsymbol{s} ,\alpha \ \sim \ \mathrm{Dir}( |\mathcal{I}_{k1} |,\cdots |\mathcal{I}_{kj_{k}} |,\alpha ).
\end{equation}
For each active atom $j=1,\cdots,j_k$ in component $k=1,\cdots, K$, draw the minimum slice variable
\begin{equation}\label{draw_rhokj}
b_{kj} \ \sim \ \mathrm{Beta}( 1,|\mathcal{I}_{kj} |) ,\ \rho _{kj}^{*} =\beta _{kj} b_{kj} ,\ i_{kj}^{*} \ \sim \ \mathrm{Uniform}(\mathcal{I}_{kj})
\end{equation}
and set the global slice threshold
\begin{equation}\label{draw_rhostar}
\rho ^{*} =\underset{1\leq k\leq K}{\min} \ \underset{1\leq j\leq j_{k}}{\min} \rho _{kj}^{*}
\end{equation}
\end{spacing}

\begin{spacing}{0.8}
\begin{algorithm}[H]
\caption{Slice Sampler for MDPM}\label{algo1}
\SetAlgoLined
\Input{$\boldsymbol{x}$, $\boldsymbol{w}^{(t)}$,$\boldsymbol{c}^{(t)}$,$\boldsymbol{r}^{(t)}$,$\boldsymbol{z}^{(t)}$,$\boldsymbol{s}^{(t)}$,$\boldsymbol{\beta}^{(t)}$,$\boldsymbol{u}^{(t)}$,$\boldsymbol{\sigma}^{(t)}$,$\{\rho_{kj}^{*(t)}\}, \{i_{kj}^{*(t)}\}$,$\rho^{*(t)}$}
\Output{$\boldsymbol{w}^{(t+1)}$,$\boldsymbol{c}^{(t+1)}$,$\boldsymbol{r}^{(t+1)}$,$\boldsymbol{z}^{(t+1)}$,$\boldsymbol{s}^{(t+1)}$,$\boldsymbol{\beta}^{(t+1)}$,$\boldsymbol{u}^{(t+1)}$,$\boldsymbol{\sigma}^{(t+1)}$,$\{\rho_{kj}^{*(t+1)}\},\{i_{kj}^{*(t+1)}\}$,$\rho^{*(t+1)}$}
\BlankLine
\For{$k \leftarrow 1$ \KwTo $K$}{
  \While{$1 - \sum_{j=1}^{j_k}\beta_{kj} \ge \rho^{*}$}{
    $j_k \gets j_k + 1$\;
    draw $\beta_{k,j_k}$,$u_{k,j_k}$ and $\sigma_{k,j_k}$ via \eqref{draw_new_beta}\;
  }
}
\BlankLine
\Fn{\FMap{$x_i$}}{
  \tcp{Global variables: $\boldsymbol{z}^{(t)}, \boldsymbol{s}^{(t)}, \boldsymbol{w}^{(t)},\boldsymbol{\beta}^{(t)},\boldsymbol{u}^{(t)}, \boldsymbol{\sigma}^{(t)}, \{\rho_{kj}^{*(t)}\}, \{i_{kj}^{*(t)}\}$}
\lIf{$i = i^\star_{z_i,s_i}$}{$\rho_i \gets \rho^{\star(t)}_{z_i,s_i}$}
\lElse{$\rho_i \gets \mathrm{Uniform}(\rho^{\star(t)}_{z_i,s_i}, \beta_{z_i,s_i}^{(t)})$}
  draw $z_i^{(t+1)}$ via (\ref{draw_zi})\;
  draw $s_i^{(t+1)}$ via (\ref{draw_si})\;
  \KwRet $\{\text{key}:(z_i^{(t+1)},s_i^{(t+1)}),\ \text{value}: x_i\}$ 
}
\BlankLine
draw $\boldsymbol{w}^{(t+1)}$ via (\ref{draw_w})\;
\BlankLine
\For{each unique key $(z^{(t+1)}, s^{(t+1)})$}{
collect all $\{x_i\}$ that have this key\;
$n=\#\{x_i\}$\;
$sx = \sum x_i$; $ssx=\sum x_i^2$\;
emit $\{\text{key}:(z^{(t+1)},s^{(t+1)}),\ \text{value}: (n,sx, ssx)\}$
}
\BlankLine
\Fn{\FReduce{$(z^{(t+1)},s^{(t+1)}),\ (n,sx, ssx)$}}{
  \tcp{Global variables: $\boldsymbol{c}^{(t)}, \boldsymbol{r}^{(t)},\boldsymbol{u}^{(t)}, \boldsymbol{\sigma}^{(t)}$}
  draw $u_{z_i,s_i}^{(t+1)}$ via (\ref{draw_ukj})\;
  draw $\sigma_{z_i,s_i}^{(t+1)}$ via (\ref{draw_sigmakj]})\;
\KwRet $u_{z_i,s_i}^{(t+1)},\ \sigma_{z_i,s_i}^{(t+1)},\ n_{z_i,s_i}$
}
\BlankLine
draw $\boldsymbol{\beta}^{(t+1)}$ via (\ref{draw_betak})\;
draw $\{\rho_{kj}^{*(t+1)}\}, \{i_{kj}^{*(t+1)}\}$ via (\ref{draw_rhokj}), update $\rho^{*(t+1)}$ via (\ref{draw_rhostar})\;
draw $\xi$ via (\ref{draw_xi})\;
draw $\boldsymbol{c}^{(t+1)}$ via (\ref{draw_ci})\;
draw $\boldsymbol{r}^{(t+1)}$ via (\ref{draw_ci})\;
\end{algorithm}
\end{spacing}

\begin{spacing}{1.2}
We follow the MapReduce framework of \cite{ge2015distributed} to design Algorithm~\ref{algo1}. Throughout, a superscript $^{(t)}$ denotes the value at the $t$-th MCMC iteration.
\end{spacing}


\begin{spacing}{1.2}
\section{Simulations}
\label{section:simulations}

In this section, we provide the implementation details of the simulation studies presented in Sections \ref{section:Mixture_of_DP} and \ref{section:Multivariate}.

\subsection{\texorpdfstring{Figure~\eqref{fig:a} and \eqref{fig:b}}{Figure (a) and (b)}}
Simulation Example 1, shown in Figures \eqref{fig:a} and \eqref{fig:b}, consists of three components. The second component is a skewed exponential-power distribution
\begin{equation*}
f( x;\mu ,\sigma ,p,\alpha ) =\begin{cases}
\frac{1}{\sigma 2p^{1/p} \Gamma ( 1+1/p)}\exp\left\{-\frac{1}{2p}\left| \frac{x-\mu }{\alpha \sigma }\right| ^{p}\right\} , & \text{if} \ x\leq \mu \\
\frac{1}{\sigma 2p^{1/p} \Gamma ( 1+1/p)}\exp\left\{-\frac{1}{2p}\left| \frac{x-\mu }{( 1-\alpha ) \sigma }\right| ^{p}\right\} , & \text{if} \ x >\mu 
\end{cases}
\end{equation*}
with $\mu=0.9$, $\sigma=1.1$, $p=0.7$, $\alpha=0.7$. The third component is a Laplace distribution
\begin{equation*}
f( x;\mu ,\theta ) =\frac{1}{2\theta }\exp\left( -\frac{|x-\mu |}{\theta }\right)
\end{equation*}
with $\mu=6.3$, and $\theta=1.1$. The first component is a random linear combination of Hermite functions (defined in \eqref{hermite_functions}), constructed as follows:
\begin{enumerate}
    \item Generate the Hermite function $\psi_j(x)$ for $j=0,\cdots, L-1$.
    \item Draw coefficient $a_j\sim N(0,\sigma_j^2)$, where $\sigma_j=\frac{\tau}{(j+1)^{\alpha/2}}$, for $j=0,\cdots, L-1$.
    \item Define $\tilde{f}(x)=(\sum_{j=0}^{L-1}a_j\psi_j(\frac{x-\mu}{s})s^{-1})^2$.
    \item Normalize to obtain the density 
    \begin{equation}\label{normalize_density}
f( x) =\begin{cases}
\frac{\tilde{f}( x)}{\int _{-T}^{T}\tilde{f}( x) dx} , & x\in [ -T,T]\\
0, & \text{otherwise}
\end{cases},
\end{equation}
with $L=14$, $\tau=0.7$, $\alpha=1.6$,  $\mu=-7.0$, $s=0.8$, and $T=15$. The histogram in \eqref{fig:a} is based on $10{,}000$ samples drawn from the mixture model.
\end{enumerate}

\subsection{\texorpdfstring{Figure~\eqref{fig:c} and \eqref{fig:d}}{Figure (c) and (d)}}
Simulation Example 2, shown in Figures \eqref{fig:c} and \eqref{fig:d}, consists of two components. The spike component is generated as follows:
\begin{enumerate}
    \item Randomly generate two orders $j_0$ and $j_1$ from the set $\{2,3,5,6\}$;
    \item Draw two coefficients $a_0\sim N(0,\sigma_0^2)$, $a_1\sim N(0,\sigma_1^2)$, where $\sigma_0=\frac{\tau}{(j_0+1)^{\alpha/2}}$, and $\sigma_1 = \frac{\tau}{(j_1+1)^{\alpha/2}}$;
    \item Define $\tilde{f}( x) =\left(\sum _{j=0}^{1} a_{j} \psi _{j}\left(\frac{x-\mu }{s}\right) s^{-1}\right)^{2\gamma }$;
    \item Normalize $\tilde{f}( x)$ to obtain the density $f(x)$ as in \eqref{normalize_density},
\end{enumerate}
where we set $\tau = 2.6$, $\alpha=1.5$, $\mu=0$, $s=1.8$, $\gamma=5.8$, and $T=20$. The slab component is constructed according to the following procedure:
\begin{enumerate}
    \item Draw $a_j\sim N(0,\sigma_j^2)$, where $\sigma_j=\frac{\tau}{(j+1)^{\alpha/2}}$ for $j=0,\cdots, 5$, and $\sigma_j=\frac{0.35\tau}{(j+1)^{\alpha/2}}$ for $j=6, 7$.
    \item Define $\tilde{f}(x)=(\sum_{j=0}^{7}a_j\psi_j(\frac{x-\mu}{s})s^{-1})^2$.
    \item Normalize $\tilde{f}( x)$ to obtain the density $f(x)$ as in \eqref{normalize_density},
\end{enumerate}
where $\tau=0.9$, $\alpha=5.2$, $\mu=0$, $s=14.5$, and $T=20$.

The histogram in \eqref{fig:c} is estimated from $10{,}000$ samples, and the density estimates in \eqref{fig:d} are obtained by MDPM with fixed intervals: $I_1=(0,0.5]$ and $I_2=[3.0,\infty]$.

\subsection{Figure \ref{fig:threecol}}
The simulation example shown in Figure \ref{fig:threecol} considers a two-component mixture model. Each component itself consists of a mixture of 100 bivariate Gaussian distributions. For the first component, the Gaussian means are equally spaced along a circle of radius 2 centered at $[-2,2]$. Similarly, for the second component, the means are equally spaced along a circle of radius 2 centered at $[2,2]$. The covariance matrices of the Gaussian distributions are generated as follows.
\begin{itemize}
    \item For each mixture component, the $11$ Gaussian distributions whose means are closest to the origin $[0,0]$ have covariance matrices independently drawn from an inverse Wishart distribution $\mathcal{W}^{-1}(\nu_1,\Psi_1)$, where $\nu_1=6$, $\Psi_1=\begin{pmatrix}
4.5 & 0\\
0 & 4.5
\end{pmatrix}$. The relatively large diagonal entries induce substantial overlap between the two components in the tails near the origin.
\item The remaining 89 Gaussian distributions in each component have covariance matrices independently drawn from \(\mathcal{W}^{-1}(\nu_2, \Psi_2)\), where \(\nu_2 = 6\) and
$\Psi_2 =
\begin{pmatrix}
0.8 & 0 \\
0 & 0.8
\end{pmatrix}$.
\end{itemize}
In this simulation example, the underlying regions $I_1$ and $I_2$ are unknown and inferred from the posterior. 


\section{Proofs for Section \ref{section:Identifiability}: Theorem \ref{Theorem:identifiability}}
\label{section:proof_identifiability}

\begin{proof} [Proof of Theorem \ref{Theorem:identifiability}]
    We first show that for any function $h(x)$ of the form $h( x) =\int g_{u,\sigma }( x) dG( u)$, where the mixing measure $G$ has compact support, if $h( x) =\int g_{u,\sigma ^{\prime}}( x) dG^{\prime}( u)$, where $G^{\prime}$ also has compact support, then we must have $\sigma=\sigma^{\prime}$ and $G=G^{\prime}$. If $\sigma^{\prime}\neq \sigma$, w.l.o.g, assume that $\sigma^{\prime}>\sigma$. Because $G$ and $G^{\prime}$ have compact support, we can find $u_0\in\mathbb{R}$ such that $\operatorname{supp}(G)\subseteq (-\infty, u_0-1]$, and $u_0^{\prime}\in (0,+\infty)$ such that $\operatorname{supp}(G^{\prime})\subseteq [-u_0^{\prime}, u_0^{\prime}]$. By definition, for any $x\in\mathbb{R}$,
\begin{equation}\label{Gaussian_conv_euqal}
\frac{1}{\sqrt{2\pi } \sigma }\int e^{-\frac{( x-u)^{2}}{2\sigma ^{2}}} dG( u) =\frac{1}{\sqrt{2\pi } \sigma ^{\prime}}\int e^{-\frac{( x-u)^{2}}{2\sigma ^{\prime 2}}} dG^{\prime}( u).
\end{equation}
Multiplying both sides by $\exp\left(\frac{ (x-u_{0})^{2}}{2\sigma ^{2}}\right)$, we get
\begin{align}
\int e^{\frac{( x-u_{0})^{2}}{2\sigma ^{2}} -\frac{( x-u)^{2}}{2\sigma ^{2}}} dG( u) & =\frac{\sigma }{\sigma ^{\prime}}\int e^{\frac{( x-u_{0})^{2}}{2\sigma ^{2}} -\frac{( x-u)^{2}}{2\sigma ^{\prime 2}}} dG^{\prime}( u)\notag\\
\int e^{\frac{( u-u_{0}) x}{\sigma ^{2}}} \cdot e^{\frac{u_{0}^{2} -u^{2}}{2\sigma ^{2}}} dG( u) & =\frac{\sigma }{\sigma ^{\prime}} \cdot e^{\left(\frac{1}{2\sigma ^{2}} -\frac{1}{2\sigma ^{\prime 2}}\right) x^{2} -\left(\frac{u_{0}}{\sigma ^{2}}\right) x+\frac{u_{0}^{2}}{2\sigma ^{2}}}\cdot\int e^{\frac{ux}{\sigma ^{\prime 2}} -\frac{u^{2}}{2\sigma ^{\prime 2}}} dG^{\prime}( u).\label{gauss_conv_expansion}
\end{align}
Because $\operatorname{supp}(G)\subseteq (-\infty, u_0-1]$, for any $x>0$,
\begin{equation*}
\text{LHS} \leq e^{-\frac{x}{\sigma ^{2}}}\int e^{\frac{u_{0}^{2} -u^{2}}{2\sigma ^{2}}} dG( u)\rightarrow 0,\ \text{as} \ x\rightarrow +\infty, 
\end{equation*}
However, because $\sigma^{\prime}>\sigma$, and $\operatorname{supp}(G^{\prime})\subseteq [-u_0^\prime,u_0^\prime]$, for any sufficiently large $x>0$,
\begin{equation*}
\text{RHS} \geq \frac{\sigma }{\sigma ^{\prime}} \cdot e^{\left(\frac{1}{2\sigma ^{2}} -\frac{1}{2\sigma ^{\prime 2}}\right) x^{2} -\left(\frac{u_{0}}{\sigma ^{2}}\right) x+\frac{u_{0}^{2}}{2\sigma ^{2}}} \cdot e^{-\frac{u_{0}^{\prime} x}{\sigma ^{\prime 2}} -\frac{u_{0}^{\prime2}}{2\sigma ^{\prime 2}}}\rightarrow \infty ,\ \text{as} \ x\rightarrow +\infty 
\end{equation*}
Therefore, there must exist $x$ sufficiently large such that the left-hand side is not equal to the right-hand side, contradicting the assumption. A similar contradiction arises if $\sigma' < \sigma$. Hence we conclude that $\sigma=\sigma^{\prime}$. Now assume $h( x) =\int g_{u,\sigma }( x) dG( u) =\int g_{u,\sigma }( x) dG^{\prime}( u)$, then for any $t\in\mathbb{R}$,
\begin{align*}
\int e^{itx} h( x) dx & =\int e^{itx}\left(\int g_{u,\sigma }( x) dG( u)\right) dx\\
 & =\int \left(\int e^{itx} g_{u,\sigma }( x) dx\right) dG( u)\\
 & =\int e^{itu-\frac{1}{2} \sigma ^{2} t^{2}} dG( u)\\
 & =e^{-\frac{1}{2} \sigma ^{2} t^{2}}\int e^{itu} dG( u).
\end{align*}
Similarly, we have $\int e^{itx} h( x) dx=e^{-\frac{1}{2} \sigma ^{2} t^{2}}\int e^{itu} dG^{\prime}( u)$, therefore, $\int e^{itu} dG( u) =\int e^{itu} dG^{\prime}( u)$, $\forall t\in \mathbb{R}$. Since the characteristic functions of $G$ and $G'$ are identical, we must have $G=G^{\prime}$.

Therefore, given any $f$ as defined in (\ref{interval_location_mixture}), we can find unique $\sigma$ and $G$, such that $f=\int g_{u,\sigma }( x) dG( u)$. Under condition \eqref{C1}, $G$ can be uniquely separated into $K$ components. Under condition \eqref{C2}, we first find the unique separation $S_{\operatorname{supp}(G)}$ as defined in Section \ref{section:preliminary}, then among all pairs of neighboring sets $(S_i,S_j)\in \mathcal{N}(S_{\operatorname{supp}(G)})$, we identify the $K-1$ pairs with the largest distances $d_c(S_i,S_j)$. The interval of length $d_c(S_i,S_j)$ between each of these pairs $S_i$ and $S_j$ will also be one of the $K-1$ intervals that separate $\operatorname{supp}(V_1),\cdots, \operatorname{supp}(V_K)$. Therefore $f_1,\cdots, f_K$ can also be uniquely identified.

\end{proof}

\section{Proofs for Section \ref{subsection: mixture_density}: Theorem \ref{thm: density_posterior}}
\label{section:proof_post_overall}
In this section, Lemma~\ref{lemma: wk_diff_bound} provides a lower bound on the probability that, under the truncated Dirichlet distribution, $(W_1,\cdots, W_K)$ lies in a neighborhood of the true weights $(w_{10},\cdots, w_{K0})$. The corresponding result for a non-truncated Dirichlet prior is given in Lemma~6.1 of \cite{ghosal2000convergence}.

\begin{lemma}\label{lemma: wk_diff_bound}
    Let $(W_1,\cdots,W_K)$ follow a truncated Dirichlet distribution under assumption \eqref{A1}. Let $(w_{10},\cdots, w_{K0})$ be any point on the $K$-simplex such that $w_{k0}\in \left[\underline{w} ,\frac{1-\underline{w}}{K-1}\right]$, for some $0<\underline{w}<1/K^2$, $\ k=1,\cdots,K$. There exist a positive constant $C$ depending only on $\mathbf{\alpha}$ and $\underline{w}$ and $K$ such that, for $\epsilon\leq \frac{1}{K\sqrt{K-1}}$,
\begin{equation}\label{ineq:wk_conctr}
\text{Pr}\left(\sum _{k=1}^{K} |W_{k} -w_{k0} |\leq 2\epsilon \right) \geq C(\underline{w},\alpha, K)\exp\left(-2(K-1)\log\frac{1}{\epsilon }\right).
\end{equation}
\end{lemma}
\begin{proof}
    Find an index $k$ such that $w_{k0}\leq 1/K$. By relabeling, we can assume that $k=K$. If $|w_k-w_{k0}|\leq \epsilon^2$ and $w_k\in \left[\underline{w} ,\frac{1-\underline{w}}{K-1}\right]$ for $k=1,\cdots, K-1$, then because $\epsilon ^{2} \leq \frac{1}{K^{2}(K-1)}$,
\begin{equation*}
\sum _{k=1}^{K-1} w_{k} \geq 1-w_{K0} -( K-1) \epsilon ^{2} \geq ( K-1)\left( 1/K-\epsilon ^{2}\right) \geq 1-\frac{K+1}{K^{2}}.
\end{equation*}
So we have $w_{K} \leq \frac{K+1}{K^{2}} < \frac{1-\underline{w}}{K-1}$. In addition, because $\sum _{k=1}^{K-1} w_{k} \leq 1-\underline{w}$, we also have $w_K\geq \underline{w}$. Hence there exists $w=(w_1,\cdots,w_K)$ in the truncated simplex with these first $K-1$ coordinates such that $\sum _{k=1}^{K} w_{k} =1,\ w_{k} \in [\underline{w} , \frac{1-\underline{w}}{K-1}] ,\forall 1\leq k\leq K$. Furthermore, $\sum _{k=1}^{K} |w_{k} -w_{k0} |\leq 2\sum _{k=1}^{K-1} |w_{k} -w_{k0} |\leq 2\epsilon ^{2}( K-1) \leq 2\epsilon $. Therefore the probability on the left-hand side of (\ref{ineq:wk_conctr}) is bounded below by
\begin{gather*}
\text{Pr}\left( |W_{k} -w_{k0} |\leq \epsilon ^{2} , W_k\in[\underline{w},\frac{1-\underline{w}}{K-1}], k=1,\cdots ,K-1\right)\\
\geq C(\underline{w},\mathbf{\alpha}, K)\cdot\prod _{k=1}^{K-1}\int _{\max\left( w_{k0} -\epsilon ^{2} ,\underline{w}\right)}^{\min\left( w_{k0} +\epsilon ^{2} ,\frac{1-\underline{w}}{K-1}\right)} w_{i}^{\alpha _{i} -1} dw_{i},
\end{gather*}
where we use that $\left( 1-\sum _{k=1}^{K-1} w_{i}\right)^{\alpha _{K} -1} \geq 1$, since $\alpha_{K}\leq 1$. Similarly, since $\alpha_k\leq 1$ for every $k$, we can lower bound the integrand by $1$. Note that, under assumption, $\underline{w}<\frac{1}{K^2}$, which implies $\frac{1-\underline{w}}{K-1} -\underline{w}  >\frac{1}{K}$. Moreover, since $\epsilon ^{2} \leq \frac{1}{K^{2}(K-1)}$, the interval of integration contains at least an interval of length $\epsilon^2$. 
We can bound the last display from below by
\begin{equation*}
C(\underline{w},\mathbf{\alpha},K)\cdot\epsilon ^{2( K-1)} \geq C(\underline{w},\mathbf{\alpha},K)\cdot\exp\left( -2( K-1)\log\frac{1}{\epsilon }\right).
\end{equation*}
This concludes the proof.
\end{proof}

Now we are ready to prove the prior concentration inequality of MDPM in the following lemma:

\begin{lemma}\label{lemma: mix_of_dp_prior_conctr}
     Let $f_{0} \in \mathcal{F}_{K,\sigma }(\boldsymbol{c} ,\boldsymbol{r} ,\underline{w})$, where $K, \sigma, \boldsymbol{c}, \boldsymbol{r}, \underline{w}$ are fixed and known. Consider the prior $\Pi$ constructed under specifications \eqref{A1}, \eqref{A2} and \eqref{A3}. Then by taking $\tilde{\epsilon }_{n} =\log n /\sqrt{n}$, when $n$ is larger than some constant, we have 
\begin{equation}
\Pi (\{f\ |\ f\in B_{2}( f_{0} ,\tilde{\epsilon }_{n})\}) \geq c_{3} e^{-c_{4} n\tilde{\epsilon }_{n}^{2}}
\end{equation}
where constants $c_3,c_4>0$.
\end{lemma}

\begin{remark*}
 Lemma~\ref{lemma: mix_of_dp_prior_conctr} shows that, for any finite number of components $K$, a mixture of $K$ Dirichlet process priors, under the specified conditions, attains the same prior concentration rate around the true mixture density $f_0$ as a single Dirichlet process prior.
\end{remark*}

\begin{proof}[Proof of Lemma~\ref{lemma: mix_of_dp_prior_conctr}]

As established in the proof of Theorem 5.1 in \cite{ghosal2001entropies}, the probability lower bound for the first set in equation~(5.12) of \cite{ghosal2001entropies} applies in our setting as well. The only difference is that $\sigma$ is fixed here; in this case, the first factor on the right-hand side of equation~(5.14) in \cite{ghosal2001entropies} can be replaced by 
1. Therefore the conclusion in (5.16) still holds, which in the proof is implied by the following: For any $k=1,\cdots,K$, $0<\epsilon<\frac{1}{K}$, and some positive constant $d_k$,
\begin{equation}\label{}
\Pi _{k}\left(\left\{V_{k} \ |\ \| f_{k} -f_{k0} \| _{1} \leq d_{k} \epsilon \left(\log\frac{1}{\epsilon} \right)^{1/2}\right\}\right) \geq C_k\exp\left[ -c_k\left(\log\frac{1}{\epsilon }\right)^{2}\right],
\end{equation}
for some positive constants $C_k$ and $c_k$. By Lemma \ref{lemma: wk_diff_bound}, when $\epsilon<\min(\frac{1}{K\sqrt{K-1}}, e^{-1})$, we have $\log\frac{1}{\epsilon}>1$, therefore
\begin{align*}
\text{Pr}\left(\sum _{k=1}^{K} |W_{k} -w_{k0} |\leq 2\epsilon \left(\log\frac{1}{\epsilon }\right)^{1/2}\right) & \geq \text{Pr}\left(\sum _{k=1}^{K} |W_{k} -w_{k0} |\leq 2\epsilon \right)\\
 & \geq C_{w}\exp\left( -2( K-1)\log\frac{1}{\epsilon }\right)
\end{align*}
for some positive constants $C_w$. Furthermore,
\begin{align*}
\Vert f-f_{0}\Vert _{1} & =\left\Vert \sum _{k=1}^{K}( W_{k} f_{k} -w_{k0} f_{k0})\right\Vert _{1}\\
 & =\left\Vert \sum _{k=1}^{K}( W_{k} -w_{k0})( f_{k} -f_{k0}) +( W_{k} -w_{k0}) f_{k0} +( f_{k} -f_{k0}) w_{k0}\right\Vert _{1}\\
 & \leq \sum _{k=1}^{K} |W_{k} -w_{k0} |\cdot \Vert f_{k} -f_{k0}\Vert _{1} +\sum _{k=1}^{K} |W_{k} -w_{k0} |+\sum _{k=1}^{K}\Vert f_{k} -f_{k0}\Vert _{1}\\
 & \leq 2\sum _{k=1}^{K}\Vert f_{k} -f_{k0}\Vert _{1} +\sum _{k=1}^{K} |W_{k} -w_{k0} |,
\end{align*}
and because $\mathbf{W}$ and $\Pi_k$'s are mutually independent, 
\begin{align*}
\Pi  & \left(\left\{f\ |\ \Vert f-f_{0}\Vert _{1} \leq 2\left(\sum _{k=1}^{K} d_{k} +1\right) \epsilon \log\left(\frac{1}{\epsilon }\right)^{1/2}\right\}\right)\\
\geq  & \text{Pr}\left(\sum _{k=1}^{K} |W_{k} -w_{k0} |\leq 2\epsilon \log\left(\frac{1}{\epsilon }\right)^{1/2} ,\text{and} \ \| f_{k} -f_{k0} \| _{1} \leq d_{k} \epsilon \left(\log\frac{1}{\epsilon }\right)^{1/2} ,k=1,\dotsc ,K\right)\\
= & \text{Pr}\left(\sum _{k=1}^{K} |W_{k} -w_{k0} |\leq 2\epsilon \log\left(\frac{1}{\epsilon }\right)^{1/2}\right) \cdot \prod _{k=1}^{K} \Pi _{k}\left( \| f_{k} -f_{k0} \| _{1} \leq d_{k} \epsilon \left(\log\frac{1}{\epsilon }\right)^{1/2}\right)\\
= & \prod _{k=1}^{K} C_{k} \cdot C_{w}\exp\left[ -\left(\sum _{k=1}^{K} c_{k}\right)\left(\log\frac{1}{\epsilon }\right)^{2} -2 (K-1)\log\frac{1}{\epsilon }\right]\\
\geq  & \tilde{C}\exp\left[ -\tilde{c}\left(\log\frac{1}{\epsilon }\right)^{2}\right]
\end{align*}
for some positive constants $\widetilde{C}$ and $\tilde{c}$. By (\ref{ineq: l1_hellinger}) and Lemma 4.1 in \cite{ghosal2001entropies} and analogous results for location-scale mixtures, we know when $\epsilon$ is sufficiently small such that\\ $2\left(\sum _{k=1}^{K} d_{k} +1\right) \epsilon \log\left(\frac{1}{\epsilon }\right)^{1/2} < \frac{1}{4}$, 
\begin{equation*}
\left\{f\ |\ \Vert f-f_{0}\Vert _{1} \leq 2\left(\sum _{k=1}^{K} d_{k} +1\right) \epsilon \log\left(\frac{1}{\epsilon }\right)^{1/2}\right\} \subset B_{2}\left( f_{0} ,c\epsilon ^{1/2}\left(\log\frac{1}{\epsilon }\right)^{5/4}\right),
\end{equation*}
for some constant $c$, therefore
\begin{equation*}
\Pi \left(\left\{f=p_{\sigma_{0},\sum _{k=1}^{K} W_{k} V_{k}} |\ f\in B_{2}\left( f_{0} ,c\epsilon ^{1/2}\left(\log\frac{1}{\epsilon }\right)^{5/4}\right)\right\}\right) \geq \tilde{C}\exp\left[ -\tilde{c}\left(\log\frac{1}{\epsilon }\right)^{2}\right]
\end{equation*}
Putting $\epsilon'=c\epsilon^{1/2}(\log \frac{1}{\epsilon})^{5/4}$ and noting that $\log\frac{1}{\epsilon '} \sim \log\frac{1}{\epsilon }$, we have
\begin{equation}\label{pi_B_epsilon'}
\Pi (\{f\ |\ f\in B_{2}( f_{0} ,\epsilon ')\}) \geq c_{3}\exp\left[ -c_{4}\left(\log\frac{1}{\epsilon '}\right)^{2}\right]
\end{equation}
for some positive constants $c_3$ and $c_4$. Take $\tilde{\epsilon }_{n} =\log n /\sqrt{n}$, we obtain
\begin{equation*}
\Pi (\{f\ |\ f\in B_{2}( f_{0} ,\tilde{\epsilon }_{n})\}) \geq c_{3} e^{-c_{4} n\tilde{\epsilon }_{n}^{2}}.
\end{equation*}
\end{proof}

We now proceed to establish the posterior contraction rate around the true density $f_0$. To this end, we apply Theorem~2.1 of \cite{ghosal2001entropies}.

\begin{proof}[Proof of Theorem \ref{thm: density_posterior}]
    We verify the conditions (2.8), (2.9) and (2.10) of Theorem 2.1 in \cite{ghosal2001entropies} are satisfied for $\Pi_n=\Pi$, 
    $\mathcal{P}=\mathcal{F}_{K,\sigma }(\boldsymbol{c} ,\boldsymbol{r} ,\underline{w})$, $\mathcal{P}_n=\mathcal{P}$ with $\bar{\epsilon}_n=\log n/\sqrt{n}$ and $\tilde{\epsilon}_n=\log n/\sqrt{n}$. Condition (2.8) follows from Theorem 3.1 in \cite{ghosal2001entropies} with $\bar{\epsilon}_n=\log n/\sqrt{n}$. Lemma \ref{lemma: mix_of_dp_prior_conctr} establishes that (2.10) holds for $\tilde{\epsilon}_n=\log n/\sqrt{n}$.
\end{proof}


\section{\texorpdfstring{Proofs for Section \ref{subsection: component_density}: Construction of $\widehat{f}_{i}$}{Proofs for Section: Construction of fi}}
\label{appendix: construction of fi_hat}

In this section, we briefly describe the point estimate $\widehat{f}_i$ given in  \cite{bryon_2023}. To this end, we first introduce Hermite functions in section \ref{subsection:hermite_function}.

\subsection{Hermite functions}\label{subsection:hermite_function}
We review some useful basics of Hermite orthonormal basis to be used in the sequel. Let  $\psi_{j}$ be the $j$th (physicist's) Hermite functions, i.e.
\begin{equation}\label{hermite_functions}
\psi _{j}( x) =( -1)^{j}\frac{1}{\sqrt{2^{j} j!\sqrt{\pi }}} h_{j}( x) e^{-\frac{1}{2} x^{2}} ,\ \forall x\in \mathbb{R},
\end{equation}
where $h_{j}( x)$ is the (physicist’s) Hermite polynomials, i.e.
\begin{equation*}
h_{j}( x) =( -1)^{j} e^{x^{2}}\frac{d^{j} e^{-\xi ^{2}}}{d\xi ^{j}}\Bigl|_{\xi =x} ,\ \forall x\in \mathbb{R}.
\end{equation*}
The Hermite functions are orthonormal in $L^2(\mathbb{R})$, i.e. $\Vert \psi_i\Vert_2=1$ and $\langle \psi_i, \psi_j\rangle =0$ $\forall i,j\in \mathbb{N}_0$, where $\langle f,g \rangle=\int_{\mathbb{R}}f(x)g(x)dx$ denotes the inner product with respect to Lebesgue measure. For any $\mu\in \mathbb{R}$ and $j\in \mathbb{N}_0$, we denote $\psi _{j,\mu }( x) =\psi _{j}( x-\mu )$. It is known that 
\begin{equation}\label{shifted_hermite}
h_j(x - \mu) = \sum_{k=0}^j \binom{j}{k} (-2\mu)^{j-k} h_k(x).
\end{equation}
This can be proved by using the generating function for Hermite polynomials: $e^{-t^2 + 2tx} = \sum_{j=0}^\infty h_j(x) \frac{t^j}{j!}$, $\forall x, t\in \mathbb{R}$. Substitute $x$ by $x - \mu$ in the generating function: $e^{-t^2 + 2t(x - \mu)} = \sum_{j=0}^\infty h_j(x - \mu) \frac{t^j}{j!}$, and rewrite $e^{-t^2 + 2t(x - \mu)} = e^{-t^2 + 2tx}\cdot e^{-2t\mu} = \left(\sum_{j=0}^\infty h_j(x) \frac{t^j}{j!}\right) \left(\sum_{k=0}^\infty \frac{(-2\mu)^k}{k!} t^k\right)$, then by comparing coefficients of $\frac{t^j}{j!}$, we get (\ref{shifted_hermite}). Based on (\ref{shifted_hermite}), it can be checked that for any $\mu\in \mathbb{R}$ and any $i,j \in \mathbb{N}_0$, the inner product (with respect to Lebesgue measure) of $\psi_{i, 0}$ and $\psi_{j, \mu}$ takes the form 
\begin{align*}
\langle \psi _{i,0} ,\psi _{j,\mu } & \rangle \notag \\
= & \int _{-\infty }^{+\infty } \psi _{i,0}( x) \psi _{j,\mu }( x) dx \notag \\
= & \frac{( -1)^{i+j}}{\sqrt{2^{i+j} i!j!\pi }}\int _{-\infty }^{+\infty } h_{i}( x) h_{j}( x-\mu ) e^{-x^{2} /2} e^{-( x-\mu )^{2} /2} dx \notag \\
= & \frac{( -1)^{i+j}}{\sqrt{2^{i+j} i!j!\pi }} e^{-\mu ^{2} /4}\int _{-\infty }^{+\infty } h_{i}( x) h_{j}( x-\mu ) e^{-( x-\mu /2)^{2}} dx \notag \\
\overset{t=x-\mu /2}{=} & \frac{( -1)^{i+j}}{\sqrt{2^{i+j} i!j!\pi }} e^{-\mu ^{2} /4}\int _{-\infty }^{+\infty } h_{i}( t+\mu /2) h_{j}( t-\mu /2) e^{-t^{2}} dt. \notag 
\end{align*}
Using the equation \eqref{shifted_hermite}, we expand both $h_{i}( t+\mu /2)$ and $h_{j}( t-\mu /2)$ in terms of $\{h_k(t)\}$. By orthogonality $\int h_{m}( x) h_{n}( x) e^{-x^{2}} dx=\sqrt{\pi } 2^{n} n!\cdot \mathbf{1}_{n=m}$,
\begin{equation}
   \langle \psi _{i,0} ,\psi _{j,\mu } \rangle =  \sum _{k=0}^{i\land j}\left( e^{-\frac{1}{8} \mu ^{2}}( -1)^{i}\sqrt{\frac{k!}{i!}}\binom{i}{k}\left(\frac{\mu }{\sqrt{2}}\right)^{i-k}\right)\left( e^{-\frac{1}{8} \mu ^{2}}( -1)^{j}\sqrt{\frac{k!}{j!}}\binom{j}{k}\left(\frac{-\mu }{\sqrt{2}}\right)^{j-k}\right).
\end{equation}
Let $g_{\mu ,\sigma }(x) =\frac{1}{\sqrt{2\pi } \sigma } e^{-\frac{( x-\mu )^{2}}{2\sigma ^{2}}}$ be the density of normal distribution with mean $\mu$ and standard deviation $\sigma$. Since $g_{\mu ,1} =\frac{1}{\sqrt{2\sqrt{\pi }}} \psi _{0,\mu }$, we have 
\begin{equation*}
\langle \psi _{j,0} ,g_{\mu ,1} \rangle =\frac{1}{\sqrt{2\sqrt{\pi }}} \langle \psi _{j,0} ,\psi _{0,\mu } \rangle =\frac{( -1)^{j}}{\sqrt{2^{j+1} j!\sqrt{\pi }}} e^{-\frac{1}{4} \mu ^{2}} \mu ^{j}.
\end{equation*}
Similarly, if we denote $\psi _{j,\mu ,\sigma } =\frac{1}{\sqrt{\sigma}}\psi _{j}\left(\frac{x-\mu }{\sigma }\right)$, then $\| \psi _{i,\mu ,\sigma } \| _{2} =1$, $\langle \psi _{i,\mu ,\sigma } ,\psi _{j,\mu ,\sigma } \rangle =0$ for $i \neq j$, and
\begin{equation}\label{inner_product_decomp}
\begin{aligned}
\langle \psi _{i,\mu _{1} ,\sigma } , & \psi _{j,\mu _{2} ,\sigma } \rangle \\
= & \sum _{k=0}^{i\land j}\left( e^{-\frac{1}{8\sigma ^{2}}( \mu _{2} -\mu _{1})^{2}}( -1)^{i}\sqrt{\frac{k!}{i!}}\binom{i}{k}\left(\frac{\mu _{2} -\mu _{1}}{\sqrt{2} \sigma }\right)^{i-k}\right) \times \\
 & \ \ \ \ \ \ \left( e^{-\frac{1}{8\sigma ^{2}}( \mu _{2} -\mu _{1})^{2}}( -1)^{j}\sqrt{\frac{k!}{j!}}\binom{j}{k}\left(\frac{\mu _{1} -\mu _{2}}{\sqrt{2} \sigma }\right)^{j-k}\right). 
\end{aligned}
\end{equation}
Here the inner product is with regard to Lebesgue measure. In particular, since $g_{\mu ,\sigma } =\frac{1}{\sqrt{2\sigma\sqrt{\pi }}} \psi _{0,\mu ,\sigma }$, we have
\begin{equation}\label{inner_psi_g}
\langle \psi _{j,\mu_{1},\sigma } ,g_{\mu_{2} ,\sigma } \rangle =\frac{( -1)^{j}}{\sqrt{2^{j+1} j!\sigma\sqrt{\pi }}} e^{-\frac{(\mu_{2}-\mu_{1} )^{2}}{4\sigma ^{2}}}\left(\frac{\mu_{2}-\mu_{1} }{\sigma }\right)^{j}.
\end{equation}

\subsection{\texorpdfstring{Construction of $\widehat{f}_{i}$}{Construction of fi}}
For any $f\in \mathcal{F}$ as defined in \eqref{C3}, assume $X_1,\cdots, X_n$ are $n$ i.i.d. samples from $P_f$. 
Let us represent each component $f_i$ in terms of the Hermite basis functions: $f_{i} =\sum _{j=0}^{\infty } \alpha _{i,j} \psi _{j,c_{i},\sigma}$, where $\alpha _{i,j} =\langle f_{i} ,\psi _{j,c_{i},\sigma} \rangle $, for $i=1,2$ and $j\in \mathbb{N}_0$. Recall $f=w_1f_1+w_2f_2$. Let $\lambda _{i,j} \coloneqq w_{i} \alpha _{i,j}$, for $i=1,2$, and $j\in \mathbb{N}_0$. It follows that
\begin{equation*}
f=\sum _{j=0}^{\infty } \lambda _{1,j} \psi _{j,c_{1},\sigma} +\sum _{j=0}^{\infty } \lambda _{2,j} \psi _{j,c_{2},\sigma},
\end{equation*}
and by projecting $f$ onto each $\psi_{k,c_i,\sigma}$, for $i=1,2$ and $k\in [\ell]$, where $\ell$ is a nonnegative integer and $[\ell] \coloneqq \{0,1,\cdots ,\ell-1\}$,
\begin{equation*}
\langle f,\psi _{k,c_{i},\sigma} \rangle =\sum _{j=0}^{\infty } \lambda _{1,j} \langle \psi _{j,c_{1},\sigma} ,\psi _{k,c_{i},\sigma} \rangle +\sum _{j=0}^{\infty } \lambda _{2,j} \langle \psi _{j,c_{2},\sigma} ,\psi _{k,c_{i},\sigma} \rangle. 
\end{equation*}
Thus, we obtain
\begin{equation*}
y=A\lambda +\sum _{j=\ell}^{\infty } \lambda _{1,j} z_{1,j} +\sum _{j=\ell}^{\infty } \lambda _{2,j} z_{2,j},
\end{equation*}
where $y$ is the $2\ell$-dimensional vector whose entries are given by $\langle f, \psi_{j, c_i, \sigma}\rangle$ for $i=1,2$; $j\in [\ell]$ (the entries indexed by $(i,j)$ are ordered by $i$, and then by $j$),
likewise, $\lambda$ is the $2\ell$-dimensional vector whose entries are given by $\lambda_{i,j}=w_i\alpha_{i,j}$ for $i=1,2$; $j\in [\ell]$,
$A$ is a $2\ell$-by-$2\ell$ matrix whose entries are given by $\langle \psi _{j_{1} ,c_{i_{1}}, \sigma} ,\psi _{j_{2} ,c_{i_{2}}, \sigma} \rangle$, for $i_1, i_2=1,2$; $j_1,j_2\in [\ell]$, e.g., when $\ell=2$:
\begin{equation*}
A=\begin{pmatrix}
1 & 0 & \langle \psi _{0,c_{2} ,\sigma } ,\psi _{0,c_{1} ,\sigma } \rangle  & \langle \psi _{1,c_{2} ,\sigma } ,\psi _{0,c_{1} ,\sigma } \rangle \\
0 & 1 & \langle \psi _{0,c_{2} ,\sigma } ,\psi _{1,c_{1} ,\sigma } \rangle  & \langle \psi _{1,c_{2} ,\sigma } ,\psi _{1,c_{1} ,\sigma } \rangle \\
\langle \psi _{0,c_{1} ,\sigma } ,\psi _{0,c_{2} ,\sigma } \rangle  & \langle \psi _{1,c_{1} ,\sigma } ,\psi _{0,c_{2} ,\sigma } \rangle  & 1 & 0\\
\langle \psi _{0,c_{1} ,\sigma } ,\psi _{1,c_{2} ,\sigma } \rangle  & \langle \psi _{1,c_{1} ,\sigma } ,\psi _{1,c_{2} ,\sigma } \rangle  & 0 & 1
\end{pmatrix},
\end{equation*}\\
and
\begin{equation*}
z_{1,j} =\begin{bmatrix}
0\\
\vdots \\
0\\
\langle \psi _{j,c_{1},\sigma} ,\psi _{0,c_{2}, \sigma} \rangle \\
\vdots \\
\langle \psi _{j,c_{1}, \sigma} ,\ \psi _{\ell-1,\ c_{2}, \sigma} \rangle 
\end{bmatrix}, \ \ z_{2,j} =\begin{bmatrix}
\langle \psi _{j,c_{2},\sigma} ,\ \psi _{0,c_{1},\sigma} \rangle \\
\vdots \\
\langle \psi _{j,c_{2}, \sigma} ,\ \psi _{\ell-1,c_{1},\sigma} \rangle \\
0\\
\vdots \\
0
\end{bmatrix}\ \ \ \ \ \ \text{for}\ j\geq \ell.
\end{equation*}
For the rest of the paper, the $2\ell$ items will be indexed by $(i,k)$, $i=1,2$, $k=0,\cdots, \ell-1$, i.e., the $1,2,\cdots, 2\ell$-th columns in $A$ will be indexed by $(1,0), (1,1), \cdots, (1,\ell-1),(2,0), (2,1),\cdots, (2,\ell-1)$, respectively.

Next, let $\widehat{f}$ denote an estimate of $f$ based on the $n$ samples $X_1,\ldots, X_n$. For instance, a kernel density estimator can be used for this purpose. 
Then, take the $2\ell$-dimensional vector $\widehat{y}$ whose entries are given by $\langle \widehat{f} ,\psi _{j,c_{i}, \sigma} \rangle$ for $i=1,2$ and $j\in[\ell]$. Now we define 
\begin{equation}\label{def_lambda_hat}
\widehat{\lambda } =A^{-1}\widehat{y}.
\end{equation}
Consider a truncation using the first $l$ Hermite basis functions: 
\begin{equation*}
\widetilde{f_{i}} \coloneq \sum _{j=0}^{\ell-1}\widehat{\lambda }_{i,j} \psi _{j,c_{i},\sigma},
\end{equation*}
and then normalizing the $\widetilde{f_{i}}$, we arrive at the following estimate for the probability density function associated with the component indexed by $i=1$ and $2$:
\begin{equation}\label{fi_hat}
\widehat{f_{i}} \coloneq \frac{\left(\widetilde{f_{i}}\right)_{+}}{\left \lVert \left(\widetilde{f_{i}}\right)_{+} \right\rVert _{1}},
\end{equation}
where for any function $q:\mathbb{R}\to \mathbb{R}$, $(q(x))_+ \coloneqq\max\{q(x), 0\}$, $\forall x\in \mathbb{R}$.

\section{\texorpdfstring{Proofs for Section \ref{subsection: component_density}: Upper bound on $\lVert\widehat{f}_{i}-f_i\rVert_1$}{Proofs for Section: Upper bound on fi error}}
\label{section:appendix_A}
The goal of this section of the Appendix is to prove Lemma \ref{lemma: bound_l1_fi_hat_fi}. 
\begin{lemma}
\label{lemma: bound_l1_fi_hat_fi}
Let $f\in\mathcal{F}$ be as defined in \eqref{C3}. Denote by $P_f$ the probability measure with density $f$. Take $n$ i.i.d. samples from $P_f$, and let $\widehat{f}$ be any density estimator of $f$ from the $n$ samples. Define $\Delta f\coloneqq\widehat{f}-f$. For $i=1,2$, if $\ell\geq \lfloor 2er_i^2/\sigma^2\rfloor +2$, then the estimator $\widehat{f_i}$ defined in (\ref{fi_hat}) satisfies
\begin{align*}
\| \widehat{f_{i}} -f_{i} \| _{1} & \leq \frac{1}{w_{i}} c_{1}( r,\sigma ,r_{1} ,r_{2}) \cdot \left(\frac{8\max( r_{1} ,r_{2})}{r}\right)^{\ell } \cdot \ell ^{5/4} +c_{2}( r_{i} ,\sigma )\left(\frac{1}{2}\right)^{\ell }\\
 & \ \ \ \ \ +\ \frac{1}{w_{i}} c_{3}( r,\sigma ) \cdot \| \Delta f\| _{2} \cdot \exp( \ell \log \ell +c_{4}( r,\sigma ) \ell +c_{5}\log \ell )
\end{align*}
for some constants $c_1(r,\sigma,r_1,r_2)>0$, $c_2(r_i,\sigma)>0$, $c_3(r,\sigma)>0$, $c_4(r,\sigma)>0$, and $c_5>0$, where the dependence on parameters is indicated in parentheses.
\end{lemma}

This Lemma is a refinement of a result of \cite{bryon_2023}, who provide the order of the convergence rate and fix $r_1=r_2=\frac{1}{2}$, and $\sigma=1$. We extend their result by making explicit the constants and their dependence on  $r_i\in (0,+\infty)$, $i=1,2$, $\sigma\in (0,+\infty)$.

\subsection{Technical lemmas}

Recall that $f_{i} =\int _{c_{i} -r_{i}}^{c_{i} +r_{i}} g_{u, \sigma}( x) dV_i(u)$, for some $\sigma>0$, where $V_i(u)$ is the mixing measure supported within $[c_i-r_i, c_i+r_i]$. We can expand $f_i$ in terms of the Hermite functions: $f_{i} =\sum _{j=0}^{\infty } \alpha _{i,j} \psi _{j,c_{i},\sigma}$, where $\alpha _{i,j} =\langle f_{i} ,\psi _{j,c_{i},\sigma} \rangle $, for $i=1,2$ and $j\in \mathbb{N}_0$. 
We obtain the following (cf. Lemma 16 of \cite{bryon_2023}).

\begin{lemma}\label{lemma16}
    For $i=1,2$, and any $j\in \mathbb{N}_0$, we have $| \alpha_{i,j}| \leq \left(\frac{\sqrt{2} r_{i}}{2\sigma }\right)^{j} \cdot \frac{1}{\sqrt{2\sigma j!\sqrt{\pi }}}$.
\end{lemma}

\begin{proof} We have
\begin{equation*}
\begin{aligned}
\alpha_{i,j} =\langle f_{i} ,\psi _{j,c_{i} ,\sigma } \rangle  & =\int _{-\infty }^{+\infty }\left(\int _{c_{i} -r_{i}}^{c_{i} +r_{i}} g_{u,\sigma }( x) dV_i(u)\right) \psi _{j,c_i,\sigma}(x) dx\\
 & =\int _{c_{i} -r_{i}}^{c_{i} +r_{i}} \left(\int _{-\infty }^{+\infty } g_{u,\sigma }( x) \psi _{j,c_i,\sigma}(x) dx\right) dV_i(u)\\
 & =\int _{-r_{i}}^{r_{i}} \langle g_{u,\sigma }( x) ,\psi _{j,0,\sigma } \rangle dV_i(c_{i}+u),
\end{aligned}
\end{equation*}
so, by (\ref{inner_psi_g}), 
\begin{equation}\label{bound_alpha_ij}
|\alpha _{i,j} |\leq \int _{-r_{i}}^{r_{i}} |\langle \psi _{j,0,\sigma }, g_{u,\sigma } \rangle |dV_{i}( c_{i} +u) \leq \frac{1}{\sqrt{2\sigma \sqrt{\pi }}} \cdot \frac{\left(\frac{r_{i}}{\sqrt{2} \sigma }\right)^{j}}{\sqrt{j!}}.
\end{equation}
\end{proof}


Recall the definition of $\widehat{\lambda}$ in (\ref{def_lambda_hat}) that
$\widehat{\lambda } =A^{-1}\widehat{y}$,
where  $\widehat{y}$ is a $2l$-dimensional vector whose entries are given by $\langle \widehat{f} ,\psi _{j,c_{i}, \sigma} \rangle$ for $i=1,2$ and $j\in[\ell]$, while
$\widehat{f}$ is a density estimator of $f$. 
Let $y^{\Delta } \coloneqq \widehat{y} -y$, then we have
\begin{align}
\widehat{\lambda } =A^{-1}\widehat{y} & =A^{-1} y+A^{-1} y^{\Delta } \notag \\
 & =A^{-1}\left( A\lambda +\sum _{j=\ell}^{\infty } \lambda _{1,j} z_{1,j} +\sum _{j=\ell}^{\infty } \lambda _{2,j} z_{2,j}\right) +A^{-1} y^{\Delta } \notag\\
 & =\lambda +\underbrace{\sum _{j=\ell}^{\infty } \lambda _{1,j} A^{-1} z_{1,j} +\sum _{j=\ell}^{\infty } \lambda _{2,j} A^{-1} z_{2,j}}_{\mathcal{E}_{t}} +\underbrace{A^{-1} y^{\Delta }}_{\mathcal{E}_{a}}. \label{lambda_hat}
\end{align}
Let
\begin{align}
\mathcal{E}_{t} & =\sum _{j=\ell}^{\infty } \lambda _{1,j} A^{-1} z_{1,j} +\sum _{j=\ell}^{\infty } \lambda _{2,j} A^{-1} z_{2,j}, \label{E_t}\\
\mathcal{E}_{a} & =A^{-1} y^{\Delta }.\label{E_a}
\end{align}
$\mathcal{E}_t$ is the truncation error and $\mathcal{E}_a$ is the approximation error. As $\ell\rightarrow \infty$ and sample size $n\rightarrow \infty$, we expect $\mathcal{E}_t$ and $\mathcal{E}_a$ to vanish. 

Recall that $\lambda_{i,j}=w_i\alpha_{i,j}$, hence the function
\begin{equation}\label{fi_tilde}
\widetilde{f_{i}} \coloneq \sum _{j=0}^{\ell-1}\widehat{\lambda }_{i,j} \psi _{j,c_{i},\sigma}
\end{equation}
is expected to be an approximation of $w_if_i$. This motivates the following lemma (which refines Lemma 17 of \cite{bryon_2023}).

\begin{lemma}\label{Lemma 17}
    Let $\Delta>0$ and $\ell$ be a nonnegative integer. If $|\lambda _{i,j} -\widehat{\lambda }_{i,j} |< \Delta ,\ \forall i=1,2,\ j\in [ \ell]$, then $\| w_{i} f_{i} -\widetilde{f_{i}} \| _{1} \leq \frac{2^{17/4} c_{\psi } \sqrt{\sigma} }{\Gamma \left(\frac{1}{4}\right)^{2}} \cdot \Delta \cdot \ell^{5/4} +\frac{2^{11/2} c_{\psi } r_{i} w_{i}}{\Gamma \left(\frac{1}{4}\right)^{2} \pi ^{1/4}\sigma} \cdot \left(\frac{1}{2}\right)^{\ell}$, for some $c_\psi>0$ and any $\ell\geq\lfloor 2er_{i}^{2}/\sigma^2 \rfloor+2$.
\end{lemma}

\begin{proof}
From \cite{Hermite_2012} and \cite{Hermite_1995} we know $\forall j\in \mathbb{N}_{+}$,
\begin{equation*}
\| \psi _{j-1} \| _{1} =\frac{2^{17/4}}{\Gamma \left(\frac{1}{4}\right)^{2}} \cdot j^{1/4}( 1+o( 1)), 
\end{equation*}
so there exists a constant $c_\psi>0$, s.t., $\| \psi _{j-1} \| _{1} \leq (2^{17/4} c_{\psi}/ \Gamma \left(\frac{1}{4}\right)^{2}) j^{1/4}$, $\forall j\in \mathbb{N}_{+}$. Because $\| \psi _{j,\mu ,\sigma } \| _{1} =\| \frac{1}{\sqrt{\sigma}} \psi _{j}\left(\frac{x-\mu }{\sigma }\right) \| _{1} =\sqrt{\sigma} \| \psi _{j} \| _{1}$, we obtain $\| \psi _{j-1,\mu ,\sigma } \| _{1} \leq \left( 2^{17/4} c_{\psi } \sqrt{\sigma} /\Gamma \left(\frac{1}{4}\right)^{2}\right) j^{1/4}$. 
It follows that
\begin{align}\label{obj1}
\| w_{i} f_{i} -\widetilde{f_{i}} \| _{1} & =\left\lVert \sum _{j=0}^{\ell-1}( \lambda _{i,j} -\widehat{\lambda }_{i,j}) \psi _{j,c_{i},\sigma} +\sum _{j=\ell}^{\infty } \lambda _{i,j} \psi _{j,c_{i},\sigma} \right\rVert _{1} \notag\\
 & \leq \sum _{j=0}^{\ell-1} |\lambda _{i,j} -\widehat{\lambda }_{i,j} |\cdot \| \psi _{j,c_{i},\sigma} \| _{1} +\sum _{j=\ell}^{\infty } |\lambda _{i,j} |\cdot \| \psi _{j,c_{i},\sigma} \| _{1}.
\end{align}
For the first sum in the RHS of \eqref{obj1},
\begin{align}
\sum _{j=0}^{\ell-1} |\lambda _{i,j} -\widehat{\lambda }_{i,j} |\cdot \| \psi _{j,c_{i},\sigma} \| _{1} & \leq \Delta \cdot \sum _{j=0}^{\ell-1} \| \psi _{j,c_{i},\sigma} \| _{1} \notag \\
 & \leq \Delta \cdot \ell\cdot \left( 2^{17/4} c_{\psi } \sqrt{\sigma}/\Gamma \left(\frac{1}{4}\right)^{2}\right) \ell^{1/4} \notag \\
 & =\frac{2^{17/4} c_{\psi }\sqrt{\sigma}}{\Gamma \left(\frac{1}{4}\right)^{2}} \cdot \Delta \cdot \ell^{5/4} \label{term1}
\end{align} 
when $\ell$ is large enough. For the second sum in the RHS of \eqref{obj1},
\begin{align*}
\sum _{j=\ell }^{\infty } |\lambda _{i,j} |\cdot \| \psi _{j,c_{i} ,\sigma } \| _{1} & \leq \sum _{j=\ell }^{\infty } w_{i} |\alpha _{i,j} |\cdot \frac{2^{17/4} c_{\psi }\sqrt{\sigma }}{\Gamma \left(\frac{1}{4}\right)^{2}} \cdot ( j+1)^{1/4}\\
 & \leq \sum _{j=\ell }^{\infty } w_{i} \cdot \left(\frac{r_{i}}{\sqrt{2} \sigma }\right)^{j} \cdot \frac{1}{\sqrt{2\sigma j!\sqrt{\pi }}} \cdot \frac{2^{17/4} c_{\psi }\sqrt{\sigma }}{\Gamma \left(\frac{1}{4}\right)^{2}} \cdot 2^{1/4} j^{1/4}\\
 & =\frac{2^{9/2} c_{\psi } w_{i}}{\Gamma \left(\frac{1}{4}\right)^{2}} \cdot \frac{1}{\sqrt{2\sqrt{\pi }}} \cdot \sum _{j=\ell }^{\infty }\frac{\left(\frac{r_{i}}{\sqrt{2} \sigma }\right)^{j}}{\sqrt{j!}} \cdot j^{1/4} .
\end{align*}

The series on the right hand side absolutely converges. Stirling's approximation yields $j!\geq \left(\frac{j}{e}\right)^{j}$, so we have
\begin{equation*} 
\begin{aligned}
\sum _{j=\ell}^{\infty }\frac{\left(\frac{r_{i}}{\sqrt{2} \sigma }\right)^{j}}{\sqrt{j!}} \cdot j^{1/4} & =\sum _{j=\ell}^{\infty }\sqrt{\frac{j^{1/2}}{j!}} \cdot \left(\frac{r_{i}}{\sqrt{2} \sigma }\right)^{j}\\
 & \leq \sum _{j=\ell}^{\infty }\frac{1}{\sqrt{( j-1) !}} \cdot \left(\frac{r_{i}}{\sqrt{2} \sigma }\right)^{j}\\
 & \leq \frac{r_{i}}{\sqrt{2} \sigma }\sum _{j=\ell}^{\infty }\frac{\left(\frac{r_{i}}{\sqrt{2} \sigma }\right)^{j-1}}{\sqrt{\left(\frac{j-1}{e}\right)^{j-1}}}\\
 & =\frac{r_{i}}{\sqrt{2} \sigma }\sum _{j=\ell-1}^{\infty }\left(\frac{r_{i}\sqrt{e}}{\sqrt{2j} \sigma }\right)^{j}
\end{aligned}
\end{equation*}
and when $j >2er_{i}^{2}/\sigma^2$, $\frac{r_{i}\sqrt{e}}{\sqrt{2j} \sigma } < \frac{1}{2}$, therefore when $\ell\geq\lfloor 2er_{i}^{2}/\sigma^2 \rfloor+2$, 
\begin{equation*}
\frac{r_{i}}{\sqrt{2} \sigma }\sum _{j=\ell-1}^{\infty }\left(\frac{r_{i}\sqrt{e}}{\sqrt{2j} \sigma }\right)^{j} \leq \frac{r_{i}}{\sqrt{2} \sigma } \cdot \left(\frac{1}{2}\right)^{\ell-2},
\end{equation*}
so the second sum in the RHS of \eqref{obj1} is upper bounded by
\begin{equation}\label{term2}
\sum _{j=\ell}^{\infty } |\lambda _{i,j} |\cdot \| \psi _{j,c_{i} ,\sigma } \| _{1} \leq \frac{2^{11/2} w_{i} r_{i} c_{\psi }}{\Gamma \left(\frac{1}{4}\right)^{2} \pi ^{1/4} \sigma} \cdot \left(\frac{1}{2}\right)^{\ell}.
\end{equation}
By combining (\ref{term1}) and (\ref{term2}), we get
\begin{equation*}
\| w_{i} f_{i} -\widetilde{f_{i}} \| _{1} \leq \frac{2^{17/4} c_{\psi } \sqrt{\sigma} }{\Gamma \left(\frac{1}{4}\right)^{2}} \cdot \Delta \cdot \ell^{5/4} +\frac{2^{11/2} c_{\psi } r_{i} w_{i}}{\Gamma \left(\frac{1}{4}\right)^{2} \pi ^{1/4}\sigma} \cdot \left(\frac{1}{2}\right)^{\ell}.
\end{equation*}
\end{proof}

Lemma \ref{Lemma 17} gives an upper bound for $\| w_{i} f_{i} -\widetilde{f_{i}} \| _{1}$ when $|\lambda _{i,j} -\widehat{\lambda }_{i,j} |< \Delta$, $\forall i=1,2,\ j\in[\ell]$. Next we want to find the upper bound $\Delta$ for all $|\lambda _{i,j} -\widehat{\lambda }_{i,j} |$'s.   Recall from (\ref{lambda_hat}),
\begin{equation*}
\widehat{\lambda } -\lambda =\underbrace{\sum _{j=\ell}^{\infty } \lambda _{1,j} A^{-1} z_{1,j} +\sum _{j=\ell}^{\infty } \lambda _{2,j} A^{-1} z_{2,j}}_{_{\mathcal{E}_{t}}} +\underbrace{A^{-1} y^{\Delta }}_{\mathcal{E}_{a}},
\end{equation*}
and we need to bound the entries of $\mathcal{E}_t$ and $\mathcal{E}_a$ respectively. 

\subsubsection{\texorpdfstring{Bounding the entries of $\mathcal{E}_t$}{Bounding the entries of Et}}
We first bound the entries of $\mathcal{E}_t$. Let $\mathcal{E}_{t,i,j}$ be the vector $A^{-1}z_{i,j}$ for $i=1,2$ and $j\geq \ell$. Namely, we have
\begin{equation*}
\mathcal{E}_{t} =\sum _{j=\ell}^{\infty } \lambda _{1,j} \mathcal{E}_{t,1,j} +\sum _{j=\ell}^{\infty } \lambda _{2,j} \mathcal{E}_{t,2,j}.
\end{equation*}
We only need to analyze the first sum $\sum _{j=\ell}^{\infty } \lambda _{1,j} \mathcal{E}_{t,1,j}$ and by symmetry we can conclude a similar bound for the second sum $\sum _{j=\ell}^{\infty } \lambda _{2,j} \mathcal{E}_{t,2,j}$. By Cramer's rule, the entry of $\mathcal{E}_{t,1,j} =A^{-1} z_{1,j}$ indexed at $(i,k)$ is given by
\begin{equation*}
\frac{\det\left( A^{( i,k)\rightarrow (1,j)}\right)}{\det( A)},
\end{equation*}
where $A^{( i,k)\rightarrow (1,j)}$ is the $2l$-by-$2l$ matrix same as $A$ except that the column indexed at $(i,k)$ is replaced with $z_{1,j}$ for $i=1,2$, $k\in [\ell]$ and $j\geq \ell$. Lemma \ref{Lemma 30} gives a bound on $|\det\left( A^{( i,k)\rightarrow ( 1,j)}\right) |$ when comparing to $\det(A)$. Before proving this,  we first state the Cauchy-Binet formula in the following lemma.

\begin{lemma}[Cauchy–Binet formula]\label{Lemma 9}
Let $b_0, b_1, \dots$ \textit{and} $c_0, c_1, \dots$ \textit{be two infinite sequences of} $n$ \textit{dimensional vectors. For any (ordered) subset} $S = \{s_0 < s_1 < \cdots < s_{n-1}\}$ \textit{of} $\mathbb{N}_0$, \textit{let} $B_S$ \textit{(resp.} $C_S$\textit{) be the matrix that the column indexed at} $i$ \textit{is} $b_{s_i}$ \textit{(resp.} $c_{s_i}$\textit{) for} $i \in [n]$\textit{. The determinant of} $A = \sum_{i=0}^{\infty} b_i c_i^\top$ is
\[
\det A = \sum_{S \subseteq \mathbb{N}_0, |S| = n} (\det B_S) \cdot (\det C_S)
\]
if the RHS converges.
\end{lemma}

\begin{lemma}[Lemma 30, \cite{bryon_2023}]\label{Lemma 30}
We can assume $c_2>c_1$ w.l.o.g, and denote $r=c_2-c_1$, Then for any $\ell\in\mathbb{N}_0$, $k\in [\ell]$ and $j\geq \ell$, we have
\begin{align*}
\left|\det\left( A^{( 1,k)\rightarrow ( 1,j)}\right) \right| & \leq e^{\frac{r^{2}}{4\sigma ^{2}}} 4^{\ell}\left(\frac{2\sqrt{2} \sigma }{r}\right)^{j}\sqrt{j!} \cdot \det( A) ,\ \ \ \text{and} \ \\
\left| \det\left( A^{( 2,k)\rightarrow ( 1,j)}\right) \right| & \leq e^{\frac{5r^{2}}{4\sigma ^{2}}} 4^{\ell}\left(\frac{2\sqrt{2} \sigma }{r}\right)^{j}\sqrt{j!} \cdot \det( A).
\end{align*}
Hence, $\forall k\in [\ell]$, the absolute values of the entries of $\lambda _{1,j}\mathcal{E}_{t,1,j}$ are bounded by,
\begin{align*}
| \lambda _{1,j}(\mathcal{E}_{t,1,j})_{1,k}|  & \leq w_{1} e^{\frac{r^{2}}{4\sigma ^{2}}}\frac{1}{\sqrt{2\sigma \sqrt{\pi }}} \cdot \left(\frac{2r_{1}}{r}\right)^{j} \cdot 4^{\ell },\\
| \lambda _{1,j}(\mathcal{E}_{t,1,j})_{2,k}|  & \leq w_{1} e^{\frac{5r^{2}}{4\sigma ^{2}}}\frac{1}{\sqrt{2\sigma \sqrt{\pi }}} \cdot \left(\frac{2r_{1}}{r}\right)^{j} \cdot 4^{\ell }.
\end{align*}
\end{lemma}

\begin{proof}
We first observe that the matrices \( A \), \( A^{(1,k)\to (1,j)} \) and \( A^{(2,k)\to (1,j)} \) can be decomposed as
\[
A = V^\top V, \quad A^{(1,k)\to (1,j)} = V^\top V^{(1,k)\to (1,j)} \quad \text{and} \quad A^{(2,k)\to (1,j)} = V^\top V^{(2,k)\to (1,j)}
\]
where we abuse the notation to define \( V \), \( V^{(1,k)\to (1,j)} \) and \( V^{(2,k)\to (1,j)} \) as follows. Let \( V \) be the \( |\mathbb{N}_0| \)-by-\( 2\ell \) matrix whose column indexed at \( (i,k) \) is the \( |\mathbb{N}_0| \) dimensional vector \( v^{(i,k)} \) for \( i = 1,2 \) and \( k \in [\ell] \). Here, for \( i = 1,2 \) and \( j \in \mathbb{N}_0 \), \( v^{(i,j)} \) is the \( |\mathbb{N}_0| \) dimensional vector whose \( k \)-th entry is \( \langle \psi_{j,c_i, \sigma}, \psi_{k,c_1, \sigma} \rangle \) for \( k \in \mathbb{N}_0 \). In particular, \( v^{(1,j)} \) is the zero vector except that the \( j \)-th entry is 1. For example, when \( \ell = 2 \),
\[
V =
\begin{bmatrix}
1 & 0 & \langle \psi_{0,r, \sigma}, \psi_{0,0, \sigma} \rangle & \langle \psi_{1,r, \sigma}, \psi_{0,0, \sigma} \rangle \\
0 & 1 & \langle \psi_{0,r, \sigma}, \psi_{1,0, \sigma} \rangle & \langle \psi_{1,r, \sigma}, \psi_{1,0, \sigma} \rangle \\
0 & 0 & \langle \psi_{0,r, \sigma}, \psi_{2,0, \sigma} \rangle & \langle \psi_{1,r, \sigma}, \psi_{2,0,\sigma} \rangle \\
\vdots & \vdots & \vdots & \vdots
\end{bmatrix}.
\]
It is easy to check that, by the orthogonality of Hermite functions,
\[
V^\top V =
\begin{bmatrix}
1 & 0 & \langle \psi_{0,r,\sigma}, \psi_{0,0,\sigma} \rangle & \langle \psi_{1,r,\sigma}, \psi_{0,0,\sigma} \rangle \\
0 & 1 & \langle \psi_{0,r,\sigma}, \psi_{1,0,\sigma} \rangle & \langle \psi_{1,r,\sigma}, \psi_{1,0,\sigma} \rangle \\
\langle \psi_{0,0,\sigma}, \psi_{0,r,\sigma} \rangle & \langle \psi_{1,0,\sigma}, \psi_{0,r,\sigma} \rangle & 1 & 0 \\
\langle \psi_{0,0,\sigma}, \psi_{1,r,\sigma} \rangle & \langle \psi_{1,0,\sigma}, \psi_{1,r,\sigma} \rangle & 0 & 1
\end{bmatrix} = A.
\]

Define \( V^{(i,k) \to (1,j)} \) to be the same matrix as \( V \) except that the column indexed at \( (i,k) \) is replaced with \( v^{(1,j)} \) for \( i = 1,2 \), \( k \in [\ell] \) and \( j \geq \ell \). For example, when \( \ell = 2 \), \( i = 1 \), \( k = 1 \) and \( j = 3 \),
\[
V^{(i,k)\to (1,j)} =
\begin{bmatrix}
1 & 0 & \langle \psi_{0,r,\sigma}, \psi_{0,0,\sigma} \rangle & \langle \psi_{1,r,\sigma}, \psi_{0,0,\sigma} \rangle \\
0 & 0 & \langle \psi_{0,r,\sigma}, \psi_{1,0,\sigma} \rangle & \langle \psi_{1,r,\sigma}, \psi_{1,0,\sigma} \rangle \\
0 & 0 & \langle \psi_{0,r,\sigma}, \psi_{2,0,\sigma} \rangle & \langle \psi_{1,r,\sigma}, \psi_{2,0,\sigma} \rangle \\
0 & 1 & \langle \psi_{0,r,\sigma}, \psi_{3,0,\sigma} \rangle & \langle \psi_{1,r,\sigma}, \psi_{3,0,\sigma} \rangle \\
\vdots & \vdots & \vdots & \vdots
\end{bmatrix}
\]
and when \( \ell = 2, i = 2, k = 1 \) and \( j = 3 \),

\[
V^{(i,k) \to (1,j)} =
\begin{bmatrix}
1 & 0 & \langle \psi_{0,r,\sigma}, \psi_{0,0,\sigma} \rangle & 0 \\
0 & 1 & \langle \psi_{0,r,\sigma}, \psi_{1,0,\sigma} \rangle & 0 \\
0 & 0 & \langle \psi_{0,r,\sigma}, \psi_{2,0,\sigma} \rangle & 0 \\
0 & 0 & \langle \psi_{0,r,\sigma}, \psi_{3,0,\sigma} \rangle & 1 \\
\vdots & \vdots & \vdots & \vdots
\end{bmatrix}.
\]
Again, we can check that \( A^{(i,k) \to (1,j)} = V^T V^{(i,k) \to (1,j)} \).

By the Cauchy-Binet formula (Lemma \ref{Lemma 9}) and expanding the determinant along the columns indexed at \( (1,k) \), we have
\[
\det(A) = \sum_{K \in \mathcal{K}} \det(U_K)^2,
\]
where \( \mathcal{K} \) is the set of subsets of \( \mathbb{N}_0 \) of size \( \ell \) whose elements are larger than or equal to \( \ell \), i.e. $\mathcal{K} = \{ K \mid K = \{\ell \leq a_0 < \cdots < a_{\ell-1}\} \}$ and \( U_K \) is the \( \ell \)-by-\( \ell \) matrix whose \( (i, j) \)-entry is \( \langle \psi_{c_j,r,\sigma}, \psi_{b_i,0,\sigma} \rangle \) for \( b_i \in K \) and \( c_j \in [\ell] \). For example, when \( \ell = 2 \),
\begin{align*}
\det(A) = \left( \det \begin{bmatrix}
\langle \psi_{0,r,\sigma}, \psi_{\textcolor{red}{2},0,\sigma} \rangle & \langle \psi_{1,r,\sigma}, \psi_{\textcolor{red}{2},0,\sigma} \rangle \\
\langle \psi_{0,r,\sigma}, \psi_{\textcolor{red}{3},0,\sigma} \rangle & \langle \psi_{1,r,\sigma}, \psi_{\textcolor{red}{3},0,\sigma} \rangle
\end{bmatrix} \right)^2 
&+ \left( \det \begin{bmatrix}
\langle \psi_{0,r,\sigma}, \psi_{\textcolor{red}{2},0,\sigma} \rangle & \langle \psi_{1,r,\sigma}, \psi_{\textcolor{red}{2},0,\sigma} \rangle \\
\langle \psi_{0,r,\sigma}, \psi_{\textcolor{red}{4},0,\sigma} \rangle & \langle \psi_{1,r,\sigma}, \psi_{\textcolor{red}{4},0,\sigma} \rangle
\end{bmatrix} \right)^2 
+ \cdots  \notag\\
&+ \left( \det \begin{bmatrix}
\langle \psi_{0,r,\sigma}, \psi_{\textcolor{red}{3},0,\sigma} \rangle & \langle \psi_{1,r,\sigma}, \psi_{\textcolor{red}{3},0,\sigma} \rangle \\
\langle \psi_{0,r,\sigma}, \psi_{\textcolor{red}{4},0,\sigma} \rangle & \langle \psi_{1,r,\sigma}, \psi_{\textcolor{red}{4},0,\sigma} \rangle
\end{bmatrix} \right)^2 
+ \cdots \notag\\
& + \cdots
\end{align*}
where the numbers in red represent the set \(K\).

We first give a bound on \( \left|\det(A^{(1,k) \to (1,j)}) \right| \). By a similar argument, we have
\[
\left|\det(A^{(1,k) \to (1,j)}) \right| \leq \sum_{K \in \mathcal{K}_j} |\det(U_{K \cup \{k\}})||\det(U_{K \cup \{j\}})|
\]
where \(\mathcal{K}_j\) is the set of subsets of \(\mathbb{N}_0\) of size \(\ell - 1\) whose elements are larger than or equal to \(\ell\) and not equal to \(j\), i.e. $\mathcal{K}_j = \{ K \mid K = \{\ell \leq a_0 < \cdots < a_{\ell -2} \text{ and } a_i \neq j\} \}$. Furthermore, by Cauchy-Schwarz inequality, we have
\begin{align}
\left| \det(A^{(1,k)\rightarrow (1,j)}) \right| & \leq \sqrt{\left( \sum_{K \in \mathcal{K}_j} \det(U_{K\cup\{k\}})^2 \right) \left( \sum_{K \in \mathcal{K}_j} \det(U_{K\cup\{j\}})^2 \right)} \notag \\ 
& \leq \sqrt{\left( \sum_{K \in \mathcal{K}_j} \det(U_{K\cup\{k\}})^2 \right) \cdot \det(A)}. \label{bound_A1kj}
\end{align}
The last line is due to the fact that the subset $K\cup \{j\} \in \mathcal{K}$, $\forall K\in \mathcal{K}_j$. By Lemma \ref{Lemma 32}, we have
\begin{equation*}
\sum _{K\in \mathcal{K}_{j}}\det( U_{K\cup \{k\}})^{2} \leq \left( e^{\frac{r^{2}}{4\sigma ^{2}}} 4^{\ell}\left(\frac{2\sqrt{2} \sigma }{r}\right)^{j}\sqrt{j!}\right)^{2} \cdot \det( A).
\end{equation*}
and, by plugging it into (\ref{bound_A1kj}), we have
\begin{equation}\label{bound_A_1k_1j}
\left|\det\left( A^{( 1,k)\rightarrow ( 1,j)}\right) \right|\leq e^{\frac{r^{2}}{4\sigma ^{2}}} 4^{\ell }\left(\frac{2\sqrt{2} \sigma }{r}\right)^{j}\sqrt{j!} \cdot \det( A).
\end{equation}

We now give a bound on $\left| \det\left( A^{( 2,k)\rightarrow ( 1,j)}\right)\right|$. By a similar argument, we have
\begin{equation*}
\left| \det\left( A^{( 2,k)\rightarrow ( 1,j)}\right)\right| \leq \sum _{K\in \mathcal{K}_{j}}\left| \det U_{K}^{( -k)}\right| \cdot \left| \det( U_{K\cup \{j\}}) \right| 
\end{equation*}
where $U_K^{(-k)}$ is the $(\ell-1)$-by-$(\ell-1)$ matrix whose $(i,j)$-entry is $\langle \psi_{c_j,r,\sigma}, \psi_{b_i,0,\sigma}\rangle$ for $b_i\in K$ and $c_j\in[\ell]\backslash \{k\}$. Furthermore, by Cauchy-Schwarz inequality, we have
\begin{align}
\left| \det\left( A^{( 2,k)\rightarrow ( 1,j)}\right)\right|  & \leq \sum _{K\in \mathcal{K}_{j}}\left| \det U_{K}^{( -k)}\right| \cdot | \det( U_{K\cup \{j\}})| \notag \\
 & \leq \sqrt{\sum _{K\in \mathcal{K}_{j}}\det\left( U_{K}^{( -k)}\right)^{2} \cdot \det( A)}. \label{bound_det_A_2k_to_1j}
\end{align}
The last line is due to the fact that the subset $K\cup \{j\}$ is in $\mathcal{K}$ for each $K\in \mathcal{K}_j$. By Lemma \ref{Lemma 33} below, we have
\begin{equation*}
\sum _{K\in \mathcal{K}_{j}}\det\left( U_{K}^{( -k)}\right)^{2} \leq \left( e^{\frac{5}{4\sigma ^{2}} r^{2}} 4^{\ell } \cdot \left(\frac{2\sqrt{2} \sigma }{r}\right)^{j}\sqrt{j!}\right)^{2} \cdot \det( A),
\end{equation*}
and, by plugging it into (\ref{bound_det_A_2k_to_1j}), we have
\begin{equation}\label{bound_A_2k_1j}
\left| \det\left( A^{( 2,k)\rightarrow ( 1,j)}\right)\right| \leq e^{\frac{5}{4\sigma ^{2}}r^2} 4^{\ell }\left(\frac{2\sqrt{2} \sigma }{r}\right)^{j}\sqrt{j!} \cdot \det( A).
\end{equation}
By combining the above results with the upper bound of $\alpha_{i,j}$ given in (\ref{bound_alpha_ij}), we obtain $\forall k\in[\ell]$
\begin{align*}
\left| \lambda _{1,j}(\mathcal{E}_{t,1,j})_{1,k} \right| & = | \lambda _{1,j}| \cdot \left| \left( A^{-1} z_{1,j}\right)_{1,k}\right| =w_{1} \cdot | \alpha _{1,j}| \cdot \frac{\left| \det\left( A^{( 1,k)\rightarrow ( 1,j)}\right)\right| }{| \det( A)| }\\
 & \leq w_{1} \cdot \left(\frac{r_{1}}{\sqrt{2} \sigma }\right)^{j} \cdot \frac{1}{\sqrt{2j!\sigma \sqrt{\pi }}} \cdot e^{\frac{r^{2}}{4\sigma ^{2}}} 4^{\ell }\left(\frac{2\sqrt{2} \sigma }{r}\right)^{j}\sqrt{j!}\\
 & =w_{1} e^{\frac{r^{2}}{4\sigma ^{2}}}\frac{1}{\sqrt{2\sigma \sqrt{\pi }}} \cdot \left(\frac{2r_{1}}{r}\right)^{j} \cdot 4^{\ell }.
\end{align*}
and
\begin{align*}
\left| \lambda _{1,j}(\mathcal{E}_{t,1,j})_{2,k} \right|& = | \lambda _{1,j}| \cdot \left| \left( A^{-1} z_{1,j}\right)_{2,k}\right| =w_{1} \cdot | \alpha _{1,j}| \cdot \frac{\left| \det\left( A^{( 2,k)\rightarrow ( 1,j)}\right)\right| }{| \det( A)| }\\
 & \leq w_{1} \cdot \left(\frac{r_{1}}{\sqrt{2} \sigma }\right)^{j} \cdot \frac{1}{\sqrt{2j!\sigma \sqrt{\pi }}} \cdot e^{\frac{5r^{2}}{4\sigma ^{2}}} 4^{\ell }\left(\frac{2\sqrt{2} \sigma }{r}\right)^{j}\sqrt{j!}\\
 & =w_{1} e^{\frac{5r^{2}}{4\sigma ^{2}}}\frac{1}{\sqrt{2\sigma \sqrt{\pi }}} \cdot \left(\frac{2r_{1}}{r}\right)^{j} \cdot 4^{\ell }.
\end{align*}
This concludes the proof of the lemma.
\end{proof}

By the similar argument in Lemma \ref{Lemma 30}, for any $\ell\in \mathbb{N}_0$, $k\in[\ell]$ and $j\geq \ell$, we have
\begin{align*}
\left|\det\left( A^{( 1,k)\rightarrow ( 2,j)}\right) \right| & \leq e^{\frac{5r^{2}}{4\sigma ^{2}}} 4^{\ell}\left(\frac{2\sqrt{2} \sigma }{r}\right)^{j}\sqrt{j!} \cdot \det( A) ,\ \ \ \text{and} \ \\
\left| \det\left( A^{( 2,k)\rightarrow ( 2,j)}\right) \right| & \leq e^{\frac{r^{2}}{4\sigma ^{2}}} 4^{\ell}\left(\frac{2\sqrt{2} \sigma }{r}\right)^{j}\sqrt{j!} \cdot \det( A).
\end{align*}
Hence, $\forall k\in [\ell]$, the absolute values of the entries of $\lambda _{2,j}\mathcal{E}_{t,2,j}$ are bounded by,
\begin{align}
| \lambda _{2,j}(\mathcal{E}_{t,2,j})_{1,k}|  & \leq w_{2} e^{\frac{5r^{2}}{4\sigma ^{2}}}\frac{1}{\sqrt{2\sigma \sqrt{\pi }}} \cdot \left(\frac{2r_{2}}{r}\right)^{j} \cdot 4^{\ell }, \label{bound_2j_1k} \\
| \lambda _{2,j}(\mathcal{E}_{t,2,j})_{2,k}|  & \leq w_{2} e^{\frac{r^{2}}{4\sigma ^{2}}}\frac{1}{\sqrt{2\sigma \sqrt{\pi }}} \cdot \left(\frac{2r_{2}}{r}\right)^{j} \cdot 4^{\ell }. \label{bound_2j_2k}
\end{align}

Now we are ready to discuss under what conditions each entry of $\mathcal{E}_t$ has an upper bound, as well as the conditions under which this upper bound approaches zero as $\ell \rightarrow \infty$.

\begin{lemma}\label{Lemma up_bouond_E_t}
Define $\mathcal{E}_{t,1} \coloneq \sum _{j=\ell }^{\infty } \lambda _{1,j}\mathcal{E}_{t,1,j}$ and $\mathcal{E}_{t,2} \coloneq \sum _{j=\ell }^{\infty } \lambda _{2,j}\mathcal{E}_{t,2,j}$. When $2\max(r_1,r_2)<r$, we have $\forall k\in [\ell]$, 
\begin{align*}
| (\mathcal{E}_{t,1})_{1,k}|  & \leq \frac{w_{1}}{\sqrt{2\sigma \sqrt{\pi }}} \cdot \frac{re^{\frac{1}{4\sigma ^{2}} r^{2}}}{r-2r_{1}} \cdot \left(\frac{8r_{1}}{r}\right)^{\ell },\\
| (\mathcal{E}_{t,1})_{2,k}|  & \leq \frac{w_{1}}{\sqrt{2\sigma \sqrt{\pi }}} \cdot \frac{re^{\frac{5}{4\sigma ^{2}} r^{2}}}{r-2r_{1}} \cdot \left(\frac{8r_{1}}{r}\right)^{\ell },\\
| (\mathcal{E}_{t,2})_{1,k}|  & \leq \frac{w_{2}}{\sqrt{2\sigma \sqrt{\pi }}} \cdot \frac{re^{\frac{5}{4\sigma ^{2}} r^{2}}}{r-2r_{2}} \cdot \left(\frac{8r_{2}}{r}\right)^{\ell },\\
| (\mathcal{E}_{t,2})_{2,k}|  & \leq \frac{w_{2}}{\sqrt{2\sigma \sqrt{\pi }}} \cdot \frac{re^{\frac{1}{4\sigma ^{2}} r^{2}}}{r-2r_{2}} \cdot \left(\frac{8r_{2}}{r}\right)^{\ell }.
\end{align*}
Furthermore, when $8\max( r_{1} ,r_{2}) < r$, we have for $i=1,2$, and $\forall k\in [\ell]$,
\begin{align*}
| (\mathcal{E}_{t})_{i,k}|  & \leq | (\mathcal{E}_{t,1})_{i,k}| +| (\mathcal{E}_{t,2})_{i,k}| \leq \frac{re^{\frac{5}{4\sigma ^{2}} r^{2}}}{\sqrt{2\sigma \sqrt{\pi }}( r-2\max( r_{1} ,r_{2}))}\left(\frac{8\max( r_{1} ,r_{2})}{r}\right)^{\ell }\rightarrow 0,\ \text{as}\ \ell\rightarrow\infty.
\end{align*}
\end{lemma}
\begin{proof}
By Lemma \ref{Lemma 30},  when $r_1<\frac{r}{2}$, $\forall k\in[\ell]$,
\begin{align*}
| (\mathcal{E}_{t,1})_{1,k}|  & \leq \sum _{j=\ell }^{\infty }| \lambda _{1,j}| \cdotp | (\mathcal{E}_{t,1,j})_{1,k}| \\
 & \leq w_{1} e^{\frac{r^{2}}{4\sigma ^{2}}}\frac{4^{\ell }}{\sqrt{2\sigma \sqrt{\pi }}} \cdotp \sum _{j=\ell }^{\infty }\left(\frac{2r_{1}}{r}\right)^{j}\\
 & =w_{1} e^{\frac{r^{2}}{4\sigma ^{2}}}\frac{4^{\ell }}{\sqrt{2\sigma \sqrt{\pi }}} \cdot \frac{r}{r-2r_{1}} \cdot \left(\frac{2r_{1}}{r}\right)^{\ell }\\
 & =\frac{w_{1}}{\sqrt{2\sigma \sqrt{\pi }}} \cdot \frac{re^{\frac{1}{4\sigma ^{2}} r^{2}}}{r-2r_{1}} \cdot \left(\frac{8r_{1}}{r}\right)^{\ell },
\end{align*}
and similarly,
\begin{equation*}
| (\mathcal{E}_{t,1})_{2,k}| \leq \sum _{j=\ell }^{\infty }| \lambda _{1,j}| \cdotp | (\mathcal{E}_{t,1,j})_{2,k}| \leq \frac{w_{1}}{\sqrt{2\sigma \sqrt{\pi }}} \cdot \frac{re^{\frac{5}{4\sigma ^{2}} r^{2}}}{r-2r_{1}} \cdot \left(\frac{8r_{1}}{r}\right)^{\ell }.
\end{equation*}
When $r_2<\frac{r}{2}$, by (\ref{bound_2j_1k}) and (\ref{bound_2j_2k}), we obtain $\forall k\in [\ell]$,
\begin{align*}
| (\mathcal{E}_{t,2})_{1,k}|  & \leq \sum _{j=\ell }^{\infty }| \lambda _{2,j}| \cdotp | (\mathcal{E}_{t,2,j})_{1,k}| =\frac{w_{2}}{\sqrt{2\sigma \sqrt{\pi }}} \cdot \frac{re^{\frac{5}{4\sigma ^{2}} r^{2}}}{r-2r_{2}} \cdot \left(\frac{8r_{2}}{r}\right)^{\ell }\\
| (\mathcal{E}_{t,2})_{2,k}|  & \leq \sum _{j=\ell }^{\infty }| \lambda _{2,j}| \cdotp | (\mathcal{E}_{t,2,j})_{2,k}| =\frac{w_{2}}{\sqrt{2\sigma \sqrt{\pi }}} \cdot \frac{re^{\frac{1}{4\sigma ^{2}} r^{2}}}{r-2r_{2}} \cdot \left(\frac{8r_{2}}{r}\right)^{\ell }.
\end{align*}
Therefore, for $i=1,2$, and $\forall k\in[\ell]$, we have
\begin{align*}
| (\mathcal{E}_{t})_{i,k}|  & \leq | (\mathcal{E}_{t,1})_{i,k}| +| (\mathcal{E}_{t,2})_{i,k}| \leq \frac{re^{\frac{5}{4\sigma ^{2}} r^{2}}}{\sqrt{2\sigma \sqrt{\pi }}( r-2\max( r_{1} ,r_{2}))}\left(\frac{8\max( r_{1} ,r_{2})}{r}\right)^{\ell },
\end{align*}
and when $8\max(r_1,r_2)<r$, the right hand side goes to zero as $\ell\rightarrow\infty$.

\end{proof}

\subsubsection{\texorpdfstring{Bounding the entries of $\mathcal{E}_a$}{Bounding the entries of Ea}}
Next, we turn to  $\mathcal{E}_a$. Recall that $\widehat{f}$ is a density estimator of the true density $f=w_1f_1+w_2f_2$, and we define $y^\Delta$ to be the $2l$-dimensional vector whose entries are given by $\langle \Delta f ,\psi _{j,c_{i}, \sigma} \rangle$ for $i=1,2$ and $j\in[\ell]$, where $\Delta f=\widehat{f}-f$. We defined in (\ref{E_a}) that:
\begin{equation*}
\mathcal{E}_{a} =A^{-1} y^{\Delta }.
\end{equation*}
By Cramer's rule, the entry of $\mathcal{E}_a$ indexed at $(i,k)$ is
\begin{equation*}
\frac{\det\left( A^{( i,k)\rightarrow \Delta }\right)}{\det( A)},
\end{equation*}
where $A^{(i,k)\rightarrow \Delta}$ is the $2\ell$-by$2\ell$ matrix same as $A$ except that the column indexed at $(i,k)$ is replaced with $y^\Delta$ for $i=1,2$ and $k\in[\ell]$. Lemma \ref{Lemma 31} gives a bound on $\left|\det(A^{(1,k)\rightarrow\Delta}) \right|$ when comparing to $\det(A)$.

\begin{lemma}[Lemma 31, \cite{bryon_2023}]\label{Lemma 31}
For any $\ell\in \mathbb{N}_{0} ,\ i=1,2\ and\ k\in [ \ell]$, we have
\begin{equation*}
| (\mathcal{E}_{a})_{i,k}| =\frac{\left| \det\left( A^{( i,k)\rightarrow \Delta }\right)\right| }{\det( A)} \leq \| \Delta f\| _{2} \cdot c_{1}( r,\sigma )\exp( \ell \log \ell +c_{2}( r,\sigma ) \ell +c_{3}\log \ell ),
\end{equation*}
for some constants $c_1(r,\sigma)>0$ and $c_2(r,\sigma)$ depending on $r$ and $\sigma$, and some independent constant $c_3>0$.
\end{lemma}

\begin{proof}
We first observe that the matrix $A^{(i,k)\rightarrow\Delta}$ can be decomposed as  
\[
A^{(i,k)\rightarrow\Delta} = V^\top V^{(i,k)\rightarrow\Delta}
\]
where $V^{(i,k)\rightarrow\Delta}$ is the $|\mathbb{N}_0|$-by-$2\ell$ matrix whose column indexed at $(i,k)$ is replaced with the $|\mathbb{N}_0|$ dimensional vector $v^\Delta$ for $i = 1,2$ and $k \in [\ell]$. Here, $v^\Delta$ is the $|\mathbb{N}_0|$ dimensional vector whose $k$-th entry is $\langle \Delta f, \psi_{k,c_{1},\sigma} \rangle$ for $k \in \mathbb{N}_0$. Recall that $\Delta f = \widehat{f} - f$. For example, when $\ell = 2$, $i = 1$ and $k = 1$,  
\[
V^{(i,k)\rightarrow\Delta} =
\begin{bmatrix}
1 & \langle \Delta f, \psi_{0,c_{1},\sigma} \rangle & \langle \psi_{0,c_{2},\sigma}, \psi_{0,c_{1},\sigma} \rangle & \langle \psi_{1,c_{2},\sigma}, \psi_{0,c_{1},\sigma} \rangle \\
0 & \langle \Delta f, \psi_{1,c_{1},\sigma} \rangle & \langle \psi_{0,c_{2},\sigma}, \psi_{1,c_{1},\sigma} \rangle & \langle \psi_{1,c_{2},\sigma}, \psi_{1,c_{1},\sigma} \rangle \\
0 & \langle \Delta f, \psi_{2,c_{1},\sigma} \rangle & \langle \psi_{0,c_{2},\sigma}, \psi_{2,c_{1},\sigma} \rangle & \langle \psi_{1,c_{2},\sigma}, \psi_{2,c_{1},\sigma} \rangle \\
0 & \langle \Delta f, \psi_{3,c_{1},\sigma} \rangle & \langle \psi_{0,c_{2},\sigma}, \psi_{3,c_{1},\sigma} \rangle & \langle \psi_{1,c_{2},\sigma}, \psi_{3,c_{1},\sigma} \rangle \\
\vdots & \vdots & \vdots & \vdots
\end{bmatrix}.
\]
Recall that $V$ is the $|\mathbb{N}_0|$-by-$2\ell$ matrix whose column indexed at $(i,k)$ is the $|\mathbb{N}_0|$ dimensional vector $v^{(i,k)}$ for $i = 1,2$ and $k \in [\ell]$. Here, for $i=1,2$ and $j \in \mathbb{N}_0$, $v^{(i,j)}$ is the $|\mathbb{N}_0|$ dimensional vector whose $k$-th entry is $\langle \psi_{j,c_i,\sigma}, \psi_{k,c_{1},\sigma} \rangle$ for $k \in \mathbb{N}_0$.  

We first give a bound on $|\det(A^{(1,k)\rightarrow\Delta})|$. By Cauchy-Binet formula (Lemma \ref{Lemma 9}) and expanding the determinant along the columns with a single 1, we have  
\[
|\det(A^{(1,k)\rightarrow\Delta})| \leq \sum_{K \in \mathcal{K}} |\det(U_K)||\det(U_K^\Delta)|,
\]
where $K\in \mathcal{K} =\{K|K=\{\ell \leq a_{0} < \cdots < a_{\ell -1}\}\}$, $U_{K}^{\Delta }$ is the $(\ell+1)$-by-$(\ell+1)$ matrix whose $(i,j)$-entry is $\begin{cases}
\langle \psi _{j-1,c_{2} ,\sigma } ,\psi _{b_{i} ,c_{1} ,\sigma } \rangle , & j\in [ \ell +1] \backslash\{0\}\\
\langle \Delta f,\psi _{b_{i} ,c_{1} ,\sigma } \rangle , & j=0
\end{cases}$ for $b_i\in \{k\}\cup K$. Furthermore, by Cauchy-Schwarz inequality, we have
\begin{equation*}
\begin{aligned}
|\det\left( A^{( 1,k)\rightarrow \Delta }\right) | & \leq \sqrt{\left(\sum _{K\in \mathcal{K} }\det( U_{K})^{2}\right)\left(\sum _{K\in \mathcal{K} }\det\left( U_{K}^{\Delta }\right)^{2}\right)}\\
 & =\sqrt{\det( A) \cdot \sum _{K\in \mathcal{K} }\det\left( U_{K}^{\Delta }\right)^{2}}.
\end{aligned}
\end{equation*}
We first expand the determinant $\det(U_K^\Delta)$ along the column indexed at $j=0$.
\begin{align}
& \det \left( U_{K}^{\Delta }\right)^{2} \notag\\
& \leq \left( |\langle \Delta f,\psi _{k,c_1,\sigma} \rangle |\cdot |\det( U_{K}) |+\sum _{c=0}^{\ell-1} |\langle \Delta f,\psi _{a_{c},c_1,\sigma} \rangle |\cdot |\det( U_{K\cup \{k\} \backslash \{a_{c}\}}) |\right)^{2}\notag \\
 & \leq ( \ell+1)\left( |\langle \Delta f,\psi _{k,c_1,\sigma} \rangle |^{2} \cdot |\det( U_{K}) |^{2} +\sum _{c=0}^{\ell-1} |\langle \Delta f,\psi _{a_{c} ,c_1,\sigma} \rangle |^{2} \cdot |\det( U_{K\cup \{k\} \backslash \{a_{c}\}}) |^{2}\right). \label{det_Uk_delta_upbound}
\end{align}
The last inequality is obtained by Cauchy-Schwarz inequality. Now we bound the second term on the right hand side.
\begin{align}
\sum _{K\in \mathcal{K} }\sum _{c=0}^{\ell-1} |\langle \Delta f,\psi _{a_{c} ,c_1,\sigma} \rangle |^{2} \cdot |\det( U_{K\cup \{k\} \backslash \{a_{c}\}}) |^{2} 
& =\sum _{K\in \mathcal{K} '}\left(\sum _{j\geq \ell,\ j\notin K } \langle \Delta f,\psi _{j,c_1,\sigma} \rangle ^{2}\right) |\det( U_{K\cup \{k\}}) |^{2} \notag \\
 & \leq \| \Delta f\|_2 ^{2}\sum _{K\in \mathcal{K} '} |\det( U_{K\cup \{k\}}) |^{2}, \label{det_Uk_bound2}
\end{align}
where $\mathcal{K}'$ is the set of subsets of $\mathbb{N}_0$ of size $\ell-1$ whose elements are larger than or equal to $\ell$, i.e., $\mathcal{K}'=\{K|K=\{\ell\leq a_{0} < \cdots < a_{\ell-2}\}\}$. by combining (\ref{det_Uk_delta_upbound}) and (\ref{det_Uk_bound2}), we obtain
\begin{align*}
\det( A) \cdot &\sum _{K\in \mathcal{K}}\det\left( U_{K}^{\Delta }\right)^{2} \\
& \leq ( \ell+1)\det( A)\left( |\langle \Delta f,\psi _{k,0} \rangle |^{2} \cdot \sum _{K\in \mathcal{K}} |\det( U_{K}) |^{2} +\| \Delta f\|_2 ^{2}\sum _{K\in \mathcal{K} '} |\det( U_{K\cup \{k\}}) |^{2}\right)\\
 & =( \ell+1)\det( A)\left( |\langle \Delta f,\psi _{k,0} \rangle |^{2}\det( A) +\| \Delta f\|_2 ^{2}\sum _{K\in \mathcal{K} '} |\det( U_{K\cup \{k\}}) |^{2}\right).
\end{align*}
Since each $K\in\mathcal{K}'$ has $\ell-1$ elements, then $\{\ell,\cdots,2\ell-1\}\nsubseteq K$. We have
\[
\sum_{K \in \mathcal{K}'} |\det(U_{K \cup \{k\}})|^2 
\leq \sum_{j=\ell}^{2\ell-1} \sum_{K \in \mathcal{K}_j} |\det(U_{K \cup \{k\}})|^2.
\]
Recall that $\mathcal{K}_j$ is the set of subsets of $\mathbb{N}_0$ of size $\ell - 1$ whose elements are larger than or equal to $\ell$ and not equal to $j$, i.e. $\mathcal{K}_j = \{K \mid K = \{\ell \leq a_0 < \cdots < a_{\ell-2} \text{ and } a_i \neq j\}\}$. By Lemma \ref{Lemma 32}, we have
\begin{equation*}
\begin{aligned}
\sum _{K\in \mathcal{K}'} |\det( U_{K\cup \{k\}}) |^{2} & \leq \sum _{j=\ell}^{2\ell-1}\left( e^{\frac{r^{2}}{4\sigma ^{2}}} 4^{\ell}\left(\frac{2\sqrt{2} \sigma }{r}\right)^{j}\sqrt{j!}\right)^{2} \cdot \det( A)
\\
 & =\left( e^{\frac{1}{4\sigma^2} r^{2}} 4^{\ell}\right)^{2} \cdot \det( A) \cdot \sum _{j=\ell}^{2\ell-1}( j!) \cdot \left(\frac{8\sigma^2}{r^{2}}\right)^{j}\\
 & \leq \left( e^{\frac{1}{4\sigma^2} r^{2}} 4^{\ell}\right)^{2}\det( A) \cdot ( 2\ell) !\cdot \sum _{j=\ell}^{2\ell-1}\left(\frac{8\sigma^2}{r^{2}}\right)^{j}.
\end{aligned}
\end{equation*}
From Stirling's approximation we have $( 2\ell) !\leq \sqrt{2\pi \cdot( 2\ell)} \cdot e^{\frac{1}{12\cdot ( 2\ell)}}\left(\frac{2\ell}{e}\right)^{2\ell}$, hence
\begin{equation*}
\begin{aligned}
\sum _{K\in \mathcal{K}'} & |\det( U_{K\cup \{k\}}) |^{2} \\
& \leq \begin{cases}
\left( e^{\frac{1}{4\sigma^2} r^{2}} 4^{\ell}\right)^{2}\det( A) \cdot \sqrt{4\pi \ell} \cdot e^{\frac{1}{24\ell} -2\ell}( 2\ell)^{2\ell} \cdot (\frac{8\sigma^2}{r^2})^\ell(1-(\frac{8\sigma^2}{r^2})^\ell)\cdot\frac{r^2}{r^2-8\sigma^2} ,& r\neq 2\sqrt{2}\sigma\\
\left( e^{\frac{1}{4\sigma^2} r^{2}} 4^{\ell}\right)^{2}\det( A) \cdot \sqrt{4\pi \ell} \cdot e^{\frac{1}{24\ell} -2\ell}( 2\ell)^{2\ell} \cdot \ell, & r=2\sqrt{2}\sigma
\end{cases}\\
 & \leq \begin{cases}
2\sqrt{\pi } e^{\frac{r^{2}}{2\sigma^2}+1}\left(\frac{r^{2}}{r^{2} -8\sigma^2}\right)\det( A) \cdot \exp\left( 2\ell\log \ell+\left(\log\frac{512\sigma^2}{r^{2}} -2\right) \ell+\frac{1}{2}\log \ell\right) , & r >2\sqrt{2}\sigma\\
2\sqrt{\pi } e^{\frac{r^{2}}{2\sigma^2}+1}\left(\frac{r^{2}}{8\sigma^2-r^{2}}\right)\det( A) \cdot \exp\left( 2\ell\log \ell+\left(\log\frac{4096\sigma^4}{r^{4}} -2\right) \ell+\frac{1}{2}\log \ell\right) , & r< 2\sqrt{2}\sigma\\
2\sqrt{\pi } e^{\frac{r^{2}}{2\sigma^2}+1}\det( A) \cdot \exp\left( 2\ell\log \ell+(\log 64-2) \ell+\frac{3}{2}\log \ell\right), & r=2\sqrt{2}\sigma.
\end{cases}
\end{aligned}
\end{equation*}
By combining all above results, we obtain
\begin{align*}
\det( A) & \cdot \sum _{K\in \mathcal{K}}\det\left( U_{K}^{\Delta }\right)^{2} \\
& \leq ( \ell+1)\det( A)^{2} \| \Delta f\|_2 ^{2} +( \ell+1)\det( A)^{2} \| \Delta f\|_2 ^{2} c_{1}( r,\sigma)\exp\left( 2\ell\log \ell+c_{2}( r,\sigma) \ell+c_3\log \ell\right)\\
 & \leq \det( A)^{2} \| \Delta f\|_2 ^{2} \cdot c_{1}( r,\sigma)\exp\left( 2\ell\log \ell+c_{2}( r,\sigma) \ell+c_3\log \ell\right),
\end{align*}
for some constants $c_1(r,\sigma)>0$ and $c_3>0$.
Therefore
\begin{align}
|\det\left( A^{( 1,k)\rightarrow \Delta }\right) | & \leq \sqrt{\det( A) \cdot \sum _{K\in \mathcal{K} }\det\left( U_{K}^{\Delta }\right)^{2}}\notag\\
 & \leq \det( A) \cdot \| \Delta f\|_2\cdot c_{1}( r,\sigma)\exp\left( \ell\log \ell+c_{2}( r,\sigma) \ell+c_3\log \ell\right).\label{ub_det_A1k_delta}
\end{align}

We now give a bound on \( | \det(A^{(2,k) \to \Delta}) | \). By a similar argument, we have
\[
| \det(A^{(2,k) \to \Delta}) | \leq \sum_{K \in \mathcal{K}} | \det(U_K) | | \det(U_K^{k \to \Delta}) |
\]
where, for any \( k \in [\ell] \), \( U_K^{k \to \Delta} \) is the \(\ell\)-by-\(\ell\) matrix whose \((i,j)\)-entry is 
\[
\begin{cases}
\langle \psi_{j,c_2,\sigma}, \psi_{b_i,c_1,\sigma} \rangle & \text{if } j \in [\ell]\backslash{k} \\
\langle \Delta f, \psi_{b_i,c_1,\sigma} \rangle & \text{if } j = k
\end{cases}
\]
for \( b_i \in K \). Furthermore, by the Cauchy-Schwarz inequality, we have
\begin{equation}\label{det_A2k_delta}
| \det(A^{(2,k) \to \Delta}) | \leq \sqrt{\left( \sum_{K \in \mathcal{K}} \det(U_K)^2 \right) \left( \sum_{K \in \mathcal{K}} \det(U_K^{k \to \Delta})^2 \right)} \leq \sqrt{\sum_{K \in \mathcal{K}} \det(U_K^{k \to \Delta})^2 \cdot \det(A)}.
\end{equation}
For each \( K = \{ \ell \leq a_0 < \cdots < a_{\ell-1} \} \in \mathcal{K} \), we expand the determinants along the column indexed at \( \Delta f \).
\begin{align*}
\det(U_K^{k \to \Delta})^2 & \leq \left( \sum_{c=0}^{\ell-1} |\langle \Delta f, \psi_{a_c, c_1,\sigma} \rangle| |\det(U^{(-k)}_{K \setminus \{a_c\}})|\right)^2\\
& \leq \ell \cdot \left( \sum_{c=0}^{\ell-1} |\langle \Delta f, \psi_{a_c, c_1,\sigma} \rangle|^2 |\det(U^{(-k)}_{K \setminus \{a_c\}})|^2 \right).
\end{align*}
We now consider the summation \( \sum_{K \in \mathcal{K}} \sum_{c=0}^{\ell-1} |\langle \Delta f, \psi_{a_c, c_1,\sigma} \rangle|^2 |\det(U^{(-k)}_{K \setminus \{a_c\}})|^2 \) and we have
\begin{align*}
\sum_{K \in \mathcal{K}} \sum_{c=0}^{\ell-1} |\langle \Delta f, \psi_{a_c, c_1,\sigma} \rangle|^2 |\det(U^{(-k)}_{K \setminus \{a_c\}})|^2 & = \sum_{K \in \mathcal{K}'} \left( \sum_{j \geq \ell, j \notin K} \langle \Delta f, \psi_{j, c_1,\sigma} \rangle^2 \right) |\det(U_K^{(-k)})|^2 \\
& \leq \| \Delta f \|_2^2 \sum_{K \in \mathcal{K}'} |\det(U_K^{(-k)})|^2,
\end{align*}
where \( \mathcal{K}' \) is the set of subsets of \( \mathbb{N}_0 \) of size \( \ell - 1 \) whose elements are larger than or equal to \( \ell \), i.e., $\mathcal{K}' = \{K \mid K = \{\ell \leq a_0 < \cdots < a_{\ell-2}\} \}$. Since each \( K \in \mathcal{K}' \) has \( \ell - 1 \) elements, then \(\{\ell, \cdots, 2\ell - 1\} \not\subseteq K\). We have
\[
\sum_{K \in \mathcal{K}'} |\det(U_K^{(-k)})|^2 \leq \sum_{j=\ell}^{2\ell-1} \sum_{K \in \mathcal{K}_j} |\det(U_K^{(-k)})|^2.
\]
Recall that \( \mathcal{K}_j \) is the set of subsets of \( \mathbb{N}_0 \) of size \( \ell - 1 \) whose elements are larger than or equal to \( \ell \) and not equal to \( j \), i.e.,$\mathcal{K}_j = \{K \mid K = \{\ell \leq a_0 < \cdots < a_{\ell-2} \text{ and } a_i \neq j\} \}$. By Lemma \ref{Lemma 33}, we have
\begin{align*}
\sum_{K \in \mathcal{K}'} |\det(U_K^{(-k)})|^2 & \leq \sum_{j=\ell}^{2\ell-1} \left( e^{\frac{5}{4\sigma ^{2}} r^{2}} 4^{\ell } \cdot \left(\frac{2\sqrt{2} \sigma }{r}\right)^{j}\sqrt{j!}\right)^{2} \cdot \det( A)\\
& \leq c_1(r,\sigma)\exp(2\ell\log\ell+c_2(r,\sigma)\ell+c_3\log\ell)\cdot\det(A),
\end{align*}
for some constants $c_1(r,\sigma)>0$ and $c_3>0$, which means
\[
\sum_{K \in \mathcal{K}} \det(U_K^{k \to \Delta})^2 \leq \|\Delta f\|^2 \cdot \det(A)\cdot c_1(r,\sigma)\exp(2\ell\log\ell+c_2(r,\sigma)\ell+c_3\log\ell)
\]
By plugging it into (\ref{det_A2k_delta}),
\begin{equation}\label{ub_det_A2k_delta}
|\det(A^{(2,k) \to \Delta})| \leq \|\Delta f\|_2 \cdot \det(A)\cdot c_1(r,\sigma)\exp(\ell\log\ell+c_2(r,\sigma)\ell+c_3\log\ell).
\end{equation}
By combining (\ref{ub_det_A1k_delta}) and (\ref{ub_det_A2k_delta}), we obtain that for $i=1,2$, and $\forall k\in[\ell]$,
\begin{equation*}
| (\mathcal{E}_{a})_{i,k}| =\frac{\left| \det\left( A^{( i,k)\rightarrow \Delta }\right)\right| }{\det( A)} \leq \| \Delta f\| _{2} \cdot c_{1}( r,\sigma )\exp( \ell \log \ell +c_{2}( r,\sigma ) \ell +c_{3}\log \ell ),
\end{equation*}
for some constants $c_1(r,\sigma)>0$ and $c_3>0$.
\end{proof}

\subsection{Completing the proof of Lemma \ref{lemma: bound_l1_fi_hat_fi}}
We are ready to complete the proof of this key lemma.
\begin{proof}[Proof of Lemma \ref{lemma: bound_l1_fi_hat_fi}]
By Lemma \ref{Lemma 17} we know if $\left|\lambda_{i,k}-\widehat{\lambda}_{i,k}\right|<\Delta$, $\forall i=1,2,\ k\in [\ell]$, and $\ell\geq \lfloor 2er_i^2/\sigma^2\rfloor+2$, then 
\begin{equation}\label{eq_wifi_minus_fi}
\| w_{i} f_{i} -\widetilde{f_{i}} \| _{1} \leq \frac{2^{17/4} c_{\psi } \sqrt{\sigma} }{\Gamma \left(\frac{1}{4}\right)^{2}} \cdot \Delta \ell^{5/4} +\frac{2^{11/2} c_{\psi } r_{i} w_{i}}{\Gamma \left(\frac{1}{4}\right)^{2} \pi ^{1/4}\sigma} \cdot \left(\frac{1}{2}\right)^{\ell}.
\end{equation}
By Lemma \ref{Lemma up_bouond_E_t}, we know when $8\max(r_1,r_2)<r$, for $i=1,2$, and $\forall k\in [\ell]$, as $\ell\rightarrow\infty$,
\begin{align}\label{eq_Et_bd}
| (\mathcal{E}_{t})_{i,k}|  & \leq | (\mathcal{E}_{t,1})_{i,k}| +| (\mathcal{E}_{t,2})_{i,k}| \leq \frac{re^{\frac{5}{4\sigma ^{2}} r^{2}}}{\sqrt{2\sigma \sqrt{\pi }}( r-2\max( r_{1} ,r_{2}))}\left(\frac{8\max( r_{1} ,r_{2})}{r}\right)^{\ell }\rightarrow 0,
\end{align}
Furthermore, by Lemma \ref{Lemma 31}, we know for any $\ell\in \mathbb{N}_{0} ,\ i=1,2\ and\ k\in [ \ell]$, we have
\begin{equation}\label{eq_Ea_bd}
| (\mathcal{E}_{a})_{i,k}| =\frac{\left| \det\left( A^{( i,k)\rightarrow \Delta }\right)\right| }{\det( A)} \leq \| \Delta f\| _{2} \cdot c_{1}( r,\sigma )\exp( \ell \log \ell +c_{2}( r,\sigma ) \ell +c_{3}\log \ell ).
\end{equation}
From (\ref{eq_Ea_bd}) and (\ref{eq_Et_bd}) we know $\forall i=1,2,\ k\in[\ell]$, 
\begin{align*}
| \lambda _{i,k} -\widehat{\lambda }_{i,k}|  & \leq |(\mathcal{E}_{t})_{i,k} |+|(\mathcal{E}_{a})_{i,k} |\\
 & \leq c_{1}( r,\sigma ,r_{1} ,r_{2}) \cdot \left(\frac{8\max( r_{1} ,r_{2})}{r}\right)^{\ell } +c_{2}( r,\sigma ) \| \Delta f\| _{2}\exp( \ell \log \ell +c_{3}( r,\sigma ) \ell +c_{4}\log \ell ).
\end{align*}
We plug this into $\Delta$ in (\ref{eq_wifi_minus_fi}), then we obtain
\begin{align*}
\| w_{i} f_{i} -\widetilde{f_{i}} \| _{1} & \leq c_{1}( r,\sigma ,r_{1} ,r_{2}) \cdot \left(\frac{8\max( r_{1} ,r_{2})}{r}\right)^{\ell } \cdot \ell ^{5/4} +c_{2}( r_{i} ,\sigma )w_{i}\left(\frac{1}{2}\right)^{\ell }\\
 & \ \ \ \ \ +\ c_{3}( r,\sigma ) \cdot \| \Delta f\| _{2} \cdot \exp( \ell \log \ell +c_{4}( r,\sigma ) \ell +c_{5}\log \ell ),
\end{align*}
Finally, by triangle inequality, we get
\begin{align*}
\| \widehat{f_{i}} -f_{i} \| _{1} & =\left\lVert \frac{\left(\widetilde{f_{i}}\right)_{+}}{\| \left(\widetilde{f_{i}}\right)_{+} \| _{1}} -\frac{\left(\widetilde{f_{i}}\right)_{+}}{w_{i}} +\frac{\left(\widetilde{f_{i}}\right)_{+}}{w_{i}} -f_{i} \right\rVert _{1}\\
 & \leq \left\lVert \frac{\left(\widetilde{f_{i}}\right)_{+}}{\| \left(\widetilde{f_{i}}\right)_{+} \| _{1}} -\frac{\left(\widetilde{f_{i}}\right)_{+}}{w_{i}} \right\rVert_{1} +\left\lVert \frac{\left(\widetilde{f_{i}}\right)_{+}}{w_{i}} -f_{i} \right\rVert _{1}\\
 & =\left| 1-\frac{\| \left(\widetilde{f_{i}}\right)_{+} \| _{1}}{w_{i}}\right| \cdot \left\lVert \frac{\left(\widetilde{f_{i}}\right)_{+}}{\| \left(\widetilde{f_{i}}\right)_{+} \| _{1}} \right\rVert _{1} +\frac{1}{w_{i}} \left\lVert \left(\widetilde{f_{i}}\right)_{+} -w_{i} f_{i} \right\rVert _{1}\\
 & \leq \frac{2}{w_{i}} \left\lVert w_{i} f_{i} -\left(\widetilde{f_{i}}\right)_{+} \right\rVert _{1}\\
 & \leq \frac{2}{w_{i}} \left\lVert w_{i} f_{i} -\widetilde{f_{i}} \right\rVert _{1}\\
 & \leq \frac{1}{w_{i}} c_{1}( r,\sigma ,r_{1} ,r_{2}) \cdot \left(\frac{8\max( r_{1} ,r_{2})}{r}\right)^{\ell } \cdot \ell ^{5/4} +c_{2}( r_{i} ,\sigma )\left(\frac{1}{2}\right)^{\ell }\\
 & \ \ \ \ \ +\ \frac{1}{w_{i}} c_{3}( r,\sigma ) \cdot \| \Delta f\| _{2} \cdot \exp( \ell \log \ell +c_{4}( r,\sigma ) \ell +c_{5}\log \ell )
\end{align*}
for some constants $c_1(r,\sigma,r_1,r_2)>0$, $c_2(r_i,\sigma)>0$, $c_3(r,\sigma)>0$, $c_4(r,\sigma)>0$, and $c_5>0$.
\end{proof}

\subsection{Auxiliary lemmas}

For completeness, in this subsection we collect technical lemmas required in the proof from the previous subsection.
\begin{lemma}[Lemma 32, \cite{bryon_2023}] \label{Lemma 32}
For any \(\ell \geq 1\), \(k \in [\ell]\) and \(j \geq \ell\), we have  
\begin{equation*}
\sum _{K\in \mathcal{K}_{j}}\det( U_{K\cup \{k\}})^{2} \leq \left( e^{\frac{r^{2}}{4\sigma ^{2}}} 4^{\ell}\left(\frac{2\sqrt{2} \sigma }{r}\right)^{j}\sqrt{j!}\right)^{2} \cdot \det( A).
\end{equation*}
Recall that, for any \(K = \{\ell \leq a_1 < \cdots < a_{\ell-1} \text{ and } a_i \neq j\} \in K_j\) and \(k \in [\ell]\), \(U_{K \cup \{k\}}\) is the \(\ell\)-by-\(\ell\) matrix whose \((i,j)\)-entry is \(\langle \psi_{c_j,r,\sigma}, \psi_{b_i,0,\sigma} \rangle\) for \(b_i \in \{k\} \cup K \) and \(c_j \in [\ell]\).    
\end{lemma}
\begin{proof}
For each \( K = \{ \ell \leq a_0 < \cdots < a_{\ell-2} \text{ and } a_i \neq j \} \in \mathcal{K}_j \), we have the following. By Lemma \ref{Lemma 34}, we have
\begin{align*}
\frac{|\det( U_{K\cup \{k\}})|}{|\det( U_{K\cup \{j\}})|}  & =\sqrt{\frac{j!}{k!}}\left(\frac{r}{\sqrt{2} \sigma }\right)^{k-j}\left(\prod _{c=0}^{\ell=2}\frac{|a_{c} -k|}{|a_{c} -j|}\right) \cdot \frac{\Gamma _{K\cup \{k\} ,\ell }}{\Gamma _{K\cup \{j\} ,\ell }}\\
 & \leq \sqrt{\frac{j!}{k!}}\left(\frac{r}{\sqrt{2} \sigma }\right)^{k-j}\left(\prod _{c=0}^{\ell=2}\frac{|a_{c} -k|}{|a_{c} -j|}\right)
\end{align*}
The second inequality follows from Lemma \ref{Lemma 38}, which states that $\Gamma_{K,b}$ is an increasing function w.r.t any $a_i\in K$.

By Lemma \ref{Lemma 36}, we have \(\prod_{c=0}^{\ell-2} \frac{|a_c - k|}{|a_c - j|} \leq \prod_{c=0}^{\ell-2} \frac{|a_c|}{|a_c - j|} \leq 2^j \cdot 4^\ell\) and hence

\begin{align*}
|\det (U_{K\cup\{k\}})| & \leq 4^\ell \cdot 2^j \sqrt{\frac{j!}{k!}} \left( \frac{r}{\sqrt{2}\sigma} \right)^{k-j} |\det (U_{K\cup\{j\}})| \\
& \leq e^{\frac{1}{4\sigma ^{2}} r^{2}} \cdot 4^{\ell } \cdot \sqrt{j!} \cdot \left(\frac{2\sqrt{2} \sigma }{r}\right)^{j} |\det( U_{K\cup \{j\}}) |
\end{align*}
since \(\frac{1}{\sqrt{k!}} \left( \frac{r}{\sqrt{2}\sigma} \right)^k \leq \sqrt{\sum_{i=0}^\infty \frac{1}{i!} \left( \frac{r^2}{2\sigma^2} \right)^i} = e^{\frac{1}{4\sigma^2}r^2}\). For each \( K \in \mathcal{K}_j \), the set \( K \cup \{j\} \) is in \( \mathcal{K} \) since \( j \notin K \). Hence, we conclude that

\begin{align*}
\sum_{K \in \mathcal{K}_j} \det (U_{K\cup\{k\}})^2 & \leq \left(e^{\frac{1}{4\sigma ^{2}} r^{2}} \cdot 4^{\ell } \cdot \sqrt{j!} \cdot \left(\frac{2\sqrt{2} \sigma }{r}\right)^{j}  \right)^2 \sum_{K \in \mathcal{K}_j} \det (U_{K\cup\{j\}})^2 \\
& \leq \left( e^{\frac{r^{2}}{4\sigma ^{2}}} 4^{\ell }\left(\frac{2\sqrt{2} \sigma }{r}\right)^{j}\sqrt{j!}\right)^{2}\cdot\det( A).
\end{align*}
\end{proof}
Before we show the lemmas below, we first define the following notations to simplify the expressions in our proof. For any \( r \in \mathbb{R} \), and $\sigma>0$, let \( W_{r,\sigma} \) be the double-indexed sequence such that
\[
(W_{r,\sigma})_{s,t} = e^{-\frac{1}{8\sigma^2}r^2}(-1)^s \sqrt{\frac{t!}{s!}} \binom{s}{t} \left( \frac{r}{\sqrt{2}\sigma} \right)^{s-t}
\]
for \( s, t \in \mathbb{N}_0 \). Here, \( \binom{s}{t} \) is the binomial coefficient which is equal to
\[
\binom{s}{t} = \begin{cases}
0 & \text{if } s < t \\
\frac{s!}{t!(s-t)!} & \text{if } s \geq t
\end{cases}
\]
For any subsets \( S, T \subset \mathbb{N}_0 \) of same sizes, let \((W_r)_{S,T}\) be the matrix whose \((i,j)\)-entry is \((W_r)_{s_i,t_j}\) for \( s_i \in S \) and \( t_j \in T \). For any (ordered) set \( K \) of \( m \) nonnegative integers \( a_0, \cdots, a_{m-1} \) such that \( 0 \leq a_0 < \cdots < a_{m-1} \), we define
\[
\Sigma_K = \sum_{c=0}^{m-1} a_c, \quad F_K = \prod_{c=0}^{m-1} a_c!, \quad C_K = \prod_{0 \leq c_2 < c_1 \leq m-1} (a_{c_1} - a_{c_2}).
\]
Finally, for any \( b \in \mathbb{N}_0 \), we define
\[
\Gamma_{K,b} = \frac{1}{b!} \sum_{d=0}^b (-1)^d \binom{b}{d} \prod_{c=0}^{m-1} (a_c - d).
\]

\begin{lemma}[Lemma 33, \cite{bryon_2023}]\label{Lemma 33}
For any $\ell \geq 1$, $k\in [\ell]$ and $j\geq \ell$, we have
\begin{equation*}
\sum _{K\in \mathcal{K}_{j}}\det\left( U_{K}^{( -k)}\right)^{2} \leq \left( e^{\frac{5}{4\sigma ^{2}} r^{2}} 4^{\ell } \cdot \left(\frac{2\sqrt{2} \sigma }{r}\right)^{j}\sqrt{j!}\right)^{2} \cdot \det( A) .
\end{equation*}
Recall that, for any $K=\{\ell\leq a_0 <\cdots < a_{\ell-2}\ \text{and}\ a_i\neq j\} \in \mathcal{K}_j$ and $k\in [\ell]$, $U_K^{(-k)}$ is the $(\ell-1)$-by-$(\ell-1)$ matrix whose $(i,j)$-entry is $\langle\psi_{c_j, r, \sigma}, \psi_{b_i, 0,\sigma} \rangle$ for $b_i\in K$ and $c_j\in [\ell]\backslash\{k\}$.
\end{lemma}
\begin{proof}
For each $K=\left\{\ell \leq a_{0} < \cdots < a_{\ell -2} \ \text{and} \ a_{i} \neq j\right\} \in \mathcal{K}_{j}$, we have the following. By Lemma \ref{Lemma 34} and \ref{Lemma 35}, we have
\begin{align*}
\frac{\left| \det\left( U_{K}^{( -k)}\right)\right| }{| \det( U_{K\cup \{j\}})| } & \leq \frac{e^{-\frac{\ell -1}{4\sigma ^{2}} r^{2}} 2^{k}\sqrt{\frac{1}{k!F_{[ \ell ]} F_{K}}}\left(\frac{r}{\sqrt{2} \sigma }\right)^{\Sigma _{K} -\ell ( \ell -1) /2+k} C_{K}\prod _{c=0}^{\ell -2} a_{c}}{e^{-\frac{\ell }{4\sigma ^{2}} r^{2}}\sqrt{\frac{1}{F_{[ \ell ]} F_{K\cup \{j\}}}}\left(\frac{r}{\sqrt{2} \sigma }\right)^{\Sigma _{K\cup \{j\}} -\ell ( \ell -1) /2} C_{K\cup \{j\}} \Gamma _{K\cup \{j\} ,\ell }}\\
 & =e^{\frac{r^{2}}{4\sigma ^{2}}} 2^{k}\sqrt{\frac{j!}{k!}}\left(\frac{r}{\sqrt{2} \sigma }\right)^{k-j}\prod _{c=0}^{\ell -2}\frac{a_{c}}{| a_{c} -j| } \cdot \frac{1}{\Gamma _{K\cup \{j\} ,\ell }}\\
 & \leq e^{\frac{r^{2}}{4\sigma ^{2}}} 2^{k}\sqrt{\frac{j!}{k!}}\left(\frac{r}{\sqrt{2} \sigma }\right)^{k-j}\prod _{c=0}^{\ell -2}\frac{a_{c}}{| a_{c} -j| }
\end{align*}
The last inequality is because $\Gamma _{K\cup \{j\} ,\ell } \geq \Gamma _{[ \ell +1] \backslash \{0\} ,\ell } =1$ by Lemma \ref{Lemma 38}. In addition, by Lemma \ref{Lemma 36}, we have
\begin{align*}
\frac{\left| \det\left( U_{K}^{( -k)}\right)\right| }{| \det( U_{K\cup \{j\}})| } & \leq e^{\frac{r^{2}}{4\sigma ^{2}}}\sqrt{j!}\left(\frac{r}{\sqrt{2} \sigma }\right)^{-j}\sqrt{\frac{\left(\frac{2r^{2}}{\sigma ^{2}}\right)^{k}}{k!}} \cdot 2^{j} \cdot 4^{\ell-1}\\
 & =e^{\frac{5r^{2}}{4\sigma ^{2}}} 4^{\ell }\sqrt{j!}\left(\frac{2\sqrt{2} \sigma }{r}\right)^{j}
\end{align*}
since $\sqrt{\left( 2r^{2} /\sigma ^{2}\right)^{k} /k!} \leq \sqrt{e^{2r^{2} /\sigma ^{2}}} =e^{r^{2} /\sigma ^{2}}$. For each $K\in\mathcal{K}_j$, the set $K\cup \{j\}$ is in $\mathcal{K}$ since $j\notin K$. Hence,
\begin{align*}
\sum _{K\in \mathcal{K}_{j}}\det\left( U_{K}^{( -k)}\right)^{2} & \leq \left( e^{\frac{5r^{2}}{4\sigma ^{2}}} 4^{\ell }\left(\frac{2\sqrt{2} \sigma }{r}\right)^{j}\sqrt{j!}\right)^{2}\sum _{K\in \mathcal{K}_{j}}\det( U_{K\cup \{j\}})^{2}\\
 & \leq \left( e^{\frac{5r^{2}}{4\sigma ^{2}}} 4^{\ell }\left(\frac{2\sqrt{2} \sigma }{r}\right)^{j}\sqrt{j!}\right)^{2}\det( A)
\end{align*}

\end{proof}

\begin{lemma}[Lemma 34, \cite{bryon_2023}]\label{Lemma 34}
Let \( a_0, \dots, a_{\ell-1} \) be \(\ell\) nonnegative integers and \( K \) be the set \(\{ 0 \leq a_0 < \dots < a_{\ell-1} \}\). Then, we have
\[
\left| \det(U_K) \right| = e^{-\frac{\ell}{4\sigma^2}r^2}\sqrt{\frac{1}{F_{[\ell]} F_K}} \left( \frac{r}{\sqrt{2}\sigma} \right)^{\Sigma_{K}- \ell(\ell-1)/2}  C_K\cdot \Gamma_{K,\ell}.
\]
Recall that, for any \( K = \{ 0 \leq a_0 < \dots < a_{\ell-1} \} \), \( U_K \) is the \(\ell\)-by-\(\ell\) matrix whose \((i,j)\)-entry is \(\langle \psi_{c_j,r,\sigma}, \psi_{b_i,0,\sigma} \rangle\) for \( b_i \in [K] \) and \( c_j \in [\ell] \).  
\end{lemma}
\begin{proof}
Recall that, from (\ref{inner_product_decomp}),
\begin{align*}
\langle \psi _{i,0 ,\sigma } , & \psi _{j,r ,\sigma } \rangle \notag \\
= & \sum _{k=0}^{i\land j}\left( e^{-\frac{r^2}{8\sigma ^{2}}}( -1)^{i}\sqrt{\frac{k!}{i!}}\binom{i}{k}\left(\frac{r}{\sqrt{2} \sigma }\right)^{i-k}\right)\left( e^{-\frac{r^2}{8\sigma ^{2}}}( -1)^{j}\sqrt{\frac{k!}{j!}}\binom{j}{k}\left(\frac{-r}{\sqrt{2} \sigma }\right)^{j-k}\right)
\end{align*}
It means that \( U_K \) can be decomposed as
\[
U_K = (W_{r,\sigma})_{K,[\ell]} (W_{-r,\sigma})_{[\ell], [\ell]}^\top.
\]
By Lemma \ref{Lemma 37}, the determinant of \((W_{-r,\sigma})_{[\ell], [\ell]}^\top \)is
\[
e^{-\frac{\ell}{8\sigma^2}r^2}(-1)^{\Sigma_{[\ell]}} \sqrt{\frac{1}{F_{[\ell]} F_{[\ell]}}} \left( -\frac{r}{\sqrt{2}\sigma} \right)^{\Sigma_{[\ell]} - \ell(\ell-1)/2} C_{[\ell]}\cdot \Gamma_{[\ell], \ell}= e^{-\frac{\ell}{8\sigma^2}r^2}(-1)^{\ell(\ell-1)/2},
\]
since \(\Sigma_{[\ell]} = \ell(\ell-1)/2\), \( C_{[\ell]} = F_{[\ell]}\) and $\Gamma_{[\ell], \ell}=1$. Also, the determinant of \((W_{r,\sigma})_{K,[\ell]}\) is
\[
\det\left((W_{r,\sigma})_{K,[\ell]}\right) = e^{-\frac{\ell}{8\sigma^2}r^2}(-1)^{\Sigma_K}\sqrt{\frac{1}{F_{[\ell]}F_K}}\left(\frac{r}{\sqrt{2}\sigma}\right)^{\Sigma_K-\ell(\ell-1)/2}C_K \cdot \Gamma_{K,\ell}.
\]
Hence, we have
\begin{align*}
\det(U_K) &= \det(( W_{r,\sigma })_{K,[ \ell ]}) \cdot \det(( W_{-r,\sigma })_{[ \ell ] ,[ \ell ]})\\
& = e^{-\frac{\ell}{8\sigma^2}r^2}(-1)^{\ell(\ell-1)/2} \cdot e^{-\frac{\ell}{8\sigma^2}r^2}(-1)^{\Sigma_{K}} \sqrt{\frac{1}{F_{[\ell]} F_K}} \left( \frac{r}{\sqrt{2}\sigma} \right)^{\Sigma_{K} - \ell(\ell-1)/2} C_K\cdot \Gamma_{K,\ell}\\
& = e^{-\frac{\ell}{4\sigma^2}r^2}(-1)^{\Sigma_{K} + \ell(\ell-1)/2} \sqrt{\frac{1}{F_{[\ell]} F_K}} \left( \frac{r}{\sqrt{2}\sigma} \right)^{\Sigma_{K} - \ell(\ell-1)/2} C_K\cdot \Gamma_{K,\ell}.
\end{align*}
\end{proof}

\begin{lemma}[Lemma 35, \cite{bryon_2023}]\label{Lemma 35}
Let \( a_0, \cdots, a_{\ell-2} \) be \( \ell - 1 \) nonnegative integers and \( K \) be the set \( \{ 0 \leq a_0 < \cdots < a_{\ell-2} \} \). Then, for any \( k \in [\ell] \), we have
\begin{equation*}
\left| \det\left( U_{K}^{( -k)}\right)\right| \leq e^{-\frac{\ell -1}{4\sigma ^{2}} r^{2}} 2^{k}\sqrt{\frac{1}{k!F_{[ \ell ]} F_{K}}}\left(\frac{r}{\sqrt{2} \sigma }\right)^{\Sigma _{K} -\ell ( \ell -1) /2+k} C_{K}\prod _{c=0}^{\ell -2} a_{c}
\end{equation*}
Recall that, for any \( K = \{ 0 \leq a_0 < \cdots < a_{\ell-2} \} \in \mathcal{K} \) and \( k \in [\ell] \), \( U_K^{(-k)} \) is the \( (\ell - 1) \)-by-\( (\ell - 1) \) matrix whose \( (i,j) \)-entry is \( \langle \psi_{c_j,r,\sigma}, \psi_{b_i,0,\sigma} \rangle \) for \( b_i \in K \) and \( c_j \in [\ell] \setminus \{ k \} \).
\end{lemma}

\begin{proof}
Recall that, from (\ref{inner_product_decomp}),
\begin{align*}
\langle \psi _{i,0 ,\sigma } , & \psi _{j,r ,\sigma } \rangle \notag \\
= & \sum _{k=0}^{i\land j}\left( e^{-\frac{r^2}{8\sigma ^{2}}}( -1)^{i}\sqrt{\frac{k!}{i!}}\binom{i}{k}\left(\frac{r}{\sqrt{2} \sigma }\right)^{i-k}\right)\left( e^{-\frac{r^2}{8\sigma ^{2}}}( -1)^{j}\sqrt{\frac{k!}{j!}}\binom{j}{k}\left(\frac{-r}{\sqrt{2} \sigma }\right)^{j-k}\right)
\end{align*}
It means that \( U_K^{(-k)} \) can be decomposed as
\begin{equation*}
U_{K}^{( -k)} =( W_{r,\sigma})_{K,[ \ell ]}( W_{-r,\sigma})_{[ \ell ] ,[ \ell ] \backslash \{k\}}.
\end{equation*}
By Cauchy-Binet formula (Lemma \ref{Lemma 9}), we have
\begin{equation}\label{det_U_K_minusk}
\det\left( U_{K}^{( -k)}\right) =\sum _{b=0}^{\ell -1}\det(( W_{r,\sigma})_{K,[ \ell ] \backslash \{b\}}) \cdot \det(( W_{-r,\sigma})_{[ \ell ] \backslash \{b\} ,\ [ \ell ] \backslash \{k\}}).
\end{equation}
By Lemma \ref{Lemma 37}, the determinant of $( W_{-r,\sigma})_{[ \ell ] \backslash \{b\} ,[ \ell ] \backslash \{k\}}$ is
\begin{gather*}
e^{-\frac{\ell -1}{8\sigma ^{2}} r^{2}}( -1)^{\frac{( \ell -1) \ell }{2} -b} \cdot \sqrt{\frac{k!}{F_{[ \ell ] \backslash \{b\}} \cdot F_{[ \ell ]}}} \cdot \left( -\frac{r}{\sqrt{2} \sigma }\right)^{-b+k} \cdot C_{[ \ell ] \backslash \{b\}} \cdot \Gamma _{[ \ell ] \backslash \{b\} ,k}\\
=\begin{cases}
0, & \text{if} \ k< b\\
e^{-\frac{\ell -1}{8\sigma ^{2}} r^{2}}( -1)^{\ell ( \ell -1) /2+k}\frac{\sqrt{k!b!}}{b!( k-b) !}\left(\frac{r}{\sqrt{2} \sigma }\right)^{k-b} , & \text{if} \ k\geq b
\end{cases}
\end{gather*}
since $C_{[ \ell ] \backslash \{b\}} =\frac{F_{[ \ell ]}}{( \ell -1-b) !b!}$ and 
$\Gamma _{[ \ell ] \backslash \{b\} ,\ k} =\begin{cases}
0 & \text{if} \ k< b\\
\frac{( \ell -1-b) !}{( k-b) !} & \text{if} \ k\geq b
\end{cases}$. Also, the determinant of $( W_{r,\sigma })_{K,[ \ell ] \backslash \{b\}}$ is
\begin{equation*}
e^{-\frac{\ell -1}{8\sigma ^{2}} r^{2}}( -1)^{\Sigma _{K}}\sqrt{\frac{b!}{F_{[ \ell ]} F_{K}}}\left(\frac{r}{\sqrt{2} \sigma }\right)^{\Sigma _{K} -\ell ( \ell -1) /2+b} \cdot C_{K} \cdot \Gamma _{K,b} .
\end{equation*}
Hence, when \( b \leq k \), we have
\begin{align*}
& \left| \det (W_{r,\sigma})_{K, [\ell] \setminus \{ b \}}) \det ((W_{-r,\sigma})_{[\ell] \setminus \{ b \}, [\ell] \setminus \{ k \}}) \right| \\
& = e^{-\frac{\ell -1}{4\sigma^2} r^2} \frac{1}{(k - b)!} \sqrt{\frac{k!}{F_{[\ell]} F_K}} \left( \frac{r}{\sqrt{2}\sigma} \right)^{\Sigma_K - \ell (\ell -1)/2 + k}  C_K \cdot \Gamma_{K,b} \\
& \leq e^{-\frac{\ell -1}{4\sigma^2} r^2} \frac{1}{(k - b)! b!} \sqrt{\frac{k!}{F_{[\ell]} F_K}} \left( \frac{r}{\sqrt{2}\sigma} \right)^{\Sigma_K - \ell (\ell -1)/2 + k} C_K \prod_{c=0}^{\ell-2} a_c
\end{align*}
The last inequality is due to Lemma \ref{Lemma 38}. By plugging it into (\ref{det_U_K_minusk}), we have
\begin{align*}
\left| \det (U_K^{(-k)}) \right| & = \sum_{b=0}^{k} \left| \det ((W_{r,\sigma})_{K, [\ell] \setminus \{ b \}}) \det ((W_{-r,\sigma}^{\top})_{[\ell] \setminus \{ b \}, [\ell] \setminus \{ k \}}) \right| \\
& \leq \sum_{b=0}^{k} e^{-\frac{\ell -1}{4\sigma^2} r^2} \frac{1}{(k - b)! b!} \sqrt{\frac{k!}{F_{[\ell]} F_K} }\left( \frac{r}{\sqrt{2}\sigma} \right)^{\Sigma_K - \ell (\ell -1)/2 + k}  C_K \prod_{c=0}^{\ell-2} a_c.
\end{align*}
Since $\sum _{b=0}^{k}\frac{1}{( k-b) !b!} =\frac{2^{k}}{k!}$, we obtain
\begin{equation*}
\left| \det\left( U_{K}^{( -k)}\right)\right| \leq e^{-\frac{\ell -1}{4\sigma ^{2}} r^{2}} 2^{k}\sqrt{\frac{1}{k!F_{[ \ell ]} F_{K}}}\left(\frac{r}{\sqrt{2} \sigma }\right)^{\Sigma _{K} -\ell ( \ell -1) /2+k} C_{K}\prod _{c=0}^{\ell -2} a_{c}
\end{equation*}
\end{proof}

\begin{lemma}[Lemma 37, \cite{bryon_2023}]\label{Lemma 37}
Let \( a_0, \dots, a_{\ell-1} \) be \(\ell\) nonnegative integers such that \( 0 \leq a_0 < \dots < a_{\ell-1} \) and \( K \) be the ordered set \(\{a_0, \dots, a_{\ell-1}\}\). Then, for any integer \( b \in [\ell+1] \), we have
\[
\det((W_{r,\sigma})_{K,[\ell+1]\setminus\{b\}}) = e^{-\frac{\ell}{8\sigma^2} r^2} (-1)^{\Sigma_K} \sqrt{\frac{b!}{F_{[\ell+1]} F_K}} \left( \frac{r}{\sqrt{2}\sigma} \right)^{\Sigma_K - \ell(\ell+1)/2+b} C_K \cdot \Gamma_{K,b},
\]
and in particular,
\[
\det\left((W_{r,\sigma})_{K,[\ell]}\right) = e^{-\frac{\ell}{8\sigma^2}r^2}(-1)^{\Sigma_K}\sqrt{\frac{1}{F_{[\ell]}F_K}}\left(\frac{r}{\sqrt{2}\sigma}\right)^{\Sigma_K -\ell(\ell-1)/2} C_K \cdot \Gamma_{K,\ell}.
\]
\end{lemma}
\begin{proof}
By factoring the common terms in each row and column, we have
\begin{align*}
\det & \left((W_{r,\sigma})_{K,[\ell+1]\setminus\{b\}} \right) \\
 & = e^{-\frac{\ell}{8\sigma^2}r^2}(-1)^{\Sigma_K} \sqrt{\frac{b!}{\prod_{c=0}^{\ell} c! \prod_{c=0}^{\ell-1} a_c !}} \left(\frac{r}{\sqrt{2}\sigma}\right)^{\Sigma_K - \ell(\ell+1)/2 + b} \det \left( (W')_{K,[\ell+1]\setminus\{b\}} \right)  \\
& = e^{-\frac{\ell}{8\sigma^2}r^2}(-1)^{\Sigma_K} \sqrt{\frac{b!}{F_{[\ell+1]} F_K}} \left(\frac{r}{\sqrt{2}\sigma}\right)^{\Sigma_K - \ell(\ell+1)/2 + b} \det \left( (W')_{K,[\ell+1]\setminus\{b\}} \right)
\end{align*}
where \( W'_{K,[\ell+1]\setminus\{b\}} \) is a \(\ell\)-by-\(\ell\) matrix that
\[
W'_{K,[\ell+1]\setminus\{b\}} = \begin{bmatrix}
1 & \cdots & \prod_{d=0}^{b-2}(a_0 - d) & \prod_{d=0}^{b}(a_0 - d) & \cdots & \prod_{d=0}^{\ell-1}(a_0 - d) \\
\vdots & \ddots & \vdots & \vdots & \ddots & \vdots \\
1 & \cdots & \prod_{d=0}^{b-2}(a_{\ell-1} - d) & \prod_{d=0}^{b}(a_{\ell-1} - d) & \cdots & \prod_{d=0}^{\ell-1}(a_{\ell-1} - d)
\end{bmatrix}
\]

We will apply column operations on \(\det(W'_{K,[\ell+1]\setminus\{b\}})\). For the column indexed at \(c \neq 0, b+1\), subtract the column indexed at \(c-1\) multiplied by \((a_0 - c)\) to it and, for the column indexed at \(b+1\), subtract the column indexed at \(b-1\) multiplied by \((a_0 - b)(a_0 - (b-1))\) to it. We have the first row to be a zero row except the first entry is 1 and expand the determinant along the first row. Then, the row indexed at \(a_c\) has a factor \(a_c - a_0\) and in particular the entry indexed at \(b+1\) has an extra factor \(a_c + a_0 - (2b-1) = (a_c - (b-1)) + (a_0 - b)\). By factoring out \(a_c - a_0\), we have
\begin{align*}
\det & (W'_{K,[\ell+1]\setminus\{b\}}) \\
& = \left( \prod_{c=1}^{\ell}(a_c - a_0) \right) \left( \det((W'(r))_{K\setminus\{a_0\},[\ell]\setminus\{b-1\}}) + (a_0 - b) \det((W'(r))_{K\setminus\{a_0\},[\ell]\setminus\{b\}}) \right).
\end{align*}
By induction, we conclude
\[
\det(W'_{K,[\ell+1]\setminus\{b\}}) = C_K \cdot \Gamma_{K,b}
\]
Therefore,
\[
\det((W_{r,\sigma})_{K,[\ell+1]\setminus\{b\}}) = e^{-\frac{\ell}{8\sigma^2} r^2} (-1)^{\Sigma_K} \sqrt{\frac{b!}{F_{[\ell+1]} F_K}} \left( \frac{r}{\sqrt{2}\sigma} \right)^{\Sigma_K - \ell(\ell+1)/2+b} C_K \cdot \Gamma_{K,b},
\]
and in particular,
\[
\det\left((W_{r,\sigma})_{K,[\ell]}\right) = e^{-\frac{\ell}{8\sigma^2}r^2}(-1)^{\Sigma_K}\sqrt{\frac{1}{F[\ell]F_K}}\left(\frac{r}{\sqrt{2}\sigma}\right)^{\Sigma_K-\ell(\ell-1)/2}C_K \cdot \Gamma_{K,\ell}.
\]
\end{proof}

\begin{lemma}[Lemma 36, \cite{bryon_2023}]\label{Lemma 36}
Let $a_0, \cdots, a_{\ell-1}$ be $\ell$ nonnegative integers such that $0 \leq a_0 < \cdots < a_{\ell-1}$ and $a_c \neq j$ for some $j \in \mathbb{N}_0$. Then, we have
\[
\prod_{c=0}^{\ell-1} \left| \frac{a_c}{a_c - j} \right| \leq 2^j \cdot 4^\ell.
\]
\end{lemma}
\begin{proof}
Suppose $a_0, \cdots, a_{b-1}$ are the integers less than $j$ and the rest of them are larger than $j$.  
For $c \leq b - 1$, we have  
\[
\left| \frac{a_c}{a_c - j} \right| = \frac{a_c}{j - a_c} = \frac{j}{j - a_c} - 1 \leq \frac{j}{b - c} - 1 = \frac{j - (b - c)}{b - c}
\]
since $a_c \leq j - (b - c)$. For $c \geq b$, we have  
\[
\left| \frac{a_c}{a_c - j} \right| = \frac{a_c}{a_c - j} = 1 + \frac{j}{a_c - j} \leq 1 + \frac{j}{c - b + 1} = \frac{j + 1 + (c - b)}{c - b + 1}
\]
since $a_c \geq j + 1 + (c - b)$. Hence, we have  
\[
\prod_{c=0}^{\ell-1} \left| \frac{a_c}{a_c - j} \right| \leq \prod_{c=0}^{b-1} \frac{j - (b - c)}{b - c} \cdot \prod_{c=b}^{\ell-1} \frac{j + 1 + (c - b)}{c - b + 1}
\]
\[
= \frac{(j - 1)!}{(j - b - 1)!b!} \cdot \frac{(j + \ell - b)!}{(\ell - b)!j!}
\]
\[
\leq \frac{(j + \ell - b)!}{(\ell - b)!b!(j - b)!} = \frac{(j + \ell - b)!}{\ell!(j - b)!} \cdot \frac{\ell!}{(\ell - b)!b!}
\]
\[
\leq 2^{j + \ell - b} \cdot 2^\ell \leq 2^j \cdot 4^\ell.
\]
\end{proof}

\begin{lemma}[Lemma 38, \cite{bryon_2023}]\label{Lemma 38}
Let \( a_0, \dots, a_{\ell-1} \) be \(\ell\) nonnegative integers such that \( 0 \leq a_0 < \dots < a_{\ell-1} \) and \( K \) be the set \(\{a_0, \dots, a_{\ell-1}\}\). Then, for any \( b \in \mathbb{N}_0 \), we have  
\[
0 \leq \Gamma_{K,b} \leq \frac{1}{b!} \prod_{\alpha=0}^{\ell-1} a_\alpha,
\]
and 
\begin{equation*}
\frac{\partial }{\partial a_{i}} \Gamma _{K,b} \geq 0, \forall a_i\in K
\end{equation*}
which means that $\Gamma_{K,b}$ is an increasing function with regard to any $a_i\in K$.
\end{lemma}
\begin{proof}
We first prove \(\Gamma_{K,b} \geq 0\) and will use induction on \(\ell\) to prove the statement. Suppose \(\ell = 1\). When \(b = 0\), we have \(\Gamma_{K,0} = a_0 \geq 0\). When \(b = 1\), we have \(\Gamma_{K,1} = a_0 - (a_0 - 1) = 1 > 0\). When \(b \geq 2\), we have \(\Gamma_{K,b} = \frac{1}{b!} \sum_{d=0}^{b} (-1)^d \binom{b}{d} (a_0 - d) = 0\) since both \(\sum_{d=0}^{b} (-1)^d \binom{b}{d}\) and \(\sum_{d=0}^{b} (-1)^d \binom{b}{d} d = b \cdot \sum_{d=1}^{b} (-1)^d \binom{b-1}{d-1}\) are zero. Suppose \(\ell \geq 2\). We view \(\Gamma_{K,b}\) as a function of \(a_0, a_1, \dots, a_{\ell-1}\). The partial derivative of \(\Gamma_{K,b}\) is
\[
\frac{\partial}{\partial a_i} \Gamma_{K,b} = \frac{1}{b!} \sum_{d=0}^{b} (-1)^d \binom{b}{d} \prod_{c=0,c \neq i}^{\ell-1} (a_c - d) = \Gamma_{K \setminus \{a_i\}, b} \geq 0
\]
by the induction assumption. It means that \(\Gamma_{K,b}\) is an increasing function and it implies \(\Gamma_{K,b} \geq \Gamma_{ [\ell], b} = 0\) by direct calculation.

Now, we will prove \(\Gamma_{K,b} \leq \frac{1}{b!} \prod_{c=0}^{\ell-1} a_c\) and will use induction on \(b\) to prove the statement. Suppose \(b = 1\). We have \(\Gamma_{K,0} = \prod_{c=0}^{\ell-1} a_c\). Suppose \(b \geq 2\). Note that \(\binom{b}{d} = \binom{b-1}{d} + \binom{b-1}{d-1}\). We have
\begin{align*}
\Gamma_{K,b} & = \frac{1}{b!} \left( \sum_{d=0}^{b-1} (-1)^d \binom{b-1}{d} \prod_{c=0}^{\ell-1} (a_c - d) - \sum_{d=1}^b (-1)^{d-1} \binom{b-1}{d-1} \prod_{c=0}^{\ell-1} (a_c - d) \right) \\
& \leq \frac{1}{b!} \sum_{d=0}^{b-1} (-1)^d \binom{b-1}{d} \prod_{c=0}^{\ell-1} (a_c - d) \leq \frac{1}{b} \frac{1}{(b-1)!} \prod_{c=0}^{\ell-1} a_c = \frac{1}{b!} \prod_{c=0}^{\ell-1} a_c.
\end{align*}
The inequalities are due to the induction assumption.
\end{proof}

\section{\texorpdfstring{Proofs for Section \ref{subsection: component_density}: Concentration inequality for $\lVert \widehat{f}_{i}-f_i\rVert_1$}{Proofs for Section: Concentration inequality for fi error}}
\label{section:appendix_B}

Lemma \ref{lemma: bound_l1_fi_hat_fi} gives an upper bound of $\| \widehat{f_{i}} -f_{i} \| _{1}$ for any density estimator $\widehat{f}$. In this section, let us consider a specific form of the estimator $\widehat{f}$, namely the standard kernel density estimator. Then we 
can further derive the tail behavior of $\widehat{f_{i}}$, as given in the following lemma:

\begin{lemma}\label{lemma: tail_bound_fi}
Let $f\in\mathcal{F}$ be as defined in \eqref{C3}. We can verify that all such densities $f$ satisfy the regularity conditions in Lemma \ref{lemma_KDE}: there exist $C_1>0$, $C_2>0$, $C_3>0$, $C_4>0$, such that $\forall f\in\mathcal{F}$, $\| f\| _{2}^{2} < C_{1} ,\ |\int f'( x) dx|< C_{2} ,\ \| f''\| _{2}^{2} < C_{3} ,\ |\int f''( x) dx|< C_{4}$. Let kernel $K$ be a standard Gaussian density. Suppose $X_1,\cdots,X_n$ are $n$ i.i.d. samples from $P_f$, and let $\widehat{f}(x;h)$ be the kernel density estimator defined as
\begin{equation*}
\widehat{f}( x;h) =n^{-1}\sum _{i=1}^{n} K_{h}( x-X_{i}),
\end{equation*}
where $h>0$ is the bandwidth, and $K_h(x)\coloneqq h^{-1}K(x/h)$. Choose $h=c_Kn^{-1/5}$, where $c_K$ is a positive constant depending only on $K$. Then by taking
\begin{equation*}
\begin{cases}
\alpha =\frac{2}{5}\log\frac{r}{8\max( r_{1} ,r_{2})} ,\ \beta =\frac{1}{2}\log\frac{r}{8\max( r_{1} ,r_{2})} , & \text{if} \ 1< \frac{r}{8\max( r_{1} ,r_{2})} \leq 2,\\
\alpha =\frac{3}{5}\log\frac{r}{16\max( r_{1} ,r_{2})} ,\ \beta =\log 2, & \text{if} \ \frac{r}{8\max( r_{1} ,r_{2})}  >2,
\end{cases}
\end{equation*}
and letting $\epsilon_\alpha$ be a constant such that $0<\epsilon_\alpha<e^{-1}$, and $\forall 0<\epsilon<\epsilon_\alpha$, $\log\left(\frac{1}{\epsilon }\right) < \left(\frac{1}{\epsilon }\right)^{\alpha }$, we have that for $i=1,2$, $\forall\ 0<\eta<\overline{c}_{1i}\epsilon_\alpha^\beta$, and $n>(2\varphi_K)^{5/2}(\frac{\overline{c}_{1i}}{\eta})^{5/(2\beta)}$, take $\ell=\max\left(\left\lfloor \beta ^{-1}\log(\overline{c}_{1i} /\eta )\right\rfloor ,\ \left\lfloor 2er_{i}^{2} /\sigma ^{2}\right\rfloor \right) +2$, then $\widehat{f}_i$ defined in \eqref{fi_hat} satisfies
\begin{equation*}
P_{f}^n\left( \| \widehat{f}_{i} - f_{i} \| _{1} \geq \eta \right)  \leq \exp\left( -\tilde{c}_K n^{4/5}\left(\frac{\eta }{\overline{c}_{1i}}\right)^{\frac{2}{\beta }\left(b_i\log\log\left(\frac{\overline{c}_{1i}}{\eta }\right) -b_i\log \beta +c_{2i}\right)}\right),
\end{equation*}
where $\varphi_K$ and $\tilde{c}_K$ are positive constants depending only on $K$, and $\overline{c}_{1i}=\overline{c}_{1i}( r,\sigma ,r_{1} ,r_{2} ,\underline{w})$ is a positive constant depending on $i,r,\sigma, r_1,r_2,\underline{w}$, $c_{2i}=c_{2i}(r,r_1,r_2,\sigma)$ is a positive constant depending on $i, r,r_1,r_2,\sigma$, and $b_i=b(r_i,\sigma)\geq 1$ is a constant depending on $r_i$ and $\sigma$.
\end{lemma}

This section of the Appendix is devoted to the proof of Lemma \ref{lemma: tail_bound_fi}. We will need the following standard tool.
\begin{lemma}[Bounded difference inequality]\label{bdd_diff_inq}
 Let $\chi$ be some set, and let $f:\chi^n\rightarrow R$ be a function such that there exist $c_1,\cdots,c_n>0$, $\forall k=1,\cdots,n$, $\forall x_1,\cdots,x_n\in\chi$, 
\begin{equation*}
\sup _{x_{k} '\in \chi } |f( x_{1} ,\cdots ,x_{k} ,\cdots x_{n}) -f( x_{1} ,\cdots ,x_{k} ',\cdots ,x_{n}) |\leq c_{k}.
\end{equation*}
Let $X_1,\cdots,X_n$ be independent random variables on $\chi$, then $\forall t>0$,
\begin{equation*}
\text{Pr}( f( X_{1} ,\cdots ,X_{n}) -Ef( X_{1} ,\cdots ,X_{n})  >t) \leq e^{-\frac{2t^{2}}{\sum _{i=i}^{n} c_{i}^{2}}}.
\end{equation*}

\end{lemma}

\begin{lemma}\label{lemma_KDE}
Let the kernel $K$ be a probability density on $\mathbb{R}$ that is symmetric around the origin, bounded and has a finite fourth moment. Let $\mathcal{F}$ be a density class on $\mathbb{R}$ such that there exist $C_1>0$, $C_2>0$, $C_3>0$, $C_4>0$, $\forall f\in\mathcal{F}$, $\| f\| _{2}^{2} < C_{1} ,\ |\int f'( x) dx|< C_{2} ,\ \| f''\| _{2}^{2} < C_{3} ,\ |\int f''( x) dx|< C_{4}$. \comment{$\int f^{2}( x) dx < C_{1}$, $|\int f'( x) dx| < C_{2}$, and $\int f''( x)^{2} dx< C_{3}$.}Suppose $X_1,\cdots,X_n$ are $n$ i.i.d samples from some density $f\in \mathcal{F}$, and let $\widehat{f}(x;h)$ be the kernel density estimator defined as
\begin{equation*}
\widehat{f}( x;h) =n^{-1}\sum _{i=1}^{n} K_{h}( x-X_{i}),
\end{equation*}
where $h>0$ is the bandwidth, and $K_h(x)\coloneqq h^{-1}K(x/h)$. Then, by taking $h=c_Kn^{-1/5}$, where $c_K$ is a positive constant depending on the kernel $K$, we have that $\forall f\in\mathcal{F}$,
\begin{equation*}
E\left[ \| \widehat{f} -f\| _{2}^{2}\right] \leq \psi _{K} n^{-4/5},
\end{equation*}
where $\psi_K$ is a positive constant that only depends on the kernel $K$
.

\end{lemma}

\begin{proof}
\begin{equation*}
E\left[ \| \widehat{f} -f\| _{2}^{2}\right] =E\int (\widehat{f}( x;h) -f( x))^{2} dx=\int E(\widehat{f}( x;h) -f( x))^{2} dx.
\end{equation*}
$E(\widehat{f}( x;h) -f( x))^{2}$ on the RHS is the mean square error (MSE) of $\widehat{f}$ at $x$, and the MSE can be decomposed into variance and squared bias
\begin{equation*}
E(\widehat{f}( x;h) -f( x))^{2} =\text{Var}(\widehat{f}( x;h)) +( E\widehat{f}( x;h) -f( x))^{2}.
\end{equation*}
Furthermore, we have (see e.g. Chapter 2 of \cite{wand1994kernel}),
\begin{equation*}
E\widehat{f}( x;h) =\int K( z) f( x-hz) dz,
\end{equation*}
and 
\begin{equation*}
\text{Var}\{\widehat{f}( x;h)\} =( nh)^{-1}\int K^{2}( z) f( x-hz) dz-n^{-1}( E\widehat{f}( x;h))^{2}.
\end{equation*}
By Taylor expansion of $f(x-hz)$, we obtain
\begin{align*}
E\widehat{f}( x;h) & =\int K( z) f( x-hz) dz\\
 & =\int K( z)\left[ f( x) +f'( x)( -hz) +h^{2} z^{2}\int _{0}^{1}( 1-t) f''( x-thz) dt\right] dz.
\end{align*}
Since $\int K( z) dz=1$, and $\int zK( z) dz=0$, 
\begin{equation}\label{eq:Ef_hat}
E\widehat{f}( x;h) =f( x) +h^{2}\int _{0}^{1}( 1-t)\left(\int f''( x-thz) z^{2} K( z) dz\right) dt.
\end{equation}
By Cauchy-Schwarz's inequality,
\begin{equation*}
\int f''( x-thz) z^{2} K( z) dz\leq \sqrt{\left(\int f''( x-thz)^{2} dz\right) \cdot \left(\int z^{4} K^{2}( z) dz\right)} < \infty, 
\end{equation*}
because $K$ is bounded and has finite fourth moment, and $\int f''( x)^{2} dx<C_3$. Thus, the second term on the RHS of \eqref{eq:Ef_hat} is finite. Let $R_{1}( f,K; h) \coloneqq \int _{0}^{1}( 1-t)\left(\int f''( x-thz) z^{2} K( z) dz\right) dt$, then 
$$E\widehat{f}( x;h) =f( x) + h^2R_{1}( f,K; h).$$ 
Similarly, 
\begin{align*}
\text{Var}(\hat{f}( x; & h))\\
= &  \begin{array}{l}
( nh)^{-1} \int K^{2}( z)\left[ f( x) +f'( x)( -hz) +h^{2} z^{2}\int _{0}^{1}( 1-t) f''( x-thz) dt\right] dz\\
\ -n^{-1}\left( f( x) +h^{2} R_{1}( f,K;h)\right)^{2}
\end{array}\\
= & ( nh)^{-1} \| K\| _{2}^{2} \cdot f( x) -n^{-1}\left( f'( x) R_{2}( K) +f^{2}( x)\right) +\frac{h}{n} R_{3}( f,K;h)\\
 & \ -\frac{2h^{2}}{n} f( x) R_{1}( f,K;h) -\frac{h^{4}}{n} R_{1}( f,K;h)^{2}
\end{align*}
where $R_{2}( K) \coloneq \int zK( z)^{2} dz,\ R_{3}( f,K;h) \coloneq \int _{0}^{1}( 1-t)\left(\int f''( x-thz) z^{2} K^{2}( z) dz\right) dt$. By the regularity conditions of $K$ and $f''$, $R_2(K)$ and $R_{3}( f,K;h)$ are both finite. Thus, for each $x$,
\begin{align*}
E(\widehat{f}( x;h) & -f( x))^{2}\\
= & \text{Var}(\hat{f}( x; h)) + (E \widehat{f}( x;h) - f(x))^2 \\
= & ( nh)^{-1} \| K\| _{2}^{2} \cdot f( x) -n^{-1}\left( f'( x) R_{2}( K) +f^{2}( x)\right) +\frac{h}{n} R_{3}( f,K;h)\\
 & -\frac{2h^{2}}{n} f( x) R_{1}( f,K;h) -\frac{h^{4}}{n} R_{1}( f,K;h)^{2} + h^{4} R_{1}( f,K;h)^{2},
\end{align*}
so, by Fubini's theorem,
\begin{align}
E\left[ \| \widehat{f} -f\| _{2}^{2}\right] & =\int E(\widehat{f}( x;h) -f( x))^{2} dx\notag\\
 & =h^{4}\int R_{1}( f,K;h)^{2} dx+( nh)^{-1} \| K\| _{2}^{2} -n^{-1}\left( R_{2}( K)\int f'( x) dx+\int f^{2}( x) dx\right)\notag\\
 & \ \ \ \ +\frac{h}{n}\int R_{3}( f,K;h) dx-\frac{2h^{2}}{n}\int f( x) R_{1}( f,K;h) dx-\frac{h^{4}}{n}\int R_{1}( f,K;h)^{2} dx.\label{E_l2}
\end{align}
Define $A_{t}( x) \coloneq \int f''( x-thz) z^{2} K( z) dz$. By Cauchy-Schwarz inequality, for any fixed $x$, 
\begin{equation*}
R_{1}( f,K;h)^{2} =\left| \int _{0}^{1}( 1-t) A_{t}( x) dt\right| ^{2} \leq \left(\int _{0}^{1}( 1-t)^{2} dt\right) \cdot \left(\int _{0}^{1} |A_{t}( x) |^{2} dt\right).
\end{equation*}
Integrate over $x$ and use Fubini's theorem,
\begin{equation}\label{ineq: R1_1}
\int R_{1}( f,K;h)^{2} dx\leq \left(\int _{0}^{1}( 1-t)^{2} dt\right) \cdot \int _{0}^{1} \| A_{t} \| _{2}^{2} dt=\frac{1}{3}\int _{0}^{1} \| A_{t} \| _{2}^{2} dt.
\end{equation}
Now we rewrite $A_t$ as a convolution. Let $k_{t,h}( u) \coloneq \frac{1}{th}\left(\frac{u}{th}\right)^{2} K\left(\frac{u}{th}\right)
$, then with the change of variable $u=thz$,
\begin{equation*}
A_{t}( x) =\int f''( x-u) k_{t,h}( u) du=( f''*k_{t,h})( x).
\end{equation*}
By Young's inequality ($L^2*L^1\rightarrow L^2$), 
\begin{align*}
\| A_{t} \| _{2} \leq \| f''\| _{2} \cdot \| k_{t,h} \| _{1} & =\| f''\| _{2} \cdot \int \frac{1}{th}\left(\frac{u}{th}\right)^{2} K\left(\frac{u}{th}\right) du\\
 & =\| f''\| _{2} \cdot \int z^{2} K( z) dz\\
 & \leq \sqrt{C_{3}}\int z^{2} K( z) dz.
\end{align*}
Plug this back to \eqref{ineq: R1_1},
\begin{equation*}
\int R_{1}( f,K;h)^{2} dx\leq \frac{1}{3} C_{3}\left(\int z^{2} K( z) dz\right)^{2} \leq \frac{1}{3} C_{3}\int z^{4} K( z) dz\eqcolon \widetilde{R}_{1}( K),
\end{equation*}
where the last inequality is by Jensen's inequality. Furthermore,
\begin{align*}
\int R_{3}( f,K;h) dx & =\int \left(\int _{0}^{1}( 1-t)\left(\int f''( x-thz) z^{2} K^{2}( z) dz\right) dt\right) dx\\
 & =\int _{0}^{1}( 1-t)\left(\int z^{2} K^{2}( z)\left(\int f''( x-thz) dx\right) dz\right) dt\\
 & \leq \frac{1}{2}C_4\int z^{2} K^{2}( z) dz\eqcolon \widetilde{R}_{3}( K),
\end{align*}
and
\begin{align*}
\int f( x) R_{1}( f,K;h) dx & =\int f( x)\left(\int _{0}^{1}( 1-t)\left(\int f''( x-thz) z^{2} K( z) dz\right) dt\right) dx\\
 & =\int _{0}^{1}( 1-t)\left(\int \left(\int f( x) f''( x-thz) dx\right) z^{2} K( z) dz\right) dt\\
 & \leq \frac{1}{2}\sqrt{C_{1} C_{3}}\int z^{2} K( z) dz\eqcolon R_{4}( K)
\end{align*}
Because $K$ is bounded and has finite fourth moment, $\widetilde{R}_{1}( K)$ and $\widetilde{R}_{3}( K)$ are both finite. Plug these two terms back into (\ref{E_l2}), we obtain that $\forall f\in \mathcal{F}$,
\begin{align*}
E[ \| \widehat{f} & -f\| _{2}^{2}]\\
\leq  & h^{4}\widetilde{R}_{1}( K) +( nh)^{-1} \| K\| _{2}^{2} +n^{-1}( R_{2}( K) C_{2} +C_{1}) +\frac{h}{n}\widetilde{R}_{3}( K) -\frac{2h^{2}}{n} R_{4}( K) -\frac{h^{4}}{n}\widetilde{R}_{1}( K).
\end{align*}
Because $E\left[ \| \widehat{f} -f\| _{2}^{2}\right] =h^{4}\widetilde{R}_{1}( K) +( nh)^{-1} \| K\| _{2}^{2} +o\left( h^{4} +( nh)^{-1}\right)$, we can find $h_{opt}$ such that the dominant order $O(h^{4} +( nh)^{-1})$ is minimized:
\begin{equation*}
h_{opt} =\left(\frac{\| K\| _{2}^{2}}{\widetilde{R}_{1}( K) n}\right)^{1/5},
\end{equation*}
and
\begin{align*}
E[ \| \hat{f} & -f\| _{2}^{2}]\\
\leq  & 2\| K\| _{2}^{8/5}\widetilde{R}_{1}( K)^{1/5} \cdot n^{-4/5} +( R_{2}( K) C_{2} +C_{1}) n^{-1} +\| K\| _{2}^{2/5}\widetilde{R}_{3}( K) \cdot \widetilde{R}_{1}( K)^{-1/5} \cdot n^{-6/5}\\
 & -2R_{4}( K) \| K\| _{2}^{4/5}\widetilde{R}_{1}( K)^{-2/5} \cdot n^{-7/5} -\widetilde{R}_{1}( K)^{1/5} \| K\| _{2}^{8/5} \cdot n^{-9/5}.
\end{align*}
Note that the coefficients on the RHS only depend on $K$, so it is a uniform upper bound $\forall f\in \mathcal{F}$. There must exist $\psi_K>0$, s.t. $E\left[ \| \widehat{f} -f\| _{2}^{2}\right] \leq \psi _{K} n^{-4/5}$,$\forall n\in \mathbb{N}$, $\forall f\in \mathcal{F}$.
\end{proof}

Now we are ready to prove Lemma \ref{lemma: tail_bound_fi}.

\begin{proof}[Proof of Lemma \ref{lemma: tail_bound_fi}]
For any $\epsilon$ such that $0<\epsilon<e^{-1}$, take $\ell = \max(\lfloor \log( 1/\epsilon )\rfloor, \lfloor 2er_i^2/\sigma^2 \rfloor )+2$, then there exists a constant $b(r_i,\sigma)\geq1$, such that $\log(1/\epsilon)<\ell<b(r_i,\sigma)\log(1/\epsilon)$. By Lemma \ref{lemma: bound_l1_fi_hat_fi}, and noting by assumption that $8\max(r_1,r_2)/r<1$,
\begin{align*}
\| \widehat{f_{i}} -f_{i} \| _{1} & \\
\leq \frac{1}{w_{i}} c_{1}( & r,\sigma ,r_{1} ,r_{2}) \cdot \left(\frac{8\max( r_{1} ,r_{2})}{r}\right)^{\log\left(\frac{1}{\epsilon }\right)} \cdot \left( b( r_{i} ,\sigma )\log\left(\frac{1}{\epsilon }\right)\right)^{5/4} +c_{2}( r_{i} ,\sigma )\left(\frac{1}{2}\right)^{\log\left(\frac{1}{\epsilon }\right)}\\
+\ \frac{1}{w_{i}} c_{3} & ( r,\sigma ) \cdot \| \Delta f\| _{2} \cdot \exp \left[ b( r_{i} ,\sigma )\log\left(\frac{1}{\epsilon }\right)\log\left( b( r_{i} ,\sigma )\log\left(\frac{1}{\epsilon }\right)\right) \right.\\
 & \ \ \ \ \ \ \ \ \ \ \ \ \ \ \ \ \ \ \ \ \ \ \ \ \ \ \ \ \ \ \ \ \ \ \ +\ \left. c_{4}( r,\sigma ) \cdot b( r_{i} ,\sigma )\log\left(\frac{1}{\epsilon }\right) +c_{5}\log\left( b( r_{i} ,\sigma )\log\left(\frac{1}{\epsilon }\right)\right)\right].
\end{align*}

By altering the bases and exponents in the right hand side, we obtain
\begin{align}
\| \widehat{f_{i}} -f_{i} \| _{1} & \leq \frac{1}{w_i}c_{1}(r,\sigma ,r_{1} ,r_{2}) \cdot \epsilon ^{\log\frac{r}{8\max( r_{1} ,r_{2})}}\left(\log\left(\frac{1}{\epsilon }\right)\right)^{5/4} +c_{2}( r_{i} ,\sigma ) \epsilon ^{\log 2}\notag\\
 & \ \ +\frac{1}{w_{i}}c_{3}(r,\sigma ,r_{i}) \cdot \| \Delta f\| _{2} \cdot \left(\frac{1}{\epsilon }\right)^{b( r_{i} ,\sigma ) \cdot \log\log\left(\frac{1}{\epsilon }\right)} \cdot \left(\frac{1}{\epsilon }\right)^{c_{4}( r,r_{i} ,\sigma )} \cdot \left(\log\left(\frac{1}{\epsilon }\right)\right)^{c_{5}}.\label{fhat_f_diff_epsilon}
\end{align}
Here, some terms involving $b(r_i,\sigma)$ are absorbed into the constants $c_1, c_3$ and $c_4$. Since $\forall\alpha>0$, there exists a small enough $0<\epsilon_\alpha<e^{-1}$, s.t. $\forall 0<\epsilon<\epsilon_\alpha$, $\log\left(\frac{1}{\epsilon }\right) < \left(\frac{1}{\epsilon }\right)^{\alpha }$, for $0<\epsilon<\epsilon_\alpha$, we can replace $\log(1/\epsilon)$ in (\ref{fhat_f_diff_epsilon}) with $(1/\epsilon)^\alpha$ for some $\alpha>0$ to have
\begin{align*}
\| \widehat{f_{i}} -f_{i} \| _{1} & \leq \frac{1}{w_{i}}c_{1}(r,\sigma ,r_{1} ,r_{2}) \cdot \epsilon ^{\log\frac{r}{8\max( r_{1} ,r_{2})} -\frac{5\alpha }{4}} +c_{2}( r_{i} ,\sigma ) \epsilon ^{\log 2}\\
 & \ \ +\frac{1}{w_{i}}c_{3}(r,\sigma ,r_{i}) \cdot \| \Delta f\| _{2} \cdot \left(\frac{1}{\epsilon }\right)^{b( r_{i} ,\sigma ) \cdot \log\log\left(\frac{1}{\epsilon }\right) +c_{4}( r,r_{i} ,\sigma ) +c_{5} \alpha }.
\end{align*}
Let $\beta:=\min\left(\log\frac{r}{8\max( r_{1} ,\ r_{2})} -\frac{5\alpha }{4} ,\ \log 2\right)$. Then, if $\| \Delta f\| _{2} \leq \epsilon ^{b( r_{i} ,\sigma ) \cdot \log\log\left(\frac{1}{\epsilon }\right) +c_{4}( r,r_{i} ,\sigma ) +c_{5} \alpha +\beta }$, we have
\begin{equation*}
\| \widehat{f_{i}} -f_{i} \| _{1} \leq \left(\frac{1}{w_{i}} c_{1}( r,\sigma ,r_{1} ,r_{2}) +c_{2}( r_{i} ,\sigma ) +\frac{1}{w_{i}} c_{3}( r,\sigma ,r_{i})\right) \epsilon ^{\beta }.
\end{equation*}
Note that by assumption, $\forall f\in \mathcal{F}$, $w_i\in [\underline{w}, 1-\underline{w}]$, for $i=1,2$, so if we denote $\overline{c}_{1i}( r,\sigma ,r_{1} ,r_{2} ,\underline{w}) =\frac{1}{\underline{w}} c_{1}( r,\sigma ,r_{1} ,r_{2}) +c_{2}( r_{i} ,\sigma ) +\frac{1}{\underline{w}} c_{3}( r,\sigma ,r_{i})$, we have $\| \widehat{f_{i}} -f_{i} \| _{1} \leq \overline{c}_{1i}( r,\sigma ,r_{1} ,r_{2} ,\underline{w}) \epsilon ^{\beta }$. 

We will select $\alpha$ in the following way so $\beta$ can be in an asymptotic sense as small as possible.
\begin{itemize}
    \item If $\frac{r}{8\max( r_{1} ,\ r_{2})} \leq 2$, we can take $\alpha$ as small as possible such that $0< \alpha < \frac{4}{5}\log\frac{r}{8\max( r_{1} ,r_{2})}$. However, the smaller $\alpha$ is, the smaller $\epsilon_\alpha$ becomes. It suffices to set $\alpha = \frac{2}{5}\log\frac{r}{8\max( r_{1} ,r_{2})}$, then $\beta=\frac{1}{2}\log\frac{r}{8\max( r_{1} ,r_{2})}$.
    \item If $\frac{r}{8\max( r_{1} ,\ r_{2})} > 2$, we take $0< \alpha < \frac{4}{5}\left(\log\frac{r}{8\max( r_{1} ,r_{2})} -\log 2\right) =\frac{4}{5}\log\frac{r}{16\max( r_{1} ,r_{2})}$, and $\beta = \log 2$.
\end{itemize}

Therefore, when $0<\epsilon<\epsilon_\alpha$, we  have that for $i=1,2$,
\begin{equation*}
P_{f}^n\left( \| \Delta f\| _{2} < \epsilon ^{b( r_{i} ,\sigma ) \cdot \log\log\left(\frac{1}{\epsilon }\right) +c_{2i}( r,r_{1} ,r_{2} ,\sigma )}\right) \leq P_{f}^n\left( \| \widehat{f_{i}} -f_{i} \| _{1} < \overline{c}_{1i}( r,\sigma ,r_{1} ,r_{2} ,\underline{w})  \epsilon ^{\beta }\right),
\end{equation*}
which is equivalent to,
\begin{equation}\label{tail_bound_interm}
P_{f}^n\left( \| \widehat{f_{i}} -f_{i} \| _{1} \geq \overline{c}_{1i}( r,\sigma ,r_{1} ,r_{2} ,\underline{w}) \epsilon ^{\beta }\right) \leq P_{f}^n\left( \| \Delta f\| _{2} \geq \epsilon ^{b( r_{i} ,\sigma ) \cdot \log\log\left(\frac{1}{\epsilon }\right) +c_{2i}( r,r_{1} ,r_{2} ,\sigma )}\right).
\end{equation}
Note that $\Delta f=\widehat{f} -f$, and now we account for the fact that $\widehat{f}$ is chosen to be a kernel density estimator:
\begin{equation*}
\widehat{f}_{X_{1:n} ,h }( x) =\frac{1}{nh}\sum _{i=1}^{n} K \left(\frac{x-X_{i}}{h }\right) ,\ \forall x\in R,
\end{equation*}
where $K(x)$ is the standard Gaussian kernel. Then we define $q_{n}( X_{1} ,\cdots ,X_{n}) =\| \widehat{f}_{X_{1:n} ,h } -f\| _{2}$, and we have
\begin{align*}
\Bigl| q_{n}( X_{1} ,\cdots ,X_{k} ,\cdots ,X_{n}) & -q_{n}( X_{1} ,\cdots ,X'_{k} ,\cdots ,X_{n})\Bigl|\\
=  & \Bigl| \| \widehat{f}_{X_{1:n} ,h } -f\| _{2} -\| \widehat{f}_{X_{\{1:n\} \backslash \{i\}} \cup X'_{k} ,h } -f\| _{2}\Bigl|\\
\leq  & \| \widehat{f}_{X_{1:n} ,h } -\widehat{f}_{X_{\{ 1:n\} \backslash \{i\}} \cup X'_{k} ,h } \| _{2}\\
= & \frac{1}{nh } \| K \left(\frac{x-X_{i}}{h }\right) -K \left(\frac{x-X'_{i}}{h }\right) \| _{2}\\
= & \frac{1}{nh }\left(\int \Bigl| K \left(\frac{x-X_{i}}{h }\right) -K \left(\frac{x-X'_{i}}{h }\right)\Bigl|^{2} dx\right)^{1/2}\\
\leq  & \frac{1}{nh }\left( 2\int \Bigl| K \left(\frac{x-X_{i}}{h }\right)\Bigl|^{2} +\Bigl| K \left(\frac{x-X'_{i}}{h }\right)\Bigl|^{2} dx\right)^{1/2} \ \text{(Cauchy-Schwarz)}\\
= & \frac{2}{n\left( 2\sqrt{\pi } h \right)^{1/2}}.
\end{align*}

Therefore by the bounded difference inequality (Lemma \ref{bdd_diff_inq}) we get:
\begin{equation*}
P_{f}^n( \| \Delta f\| _{2} -E[ \| \Delta f\| _{2}] \geq \epsilon ) =P_{f}^n( q_{n} -E[ q_{n}] \geq \epsilon ) \leq \exp\left( -\sqrt{\pi } h n\epsilon ^{2}\right).
\end{equation*}
By Lemma \ref{lemma_KDE}, if the density class $\mathcal{F}$ and the choice of kernel $K$ satisfy the regularity conditions, then choosing bandwidth $h=c_Kn^{-1/5}$, where $c_K$ is a constant depending only on $K$, yields $\forall f\in\mathcal{F}$, $\forall n$, $E\left[ \| \Delta f\| _{2}^{2}\right] \leq \psi _{K} n^{-4/5}$, for some constant $\psi_K$ that only depends on the kernel $K$. Because $E[\Vert \Delta f\Vert _{2}]^{2} \leq E\left[\Vert \Delta f\Vert _{2}^{2}\right]$, we have $E[\Vert \Delta f\Vert _{2}] \leq \varphi_{K} n^{-2/5}$ for some constant $\varphi_K>0$. When $n >\left(\frac{2\varphi_K}{\epsilon }\right)^{5/2}$, we have $E[\Vert \Delta f\Vert _{2}] < \frac{\epsilon }{2}$, $\forall f\in \mathcal{F}$, then
\begin{equation}\label{tail_bound_deltaf}
P_{f}^n( \| \Delta f\| _{2} \geq \epsilon ) \leq P_{f}^n\left( \| \Delta f\| _{2} -E[ \| \Delta f\| _{2}] \geq \frac{\epsilon }{2}\right) \leq \exp\left( -\tilde{c}_{K} n^{4/5} \epsilon ^{2}\right),
\end{equation}
for some positive constant $\tilde{c}_K$ depending only on $K$. 

Now we combine (\ref{tail_bound_interm}) and (\ref{tail_bound_deltaf}). We can take $\eta=\overline{c}_{1i}\epsilon^\beta$, and plug $\epsilon=(\frac{\eta}{\overline{c}_{1i}})^{1/\beta}$ in the right hand side of (\ref{tail_bound_interm}). For $\epsilon$ small enough such that $0<\epsilon<\epsilon_\alpha$, and $n$ large enough such that $n>(\frac{2\varphi_K}{\epsilon})^{5/2}$, or equivalently, for $\eta$ small enough such that $\eta<\overline{c}_{1i}\epsilon_\alpha^\beta$, and $n>(2\varphi_K)^{5/2}(\frac{\overline{c}_{1i}}{\eta})^{5/(2\beta)}$, take $\ell=\max\left(\left\lfloor \beta ^{-1}\log(\overline{c}_{1i} /\eta )\right\rfloor ,\ \left\lfloor 2er_{i}^{2} /\sigma ^{2}\right\rfloor \right) +2$, then we have
\begin{align*}
P_{f}^n\left( \| \widehat{f}_{i} - f_{i} \| _{1} \geq \eta \right) & \leq P_{f}^n\left( \| \Delta f\| _{2} \geq \left(\frac{\eta }{\overline{c}_{1i}}\right)^{\frac{1}{\beta }\left(b_i\log\log\left(\frac{\overline{c}_{1i}}{\eta }\right)^{1/\beta } +c_{2i}\right)}\right)\\
 & \leq P_{f}^n\left( \| \Delta f\| _{2} \geq \left(\frac{\eta }{\overline{c}_{1i}}\right)^{\frac{1}{\beta }\left(b_i\log\log\left(\frac{\overline{c}_{1i}}{\eta }\right) -b_i\log \beta +c_{2i}\right)}\right)\\
 & \leq \exp\left( -\tilde{c}_K n^{4/5}\left(\frac{\eta }{\overline{c}_{1i}}\right)^{\frac{2}{\beta }\left(b_i\log\log\left(\frac{\overline{c}_{1i}}{\eta }\right) -b_i\log \beta +c_{2i}\right)}\right),
\end{align*}
where $\tilde{c}_K$ is a positive constant depending only on $K$, and $\overline{c}_{1i}=\overline{c}_{1i}( r,\sigma ,r_{1} ,r_{2} ,\underline{w})$ is a positive constant depending on $i,r,\sigma, r_1,r_2,\underline{w}$, $c_{2i}=c_{2i}(r,r_1,r_2,\sigma)$ is a positive constant depending on $i, r,r_1,r_2,\sigma$, and $b_i=b(r_i,\sigma)\geq 1$ is a constant depending on $r_i$ and $\sigma$.
\end{proof}

\section{Proofs for Section \ref{subsection: component_density}:
Lemma \ref{lemma: E_phi_maintext} and Theorem \ref{thm:fi_post_rate}}
\label{section:appendix_C}

The last piece of the argument is to establish the existence of a test function with desired controls on type I and type II errors. Suppose that $f_0$ satisfies condition \eqref{C4}. Define the test function $\phi _{n} :\mathbb{R}^{n}\rightarrow \{ 0,1\}$ by
\begin{equation}\label{phi_n}
\mathbf{\phi _{n} =1}\{\Vert \widehat{f}_{i} -f_{i0}\Vert _{1}  >\epsilon _{n}\},
\end{equation}
where $\widehat{f}_i$ is as defined in \eqref{fi_hat}, and we take $\ell=\max\left(\left\lfloor \beta ^{-1}\log(\overline{c}_{1i} /\epsilon_n )\right\rfloor ,\ \left\lfloor 2er_{i}^{2} /\sigma ^{2}\right\rfloor \right) +2$ as specified in Lemma \ref{lemma: tail_bound_fi}. Although $\phi_n$ depends on $i$. We suppress this dependence in the notation for simplicity. The following lemma establish upper bounds of type I error, $E_{f_{0}}[ \phi _{n}]$, and type II error, $ E_{f}[ 1-\phi _{n}] $ when $n$ is large enough and $\epsilon_n$ is small enough. Note that $f_{i0}$ is unknown, thus this test function cannot be used in practice, but the existence of such a test function is sufficient for the proof of the main theorem (Theorem \ref{thm:fi_post_rate}) on the posterior contraction of the component density $f_i$.

\begin{lemma}[Restatement of Lemma \ref{lemma: E_phi_maintext}] \label{lemma: E_phi}
   Let $f\in\mathcal{F}$ and $f_{0}$ be as defined in \eqref{C3} and \eqref{C4}. Fix a constant $M>2$, and let $\{\epsilon_n\}$ be a positive constant sequence such that $\lim _{n\rightarrow \infty } \epsilon _{n} =0$. Let $\widehat{f}_i$ be the component density estimator defined in \eqref{fi_hat}. Let $X_1,\cdots,X_n$ be $n$ i.i.d. samples from $P_f$. For a component $i=1$ (or $2$), consider the test function $\phi _{n} :\mathbb{R}^{n}\rightarrow \{ 0,1\}$ as given by \eqref{phi_n}. Then for $\epsilon _{n} < \frac{\overline{c}_{1i} \epsilon _{\alpha }^{\beta }}{M-1}$ and $n >( 2\varphi _K)^{5/2} \left(\frac{\overline{c}_{1i}}{\epsilon _{n}}\right)^{5/( 2\beta )}$,
\begin{equation*}
E_{f_{0}}[ \phi _{n}] \leq \exp\left( -\tilde{c}_K n^{4/5}\left(\frac{\epsilon _{n}}{\overline{c}_{1i}}\right)^{\frac{2}{\beta }\left( b_i\log\log\left(\frac{\overline{c}_{1i}}{\epsilon _{n}}\right) -b_i\log \beta +c_{2i}\right)}\right),
\end{equation*}
and
\begin{equation*}
\sup _{f\in \mathcal{F} :\ \Vert f_{i} -f_{i0}\Vert _{1}  >M\epsilon _{n}} E_{f}[ 1-\phi _{n}] \leq \exp\left( -\tilde{c}_K n^{4/5}\left(\frac{( M-1) \epsilon _{n}}{\overline{c}_{1i}}\right)^{\frac{2}{\beta }\left( b_i\log\log\left(\frac{\overline{c}_{1i}}{( M-1) \epsilon _{n}}\right) -b_i\log \beta +c_{2i}\right)}\right),
\end{equation*}
where $\varphi_K$, $\tilde{c}_K$,$\alpha$, $\beta$, $\overline{c}_{1i}$, $b_i$ and $c_{2i}$ are as defined in Lemma \ref{lemma: tail_bound_fi}.
\end{lemma}

\begin{proof}[Proof of Lemma \ref{lemma: E_phi}]
    By Lemma \ref{lemma: tail_bound_fi}, if we assume $\epsilon_n<\overline{c}_{1i}\epsilon_\alpha^\beta$, and $n>(2\varphi_K)^{5/2}(\frac{\overline{c}_{1i}}{\epsilon_n})^{5/(2\beta)}$, then
\begin{align*}
E_{f_{0}}[ \phi _{n}] & =P_{f_0}^n(\Vert \widehat{f}_{i} -f_{i0}\Vert _{1}  >\epsilon _{n})\\
 & \leq \exp\left( -\tilde{c}_K n^{4/5}\left(\frac{\epsilon _{n}}{\overline{c}_{1i}}\right)^{\frac{2}{\beta }\left( b_i\log\log\left(\frac{\overline{c}_{1i}}{\epsilon _{n}}\right) -b_i\log \beta +c_{2i}\right)}\right).
\end{align*}
For any $f\in \mathcal{F}$ such that $\Vert f_{i} -f_{i0}\Vert _{1}  >M\epsilon _{n}$, because
\begin{align*}
\Vert \widehat{f}_{i} -f_{i0}\Vert _{1} & =\Vert \widehat{f}_{i} -f_{i} +f_{i} -f_{i0}\Vert _{1}\\
 & \geq \Vert f_{i} -f_{i0}\Vert _{1} -\Vert \widehat{f}_{i} -f_{i}\Vert _{1},
\end{align*}
we have
\begin{align}
E_{f}[ 1-\phi _{n}] & =P_{f}^n(\Vert \widehat{f}_{i} -f_{i0}\Vert _{1} \leq \epsilon _{n})\notag\\
 & \leq P_{f}^n(\Vert f_{i} -f_{i0}\Vert _{1} -\Vert \widehat{f}_{i} -f_{i}\Vert _{1} \leq \epsilon _{n})\notag\\
 & =P_{f}^n(\Vert \widehat{f}_{i} -f_{i}\Vert _{1} \geq \Vert f_{i} -f_{i0}\Vert _{1} -\epsilon _{n})\notag\\
 & \leq P_{f}^n(\Vert \widehat{f}_{i} -f_{i}\Vert _{1} \geq ( M-1) \epsilon _{n})\notag\\
 & \leq \exp\left( -\tilde{c}_K n^{4/5}\left(\frac{( M-1) \epsilon _{n}}{\overline{c}_{1i}}\right)^{\frac{2}{\beta }\left( b_i\log\log\left(\frac{\overline{c}_{1i}}{( M-1) \epsilon _{n}}\right) -b_i\log \beta +c_{2i}\right)}\right) \label{ineq: alternative_bound},
\end{align}
if $(M-1)\epsilon_n<\overline{c}_{1i}\epsilon_\alpha^\beta$, and $n >( 2\varphi _{K})^{5/2}\left(\frac{\overline{c}_{1i}}{( M-1) \epsilon _{n}}\right)^{5/( 2\beta )}$. Note that (\ref{ineq: alternative_bound}) holds for all $f\in \mathcal{F}$ such that $\Vert f_{i} -f_{i0}\Vert _{1}  >M\epsilon _{n}$. Therefore we obtain for $i=1,2$, 
\begin{equation*}
\sup _{f\in \mathcal{F} :\ \Vert f_{i} -f_{i0}\Vert _{1}  >M\epsilon _{n}} E_{f}[ 1-\phi _{n}] \leq \exp\left( -\tilde{c}_K n^{4/5}\left(\frac{( M-1) \epsilon _{n}}{\overline{c}_{1i}}\right)^{\frac{2}{\beta }\left( b_i\log\log\left(\frac{\overline{c}_{1i}}{( M-1) \epsilon _{n}}\right) -b_i\log \beta +c_{2i}\right)}\right).
\end{equation*}

\end{proof}

\begin{proof}[Proof of Theorem \ref{thm:fi_post_rate}]
    Let $X_1,\cdots, X_n$ be $n$ i.i.d. samples from $P_{f_0}$. Let $\tilde{\epsilon }_{n}$ be a positive constant sequence that goes to zero as $n\rightarrow\infty$. Define the event $A_{n}$ as
    \begin{equation}\label{event_An}
A_{n} =\left\{(X_{1} ,\cdots ,X_{n})\ |\int \frac{p_f( X_{1} ,\cdots ,X_{n})}{p_{0}( X_{1} ,\cdots ,X_{n})} d\Pi( f) \geq e^{-( 2+D) n\tilde{\epsilon }_{n}^2}\right\},
\end{equation}
where $p_f( X_{1} ,\cdots ,X_{n})$ is the density of $X_1,\cdots, X_n$ under $P_f$, and $p_{0}( X_{1} ,\cdots ,X_{n})$ is the density of $X_1,\cdots, X_n$ under $P_{f_0}$. $D$ is a positive constant. Let $\phi_n$ be given by \eqref{phi_n}. Let $M>2$ be a constant. Then, the posterior probability of the set $\{f_{i} |\Vert f_{i} -f_{i0}\Vert _{1}  >M\epsilon _{n}\}$ can be expressed as follows
\begin{align*}
\Pi ^{( n)} & ( \| f_{i} -f_{i0} \| _{1}  >M\epsilon _{n} |X_{1} ,\cdots ,X_{n}) =\frac{\int _{f:\ \| f_{i} -f_{i0} \| _{1}  >M\epsilon _{n}} p_{f}( X_{1} ,\cdots ,X_{n}) d\Pi ( f)}{\int p_{f}( X_{1} ,\cdots ,X_{n}) d\Pi ( f)}\\
 & =( \phi _{n} +1-\phi _{n})\frac{\int _{f :\ \| f_{i} -f_{i0} \| _{1}  >M\epsilon _{n}}\frac{p_{f}( X_{1} ,\cdots ,X_{n})}{p_{0}( X_{1} ,\cdots ,X_{n})} d\Pi ( f)}{\int \frac{p_{f}( X_{1} ,\cdots ,X_{n})}{p_{0}( X_{1} ,\cdots ,X_{n})} d\Pi ( f)}\\
 & =( \phi _{n} +\mathbf{1}_{A_{n}^{c}} \cdot ( 1-\phi _{n}) +\mathbf{1}_{A_{n}} \cdot ( 1-\phi _{n}))\frac{\int _{f:\ \| f_{i} -f_{i0} \| _{1}  >M\epsilon _{n}}\frac{p_{f}( X_{1} ,\cdots ,X_{n})}{p_{0}( X_{1} ,\cdots ,X_{n})} d\Pi ( f)}{\int \frac{p_{f}( X_{1} ,\cdots ,X_{n})}{p_{0}( X_{1} ,\cdots ,X_{n})} d\Pi ( f)}\\
 & \leq \phi _{n} +\mathbf{1}_{A_{n}^{c}} +\mathbf{1}_{A_{n}} \cdot ( 1-\phi _{n})\frac{\int _{f:\ \| f_{i} -f_{i0} \| _{1}  >M\epsilon _{n}}\frac{p_{f}( X_{1} ,\cdots ,X_{n})}{p_{0}( X_{1} ,\cdots ,X_{n})} d\Pi ( f)}{\int \frac{p_{f}( X_{1} ,\cdots ,X_{n})}{p_{0}( X_{1} ,\cdots ,X_{n})} d\Pi ( f)}\\
 & \leq \phi _{n} +\mathbf{1}_{A_{n}^{c}} +e^{( 2+D) n\tilde{\epsilon }_{n}^{2}}( 1-\phi _{n})\int _{f:\ \| f_{i} -f_{i0} \| _{1}  >M\epsilon _{n}}\frac{p_{f}( X_{1} ,\cdots ,X_{n})}{p_{0}( X_{1} ,\cdots ,X_{n})} d\Pi ( f)
\end{align*}

The last inequality is due to the definition of the event $A_n$. It follows that
\begin{align*}
 & E_{f_{0}}[ \Pi^{(n)}(\Vert f_{i} - f_{i0}\Vert _{1}  >M\epsilon _{n} |X_{1} ,\cdots ,X_{n})]\\
\leq E_{f_{0}} & [ \phi _{n}] +E_{f_{0}}\left[\mathbf{1}_{A_{n}^{c}}\right] +e^{( 2+D) n\tilde{\epsilon }_{n}^{2}} E_{f_{0}}\left[( 1-\phi _{n})\int _{f:\Vert f_{i} -f_{i0}\Vert _{1}  >M\epsilon _{n}}\frac{p_f( X_{1} ,\cdots X_{n})}{p_{0}( X_{1} ,\cdots X_{n})} d\Pi( f)\right]
\end{align*}
By Lemma \ref{lemma: E_phi}, we have if $\epsilon _{n} < \frac{\overline{c}_{1i} \epsilon _{\alpha }^{\beta }}{M-1}$ and $n >( 2\varphi _{K})^{5/2} \left(\frac{\overline{c}_{1i}}{\epsilon _{n}}\right)^{5/( 2\beta )}$, then
\begin{equation}\label{term1a}
E_{f_{0}}[ \phi _{n}] \leq \exp\left( -\tilde{c}_K n^{4/5}\left(\frac{\epsilon _{n}}{\overline{c}_{1i}}\right)^{\frac{2}{\beta }\left( b_i\log\log\left(\frac{\overline{c}_{1i}}{\epsilon _{n}}\right) -b_i\log \beta +c_{2i}\right)}\right).
\end{equation}
Furthermore, by Lemma 8.1 in \cite{ghosal2000convergence}, and a similar argument used in Lemma 8.10 of \cite{ghosal2017fundamentals}, for any $\epsilon>0$, and any $C>0$,
\begin{equation*}
P_{f_{0}}^{n}\left(\int \frac{p_f( X_{1} ,\cdots X_{n})}{p_{0}( X_{1} ,\cdots X_{n})} d\Pi ( f)  \geq\Pi ( B_{2}( f_{0} ,\epsilon )) e^{-( 1+C) n\epsilon ^{2}}\right) \geq 1-\frac{1}{C^{2} n\epsilon ^{2}}.
\end{equation*}
By Lemma \ref{lemma: mix_of_dp_prior_conctr}, if we take $\epsilon=\tilde{\epsilon}_n=\log n/\sqrt{n}$, when $n$ is larger than a constant threshold, we have $\Pi \left( B_{2}\left( f_{0} ,\tilde{\epsilon }_{n}\right)\right) \geq c_{3} e^{-c_{4} n\tilde{\epsilon }_{n}^{2}}$, therefore
\begin{equation*}
P_{f_{0}}^{n}\left(\int \frac{p_f( X_{1} ,\cdots X_{n})}{p_{0}( X_{1} ,\cdots X_{n})} d\Pi ( f)  \geq c_{3} e^{-c_{4} n\tilde{\epsilon }_{n}^{2}} e^{-( 1+C) n\tilde{\epsilon }_{n}^{2}}\right) \geq 1-\frac{1}{C^{2} n\tilde{\epsilon }_{n}^{2}}.
\end{equation*}
Take $C=1$, we can find some $D>1$ such that
\begin{equation*}
P_{f_{0}}^{n}\left(\int \frac{p_f( X_{1} ,\cdots X_{n})}{p_{0}( X_{1} ,\cdots X_{n})} d\Pi ( f)  \geq e^{-( 2+D) n\tilde{\epsilon }_{n}^{2}}\right) \geq 1-\frac{1}{n\tilde{\epsilon }_{n}^{2}}.
\end{equation*}
Recall the definition of $A_{n}$ in (\ref{event_An}), then we obtain
\begin{equation}\label{term2a}
E_{f_{0}}\left[\mathbf{1}_{A_{n}^{c}}\right] \leq \frac{1}{n\tilde{\epsilon }_{n}^{2}} =\frac{1}{(\log n)^{2}}\rightarrow  0,\ \text{as} \ n\rightarrow \infty. 
\end{equation}
Finally, by Lemma \ref{lemma: E_phi},
\begin{align}
 & E_{f_{0}}\left[( 1-\phi _{n})\int _{f :\Vert f_{i} -f_{i0}\Vert _{1}  >M\epsilon _{n}}\frac{p_{f}( X_{1} ,\cdots X_{n})}{p_{0}( X_{1} ,\cdots X_{n})} d\Pi ( f)\right]\notag\\
\overset{\text{Fubini}}{=} & \int _{f :\Vert f_{i} -f_{i0}\Vert _{1}  >M\epsilon _{n}}\left(\int p_{0}( X_{1} ,\cdots X_{n}) \cdot \frac{p_{f}( X_{1} ,\cdots X_{n})}{p_{0}( X_{1} ,\cdots X_{n})}( 1-\phi _{n}) dX_{1:n}\right) d\Pi ( f)\notag\\
= & \int _{f :\Vert f_{i} -f_{i0}\Vert _{1}  >M\epsilon _{n}} E_{f}[ 1-\phi _{n}] d\Pi ( f)\notag\\
\leq  & \exp\left( -\tilde{c}_{K} n^{4/5}\left(\frac{( M-1) \epsilon _{n}}{\overline{c}_{1i}}\right)^{\frac{2}{\beta }\left( b_{i}\log\log\frac{\overline{c}_{1i}}{( M-1) \epsilon _{n}} -b_{i}\log \beta +c_{2i}\right)}\right).\label{term3}
\end{align}

Combining \eqref{term1a}, \eqref{term2a} and \eqref{term3}, 
\begin{align*}
E_{f_{0}}[ \Pi^{(n)}( & \Vert f_{i} - f_{i0}\Vert _{1}  >M\epsilon _{n} |X_{1} ,\cdots ,X_{n})]\\
\leq  & \exp\left( -\tilde{c}_{K} n^{4/5}\left(\frac{\epsilon _{n}}{\overline{c}_{1i}}\right)^{\frac{2}{\beta }\left( b_i\log\log\frac{\overline{c}_{1i}}{\epsilon _{n}} -b_i\log \beta +c_{2i}\right)}\right) +\frac{1}{(\log n)^{2}}\\
 & +\ \exp\left(( 2+D) n\tilde{\epsilon }_{n}^{2} -\tilde{c}_{K} n^{4/5}\left(\frac{( M-1) \epsilon _{n}}{\overline{c}_{1i}}\right)^{\frac{2}{\beta }\left( b_i\log\log\frac{\overline{c}_{1i}}{( M-1) \epsilon _{n}} -b_i\log \beta +c_{2i}\right)}\right).
\end{align*}
For the the first and third term to go to zero as $n\rightarrow \infty$, there must exist a constant $C_M>(2+D)/\tilde{c}_{K}$ such that when $n$ is large enough,
\begin{equation*}
n^{4/5}\left(\frac{( M-1) \epsilon _{n}}{\overline{c}_{1i}}\right)^{\frac{2}{\beta }\left( b_i\log\log\frac{\overline{c}_{1i}}{( M-1) \epsilon _{n}} -b_i\log \beta +c_{2i}\right)} \geq C_{M }n\tilde{\epsilon }_{n}^{2} .
\end{equation*}
Take logarithm on both sides, we obtain
\begin{equation*}
\frac{2}{\beta }\left( b_i\log\log\frac{\overline{c}_{1i}}{( M-1) \epsilon _{n}} -b_i\log \beta +c_{2i}\right) \cdot \left(\log\frac{\overline{c}_{1i}}{( M-1) \epsilon _{n}}\right) \leq \frac{4}{5}\log n-2\log\log n-\log C_{M }.
\end{equation*}
Denote $x=\frac{\overline{c}_{1i}}{(M-1)\epsilon_n}$. When $\epsilon _{n} < \frac{\overline{c}_{1i}}{( M-1) e^{e}}$, we have $x >e^{e}$, and hence $\log\log x >1$. Therefore it suffices to solve the following

\begin{align*}
\frac{2}{\beta }\left(b_i -b_i\log \beta +c_{2i}\right) \cdot \left(\log\log\frac{\overline{c}_{1i}}{( M-1) \epsilon _{n}}\right) \cdot \left(\log\frac{\overline{c}_{1i}}{( M-1) \epsilon _{n}}\right) & \leq \frac{4}{5}\log n- 2\log\log n -\log C_M.
\end{align*}
Let $c_{\beta ,i} =2( b_i-b_i \log \beta +c_{2i}) /\beta $, which is positive by Lemma \ref{lemma: tail_bound_fi}, and let $y=\log\log x$, then we obtain
\begin{equation}\label{eq_y_exp_y}
ye^{y} \leq \frac{4}{5c_{\beta ,i}}\log n-\frac{2}{c_{\beta ,i}}\log\log n-\frac{1}{c_{\beta,i}}\log C_M,\ y >0.
\end{equation}
The function $W(z)$, defined as the solution to $W(z)e^{W(z)}=z$, is known as the Lambert W function (see, e.g., \cite{corless1996lambert}). We consider the portion of the principle branch of $W(z)$ where $W(z)$ is real and positive, then (\ref{eq_y_exp_y}) is equivalent to:
\begin{equation}\label{sol_lambert}
y\leq W\left(\frac{4}{5c_{\beta ,i}}\log n-\frac{2}{c_{\beta ,i}}\log\log n - \frac{1}{c_{\beta,i}}\log C_M\right).
\end{equation}
From equation (4.19) in \cite{corless1996lambert}, we know as $z\rightarrow\infty$,
\begin{equation*}
W( z) \geq\log z-\log\log z+\frac{\log\log z}{2\log z},
\end{equation*}
so applying this lower bound to the RHS of (\ref{sol_lambert}) and recalling that $y=\log \log x$, it is sufficient to have
\begin{align*}
\log\log x & \leq \log\frac{z}{\log z} +\frac{\log\log z}{2\log z}\\
\log x & \leq \exp\left(\log\frac{z}{\log z}\right) \cdot \exp\left(\frac{\log\log z}{2\log z}\right)\\
 & =\frac{z}{\log z} \cdot (\log z)^{\frac{1}{2\log z}}\\
x & \leq \exp( z)^{(\log z)^{\frac{1}{2\log z}} \cdot \frac{1}{\log z}}.
\end{align*}
Because $(\log z)^{\frac{1}{2\log z}}\downarrow 1$ as $z\rightarrow \infty$, it suffices to have that $x\leq \exp( z)^{\frac{1}{\log z}}$. Now plug in $z$ and $x$. Because there must exist some constant $c_{M,\beta,i}>1$ such that $z<c_{M,\beta,i}\log n$, we have $\log z< c_{i1}\log\log n$ for some constant $c_{i1}>0$, so it is sufficient to have
\begin{align*}
\frac{\overline{c}_{1i}}{( M-1) \epsilon _{n}} & \leq \left( n^{\frac{4}{5c_{\beta ,i}}} \cdot (\log n)^{-\frac{2}{c_{\beta ,i}}} \cdot C_{M}^{-\frac{1}{c_{\beta ,i}}}\right)^{\frac{1}{c_{i1}\log\log n}}\\
 & =n^{\frac{4}{5c_{\beta ,i} c_{i1}\log\log n}} \cdot (\log n){^{\frac{1}{\log\log n} \cdot \left( -\frac{2}{c_{\beta ,i} c_{i1}}\right)}} \cdot C_{M}^{-\frac{1}{c_{\beta ,i} c_{i1}\log\log n}}.
\end{align*}
Observe that $(\log n)^{\frac{1}{\log\log n}} =\exp\left(\frac{1}{\log\log n} \cdot \log\log n\right) =e$. Moreover, since $C_{M}^{-\frac{1}{c_{\beta ,i} c_{i1}\log\log n}} \rightarrow 1$ as $n\rightarrow \infty$, there exists a constant $0<c_{i2}<1$ such that $C_{M}^{-\frac{1}{c_{\beta ,i} c_{i1}\log\log n}} >c_{i2}$ for all $n>10$. Consequently, it suffices to solve
\begin{align*}
\frac{\overline{c}_{1i}}{( M-1) \epsilon _{n}} & \leq c_{i2} n^{\frac{c_{i1}}{\log\log n}}\\
\epsilon _{n} & \geq c_{i2} \cdot n^{-\frac{c_{i1}}{\log\log n}},
\end{align*}
for some positive constants $c_{i1}$ and $c_{i2}$.

\end{proof}

\end{spacing}

\end{document}